\documentclass[preprint,12pt]{elsarticle}

\usepackage{amssymb}
\usepackage{bm}

\journal{Nuclear Physics A}

\newcommand{\nn}\nonumber

\begin{document}

\begin{frontmatter}



\title{Comprehensive application of a coupled-channel complex scaling method 
to the $\bar{K}N$-$\pi Y$ system}


\author[KEK,JPARCb]{A. Dot\'e}

\author[NU]{T. Inoue}

\author[OIT]{T. Myo}

\address[KEK]{KEK Theory Center,
Institute of Particle and Nuclear Studies (IPNS), 
High Energy Accelerator Research Organization (KEK), 1-1 Oho, Tsukuba, Ibaraki, 305-0801, Japan}

\address[JPARCb]{J-PARC Branch, KEK Theory Center, IPNS, KEK,
203-1, Shirakata, Tokai, Ibaraki, 319-1106, Japan}

\address[NU]{Nihon University, College of Bioresource Sciences, Fujisawa 252-0880, Japan}

\address[OIT]{Osaka Institute of Technology, Osaka, Osaka 535-8585, Japan}

\begin{abstract}
We have applied the coupled-channel complex scaling method (ccCSM) to $\bar{K}N$-$\pi Y$ system.
One advantage of ccCSM is that resonant states as well as scattering states can be treated in the same framework. 
For the interactions in the system, 
we have constructed a meson-baryon potential-matrix by basing on the chiral SU(3) theory
and respecting the $\bar{K}N$ scattering length obtained in the Martin's analysis.
For future purpose to apply it more complicated system such as $\bar{K}NN$, 
we adopt a local Gaussian form in the $r$-space. 
We have investigated both the non-relativistic (NR) and 
the semi-relativistic (SR) kinematics.
In the SR case, two types of the potentials are obtained. 
To test the constructed potentials,
we have calculated scattering amplitudes and searched resonances. 
One resonance pole, corresponding to $\Lambda(1405)$, is found 
in isospin $I=0$ system
around (1419, $-$20) MeV ((1425, $-$25) or (1419, $-$13) MeV) 
on complex-energy plane with the NR (SR) kinematics. 
Mean distance between meson and baryon in the resonant state 
is $1.3 - i0.3$ fm ($1.2 - i0.5$ fm) for NR (SR), 
in which the states are treated as Gamow states.  
In addition, we have observed a signature of another pole 
in lower-energy region involving large decay width, 
although they are unstable against the change of scaling angle $\theta$. 
This may correspond to the lower pole of the double-pole of $\Lambda(1405)$ discussed in literature to date. 

\end{abstract}


\end{frontmatter}


\section{Introduction}
\label{}

$\bar{K}$-nuclear system has been a hot topic in nuclear and hadron 
physics for a long time.
Due to strongly attractive $\bar{K}N$ interaction in isospin $I=0$ channel, 
finite nuclear systems with anti-kaons are expected to have exotic 
properties such as deeply bound and quasi-stable states with high density 
\cite{AY_2002, AMDK}.
Such kaonic nuclei have been investigated with various many-body 
treatments \cite{AMDK, RMF:Mares, RMF:Muto}. 
In particular, to clarify the property of kaonic nuclei, 
great efforts have been devoted to investigate $K^-pp$\footnote
{Actually, this system is 
a $\bar{K}NN$-$\pi YN$ coupled system with quantum numbers $J^\pi=0^-$, 
$(T,T_z)=(1/2,1/2)$. It is expressed symbolically as $K^-pp$.}, 
a prototype of kaonic nuclei,  
in both of theoretical and experimental studies. 
From the theoretical studies, it has been claimed that 
$K^-pp$ will not be so deeply bound and its decay width will be large 
(total binding energy $<$ 100 MeV and decay width $>$ 50 MeV) 
\cite{AY_2002, Faddeev:Ikeda, Faddeev:Shevchenko, Kpp:DHW, Kpp:Wycech, Kpp:Barnea}, 
although there are quantitative discrepancies between calculations \cite{SummaryKpp}. 
On the other hand, experimental results indicate 
deeper binding of $K^-pp$ than theoretical predictions 
if the observed state is the bound $K^-pp$ 
\cite{Kpp:exp_FINUDA, Kpp:exp_DISTO}, although there are 
some objections to the experimental result \cite{Kpp_Criticism:Magas}. 
Thus, the consensus for $K^-pp$ has not been achieved yet. 


For study of kaonic nuclei, 
$\bar{K}N$ (involving $\pi Y$) interaction is a basic input, 
and a $\Lambda(1405)$ resonance is an essential building block because 
it can be reasonably interpreted as a quasi-bound $I=0$ state 
of $\bar{K}N$ with $s$-wave, rather than a three-quark state \cite{QM:Isgur}. 
One approach to $\bar{K}N$ system is a study based on the chiral SU(3) 
coupled-channel dynamics \cite{ChU:KSW}. 
This approach (called the chiral unitary model by Oset and Ramos 
\cite{ChU:OR}) has been succeeded in studies of $s$-wave meson-baryon 
systems including $S=-1$ sector \cite{ChU:Review}.
The double pole nature of $\Lambda(1405)$ pointed out 
within this model, is interesting \cite{ChU:Jido, ChU:HW}. 
According to further studies along this model,
some experimental data seem to support the double-pole nature 
\cite{ChU_DP:Magas_Sekihara_Hyodo}. 
Ref. \cite{L1405:Shevchenko}, however, shows that the current 
low-energy observables are not so precise to distinguish 
whether $\Lambda(1405)$ is single pole or double pole. 
Thus, the $\Lambda(1405)$-pole problem has not been solved yet. 
Recently, accurate data at the $\bar{K}N$ threshold are given by 
a precise measurement of $1s$ level shift of kaonic hydrogen 
atom \cite{KpX:SIDDHARTA}, in addition to experimental data 
on $\bar{K}N$ subthreshold region \cite{Exp:L1405} and 
old data on $K^-p$ branching ratio \cite{Exp:K-p-branch}. 
Due to such precise data, the physical 
quantities near the $\bar{K}N$ threshold are strictly constrained and 
those uncertainties below the $\bar{K}N$ threshold are also 
expected to be decreased \cite{ChU:IHW}.  

In such a current situation, 
we start a study of kaonic nuclei with a coupled-channel complex 
scaling method (ccCSM), 
keeping in mind the following three points: 
1. Simple and adequate treatment of resonant states in many-body system, 
2. Explicit inclusion of all channels in a coupled-channel problem, and 
3. Accessible to structure of kaonic nuclei. 
Looking back past studies on $K^-pp$, in variational studies 
\cite{AY_2002, Kpp:DHW} it is treated as a bound state in $\bar{K}NN$ channel 
as a consequence of the elimination of $\pi Y$ channels. 
In Faddeev-AGS studies \cite{Faddeev:Ikeda, Faddeev:Shevchenko} 
certainly a coupled-channel calculation is fully performed and resonance poles 
are searched. However, there 
a separable form is assumed to the $\bar{K}N$-$\pi Y$ potential 
which is a key ingredient in the $\bar{K}$-nuclear study, and 
a wave function is not obtained explicitly in this approach 
though it is important to investigate the nature of kaonic nuclei. 
Thus, each approach involves advantages and disadvantages. 
Since the ccCSM is expected to overcome above disadvantages, 
we employ this method to study the $\bar{K}$-nuclear system. 

The complex scaling method (CSM) has been applied 
to various nuclear physics, and it has been greatly 
succeeded in particular in the study of resonant states 
of unstable nuclei \cite{CSM:Myo}. 
The CSM is a practical tool for the study of nuclear many-body systems. 
Indeed, a resonance nature in unstable $^8$He is revealed with the CSM in 
Ref. \cite{CSM-8He:Myo} where  a five-body system of $^4{\rm He}+n+n+n+n$ 
is solved. 
The CSM is suitable for resonant states, 
because we can handle them in the same way as bound states. 
Though a resonant wave function is originally divergent 
at infinite distance, in the CSM it is transformed 
to a square-integrable function by a complex rotation 
for the coordinate and then it can be represented with 
{\it e.g.} a Gaussian base which is familiar for ones study bound states. 
In addition, by an advanced used of complex-rotated wave functions, 
the scattering amplitude can also be calculated with 
the Gaussian base \cite{CSM:Kruppa}. 
Thus, all of bound, resonant and scattering 
states can be handled in a single framework of the CSM 
with Gaussian base.

Since it is our first attempt applying the ccCSM to $\bar{K}$-nuclear systems, 
in this article we investigate the $s$-wave two-body system of 
$\bar{K}N$-$\pi Y$ coupled channels. We examine semi-relativistic 
kinematics as well as non-relativistic one to be careful of 
a pion which is a light-mass particle. 
First, we will check how the ccCSM works in the present system. 
Then, we will construct a meson-baryon potential for the $\bar{K}N$-$\pi Y$ 
coupled system, based on a chiral SU(3) theory. 
We adopt a Gaussian-form potential in the coordinate space, 
because of convenience for our further study of $\bar{K}$-nuclear system 
with Gaussian base. Our potential is constrained 
by the $\bar{K}N$ scattering lengths for both isospin states obtained 
by the Martin's analysis of old $\bar{K}N$ scattering data 
\cite{Exp:ADMartin}. 
Using the constructed potential, we will investigate 
the behavior of the scattering amplitude. 
In the $I=0$ sector, the pole on the complex-energy plane 
also will be investigated in detail 
because there should be a resonance corresponding to the $\Lambda(1405)$. 

The contents of the article are as follows: 
In the section 2, we will explain a meson-baryon potential used 
in our study and the formalism of ccCSM for the study of scattering 
state as well as resonant state in detail. In the section 3, 
the obtained results will be given. The scattering amplitudes will 
be shown for both isospin states. For the $I=0$ channel, the structure 
of a resonant state will be investigated. In the last section, 
we will mention our summary and future plans including some discussions.

\section{Formalism}
\label{}

\subsection{Kinematics and interaction}
\label{Sec:Kin_Int}

We are considering a $\bar{K}N$-$\pi Y$ coupled system in $s$-wave, where 
$Y$ indicates a hyperon which is $\Sigma$ ($\Lambda$ and $\Sigma$) for $I=0$ ($I=1$) case. We investigate such a two-body system 
in semi-relativistic kinematics as well as non-relativistic one, 
because a pion joins in our calculation and its mass is very small.  
The Hamiltonian for non-relativistic kinematics is 
\begin{eqnarray}
\hat{H}=\sum_{\alpha}  
\left( M_\alpha+m_\alpha \, + \, \frac{\hat{\bm{p}}_\alpha^2}{2\mu_\alpha}\right) 
|\alpha\rangle \langle \alpha| 
\; + \; \hat{V}^{NR}_{MB}, \label{Ham} 
\end{eqnarray} 
and that for semi-relativistic kinematics is 
\begin{eqnarray}
\hat{H}=\sum_{\alpha}  
\left( \sqrt{M_\alpha^2 + \hat{\bm{p}}_\alpha^2} 
\, + \, \sqrt{m_\alpha^2 + \hat{\bm{p}}_\alpha^2}\right) 
|\alpha\rangle \langle \alpha| 
\; + \; \hat{V}_{MB}^{SR}. \label{HamSR} 
\end{eqnarray} 
Here $M_\alpha$, $m_\alpha$ and $\mu_\alpha$ are baryon, 
meson and a reduced mass in the channel $\alpha$, respectively. 
$\hat{\bm{p}}_\alpha$ is the relative momentum between 
a meson and a baryon in the channel $\alpha$. 

The last term $\hat{V}_{MB}^{NR}$ ($\hat{V}_{MB}^{SR}$) represents 
a meson-baryon potential for non-relativistic (semi-relativistic) 
kinematics. 
In Ref. \cite{ChU:KSW}, Kaiser, Siegel and Weise (KSW) proposed 
a pseudo-potential for $s$-wave meson-baryon system with $S=-1$
derived from an effective chiral SU(3) Lagrangian,
and used it in the Lippmann-Schwinger equation to obtain $\Lambda(1405)$. 
The pseudo-potential consists of all terms up to order $q^2$ allowed by chiral symmetry. 
They introduced two kinds of cut-off; 
one is projection to a Yukawa form local potential in the $r$-space
and another is casting to a separable potential in the $p$-space. 
In this paper, we follow KSW work and adopt the leading Weinberg-Tomozawa term of the pseudo-potential:
\begin{eqnarray}
\hat{V}_{MB}^{SR} & = & \sum_{\alpha, \beta} 
\; - \frac{C^I_{\alpha \beta}}{8 f_\pi^2} (\omega_\alpha+\omega_\beta) 
\sqrt{\frac{M_\alpha M_\beta}{s\, \tilde{\omega}_\alpha \tilde{\omega}_\beta}}
\; g^I_{\alpha \beta} (r) 
\; |\alpha\rangle \langle \beta |, \label{PotMB_SR}
\\
& & g^I_{\alpha \beta} (r) = (\sqrt{\pi} \, d_{\alpha \beta}^{I})^{-3} \; \exp[-(r/d^I_{\alpha \beta})^2].  
\end{eqnarray} 
For a cut-off form factor, we introduce the local Gaussian function $g^I_{\alpha \beta} (r)$
in $r$-space with a range parameter $d^I_{\alpha \beta}$. 
Hereafter, we denote this potential 
as {\it ``KSW-type potential''}. 
Note that this potential is energy dependent because meson energy 
$\omega_\alpha$, baryon energy $E_\alpha$ and the reduced energy 
$\tilde{\omega}_\alpha = \omega_\alpha E_\alpha/(\omega_\alpha + E_\alpha )$ 
are given by function of CM energy $\sqrt{s}$ as follows: 
\begin{eqnarray}
E_\alpha = \frac{(\sqrt{s})^2 -m_\alpha^2 +M_\alpha^2}{2\sqrt{s}} 
\quad {\rm and} \quad \omega_\alpha = \frac{(\sqrt{s})^2 +m_\alpha^2 -M_\alpha^2}{2\sqrt{s}}.  
\end{eqnarray}
Note also that structure of this potential is determined by 
the coefficients $\{ C_{\alpha \beta}^{I} \}$ 
which are computed by Clebsch-Gordan coefficients of SU(3) and given as 
\begin{eqnarray}
C^{(I=0)} = \bordermatrix{
 & \bar{K}N & \pi\Sigma   \cr
 & 3        & -\sqrt{3/2} \cr
 &          & 4           \cr
}
, \quad
C^{(I=1)} = \bordermatrix{
 & \bar{K}N & \pi\Sigma   & \pi\Lambda  \cr
 & 1        & -1          & -\sqrt{3/2} \cr
 &          & 2           & 0           \cr
 &          &             & 0           \cr
}. \label{Eq:SU(3)CG}
\end{eqnarray} 
Strength of this potential depends on the pion decay constant $f_{\pi}$.
It is noted that 
the range parameter $d^I_{\alpha \beta}$ can be a different value 
in each channel set $(\alpha, \beta)$. For the range parameter 
of the transition potential, we assume that 
$d^I_{\alpha \beta} = (d^I_{\alpha \alpha}+d^I_{\beta \beta})/2$ 
to reduce the number of parameters.

Since the flux factor $\sqrt{\frac{M_\alpha M_\beta}{s\, \tilde{\omega}_\alpha \tilde{\omega}_\beta}}$ in Eq. (\ref{PotMB_SR}) is based on the relativistic kinematics, the KSW-type potential had better be used in the semi-relativistic framework in our calculation. When we adopt the non-relativistic kinematics, we make non-relativistic reduction to the flux factor 
in two prescriptions, denoted as ``NRv1'' and ``NRv2''. 
In the first prescription, comparing the expression of the differential 
cross section in both kinematics, we replace the reduced energy 
in Eq. (\ref{PotMB_SR}) with the reduced mass; 
 \begin{eqnarray}
{\rm NRv1: } \quad 
\hat{V}_{MB}^{NRv1}  =  \sum_{\alpha, \beta} 
\; - \frac{C^I_{\alpha \beta}}{8 f_\pi^2} (\omega_\alpha+\omega_\beta) 
\sqrt{\frac{M_\alpha M_\beta}{s \; \mu_\alpha \mu_\beta}}
\; g^I_{\alpha \beta} (r) 
\; |\alpha\rangle \langle \beta |. \label{PotMB_NRv1}
\end{eqnarray} 
In the second prescription, by considering the small momentum limit, 
we replace the meson and baryon energies with the meson and baryon masses, 
respectively; 
\begin{eqnarray}
{\rm NRv2: } \quad 
\hat{V}_{MB}^{NRv2}  =  \sum_{\alpha, \beta} 
\; - \frac{C^I_{\alpha \beta}}{8 f_\pi^2} (\omega_\alpha+\omega_\beta) 
\sqrt{\frac{1}{m_\alpha m_\beta}}
\; g^I_{\alpha \beta} (r) 
\; |\alpha\rangle \langle \beta |. \label{PotMB_NRv2}
\end{eqnarray} 
In both non-relativistic approximations, 
we keep remaining the meson-energy part $(\omega_\alpha+\omega_\beta)$ 
as original, because this energy dependence is attributed 
to the chiral dynamics which we respect in our study. 
Of course, energies such as $\sqrt{s}$ and $\omega_\alpha$ themselves, 
are calculated by the non-relativistic formula. 

We remark the normalization of the pseudo-potentials used in this article.
All pseudo-potentials, given in Eqs. (\ref{PotMB_SR}, \ref{PotMB_NRv1},
\ref{PotMB_NRv2}),
are normalized so that we obtain $3\mu/(8\pi f_\pi^2)$ for the scattering
length of isoscalar $\bar{K}N$
in the Born approximation and the formula of scattering amplitude 
(\ref{Eq:flcc0})
which will be explained in the next section. Here, $\mu$ is the reduced
mass of anti-kaon and a nucleon.

\subsection{Coupled-channel complex scaling method}
\label{CSM:usual}

As mentioned in the introduction, the $s$-wave $\bar{K}N$-$\pi\Sigma$ 
system has a resonant state in $I=0$ channel which corresponds to 
the $\Lambda(1405)$. Such a resonant state of a meson-baryon system 
can be investigated with the complex scaling method (CSM), in the same 
way as resonant states of unstable nuclei.
Here, we give a brief explanation of the coupled-channel 
complex scaling method (ccCSM). 
Details of the CSM and its successful application to 
unstable nuclear physics are summarized in Ref. \cite{CSM:Myo}

In the CSM, a relative coordinate ${\bm r}$ and a conjugate 
wave number ${\bm k}$ 
in Hamiltonian $\hat{H}$ and a wave function $|\Phi\rangle$ are 
complex-scaled as 
\begin{eqnarray}
U(\theta): {\bm r} \rightarrow {\bm r} e^{i\theta}, \; 
{\bm k} \rightarrow {\bm k} e^{-i\theta}.   
\end{eqnarray}
Then, transformed Hamiltonian and wave function are defined as 
$\hat{H}_\theta \equiv U(\theta) \hat{H} U^{-1}(\theta)$ and  
$|\Phi_\theta \rangle \equiv U(\theta) |\Phi \rangle$, respectively. 
We expand the wave function $\Phi({\bm r})$ in partial waves as 
$\Phi ({\bm r}) = \sum_{lm} \phi_l(r)/r \, Y_{lm} (\Omega)$ as usual. 
In our definition, all radial wave functions are transformed 
by the complex scaling operator as 
\begin{eqnarray}
\phi_l^\theta (r) \equiv U(\theta) \phi_l(r) 
= e^{i\theta/2} \phi_l(re^{i\theta}), \label{Eq:CSforWfnc}
\end{eqnarray} 
taking into account the Jacobian in integration to calculate 
expectation value of operators.
An expectation value of operator $\hat{O}$ is calculated 
with bi-orthogonal set $\{\tilde{\Phi}_\theta, {\Phi}_\theta\}$: 
\begin{eqnarray}
\langle \hat{O} \rangle_\theta \equiv \langle \tilde{\Phi}_\theta | \hat{O}_\theta | \Phi_\theta \rangle, 
\end{eqnarray}
where $\tilde{\Phi}_\theta (r) = \Phi^*_\theta (r)$ in terms of 
radial part of wave function for bound and resonant states. 
The complex scaled wave function is generally normalized as 
$\langle \tilde{\Phi}_\theta | \Phi_\theta \rangle =1$. 
With the definition of the bi-orthogonal state, the radial part of 
the complex scaled wave function is normalized as 
\begin{eqnarray}
\sum_l\int^\infty_0 dr \; 
 \left\{\phi^{\theta}_{l}(r) \right\}^2  = 1 \label{Eq:Norm_CSwfnc}
\end{eqnarray}
for bound and resonant states \cite{Resonance:Peierls}. 

It is known that energies of bound and resonant states 
are independent of the scaling angle $\theta$ 
while those of continuum states vary with $\theta$ as 
$\frac{{\bm k}^2 }{2m} e^{-2i\theta}$ in case of 
non-relativistic kinematics ({\it ABC theorem} \cite{CSM:Myo}). 
In addition, it is easily understood that a resonant wave function 
is transformed from a divergent function to a damped function 
by the complex scaling $U(\theta)$ with adequate values of $\theta$. 
In other words, the boundary condition for resonant states 
is modified to the same one for bound states.   
Therefore, we can obtain resonant states as follows: 
For various $\theta$'s, complex eigenvalues are calculated 
by diagonalizing the complex-scaled Hamiltonian $\hat{H}_\theta$ 
with Gaussian base as done in usual studies of bound states. 
Among obtained eigenstates, 
the states with the eigenvalues independent of $\theta$ 
are recognized as resonant states. Since the continuum states appear 
along a line $\tan^{-1} ({\rm Im} \, E \, / \, {\rm Re} \, E) = -2\theta$ 
on the complex-energy plane, resonant states can be separated 
from the continuum states when we set the scaling angle 
$\theta$ appropriately. 
Similarly, resonant states can be found also in case 
of semi-relativistic kinematics, although complex eigenvalues 
of continuum states have different $\theta$ dependence 
from non-relativistic case. 

%
%

In the present study, the CSM is applied to a coupled-channel 
problem, because the $I=0$ wave function contains the $\bar{K}N$ and 
$\pi\Sigma$ components. 
The spatial parts of these wave functions with $s$-wave, 
$\phi_{\bar{K}N}(r)$ and $\phi_{\pi\Sigma}(r)$, 
are expanded with Gaussian base $G^\alpha_n(r)$: 
\begin{eqnarray}
| \Phi^{I=0, l=0}_{\bar{K}N-\pi\Sigma} \rangle 
& = & \frac{1}{r}\phi^{l=0}_{\bar{K}N}(r) \, Y_{00}(\Omega) | \bar{K}N \rangle  \; + \;  
\frac{1}{r} \phi^{l=0}_{\pi\Sigma}(r) \, Y_{00}(\Omega) | \pi\Sigma \rangle 
\label{Eq:I=0wfnc}\\
\phi^{l=0}_{\alpha}(r) & = & \sum_j C^\alpha_j G_j(r) =
\sum_j C^\alpha_j \; N_{l=0}(b_j) \,r \exp[-r^2/2b^2_j],  \label{Eq:GaussE_CSM}
\end{eqnarray}
where coefficients $\{C^\alpha_n\}$ are complex parameters to be determined. 
As explained above, by diagonalizing the $\hat{H}_\theta$ with 
the basis $\{ G^\alpha_n(r) |\alpha \rangle \}$ 
and investigating the $\theta$ dependence of eigenvalues, 
we can find resonant states. As explained in Eq. (\ref{Eq:Norm_CSwfnc}), 
the radial part of the $I=0$ wave function (Eq. (\ref{Eq:I=0wfnc})) 
is normalized as 
\begin{eqnarray}
\int^\infty_0 dr \; 
\left[
\left\{\phi^{l=0}_{\bar{K}N, \,\theta}(r) \right\}^2 + 
\left\{\phi^{l=0}_{\pi\Sigma, \,\theta}(r) \right\}^2 
\right]
= 1,
\end{eqnarray}
when it is complex-scaled. 

Note that the self-consistency for the complex energy 
is needed to be considered, since the meson-baryon potentials 
(Eqs. (\ref{PotMB_SR}), (\ref{PotMB_NRv1}) and (\ref{PotMB_NRv2})) 
have an energy dependence attributed to chiral dynamics. 
The eigen-energy $E_{calc}$ of Hamiltonian calculated with the ccCSM should 
coincide with the energy $E_{int}$ input of the meson-baryon potential 
$\hat{V}_{MB} (\sqrt{s}=E_{int})$. Here $E_{calc}$ and $E_{int}$ are complex 
values because they are the energies of resonant states, $E_R-i\Gamma/2$. 
($E_R$ and $\Gamma$ are energy and decay width, respectively.) 
At the $n$-th iteration, we set a value of $E^{(n)}_{int}$ 
as an input for $\hat{V}_{MB}$ and then obtain an eigen-energy 
$E^{(n)}_{calc}$ of resonant state with ccCSM. 
We use $E^{(n)}_{calc}$ as an input $E^{(n+1)}_{int}$ for the next turn. 
Such iterations are repeated until the self-consistent condition is satisfied; 
$E^{(n)}_{calc}=E^{(n)}_{int}$. 
In case of the present system, the self-consistency is achieved in 
five-times iterations \cite{procBARYONS10} $(n \leq 5)$. 

\subsection{Scattering amplitude calculated with ccCSM --- ``CS-WF'' method}

We are interested in the scattering state as well as 
the resonant state which can be investigated with 
the usual CSM as explained in the previous section. 
In this article, we solve a scattering problem also 
with a square-integrable base such as Gaussian base 
with the help of the CSM, following a method called 
``CS-WF'' which was developed by Kruppa, Suzuki and Kato \cite{CSM:Kruppa}. 
Here, the detailed formalism of CS-WF is shown for 
a multi-channel case such as the $\bar{K}N$-$\pi Y$ system. 

We assume a system involving $n$ channels in which 
the channel $c_0$ is the incident channel with the energy $E$. 
In a non-relativistic case, 
the radial Schr\"odinger equation for such a multi-channel system is 
\begin{eqnarray}
&& (E - H^l_c) \Phi^{(c_0)}_{l,c} (r) = \sum_{c'=1}^n V_{cc'} \Phi^{(c_0)}_{l,c'} (r), \label{SchEq_m} \\ 
&& H^l_c = -\frac{\hbar^2}{2\mu_c} \frac{d^2}{dr^2} 
+ \frac{\hbar^2}{2\mu_c} \frac{l(l+1)}{r^2} + V_{D,c}(r) + M_{T, c}, 
\end{eqnarray}
where $\mu_c$ and $M_{T, c}$ are the reduced mass and the total mass 
in the channel $c$, respectively. $V_{D,c}(r)$ is a direct potential 
for the channel $c$ and $V_{cc'}$ is a transition potential between channels 
$c$ and $c'$. We assign closed (open) channels for the incident energy $E$ to 
channel numbers $c=1, \ldots ,n_B$ ($c=n_B+1, \ldots ,n$). The wave function of the channel $c$ can be written as 
\begin{eqnarray}
\Phi^{(c_0)}_{l,c} (r) = 
\left\{
\begin{array}{ll}
\psi^{(c_0), B}_{l,c} (r)  & (c=1, \ldots, n_B) \\ 
\hat{j}_l(k_{c_0}r) \, \delta_{c,c_0} \; + \; \psi^{(c_0), sc}_{l,c} (r) & (c=n_B+1, \ldots, n)
\end{array}
\right. , \label{Eq:Phi_cc0}
\end{eqnarray}
where $\hat{j}_l (kr)$ is the Riccati-Bessel function. 
The incident wave number $k_{c_0}$ satisfies $\frac{\hbar^2}{2\mu_{c_0}} k_{c_0}^2 + M_{T,c_0} =E$. 

Inserting this equation into Eq. (\ref{SchEq_m}) and applying the complex scaling, then we obtain the coupled equations as follows: 
\begin{eqnarray}
&& (E - H_c^{l,\theta}) 
\left\{
\begin{array}{c}
\psi_{l,c}^{(c_0), B,\theta} (r) \\
\psi_{l,c}^{(c_0), sc,\theta} (r)
\end{array}
\right\} \nonumber \\
&& \quad -\sum_{c'=1}^{n_B} V_{cc'}(re^{i\theta}) 
\psi_{l,c'}^{(c_0),B,\theta} (r)
-\sum_{c'=n_B+1}^{n} V_{cc'}(re^{i\theta}) \psi_{l,c'}^{(c_0),sc,\theta} (r) 
\nonumber \\
&& = e^{i\theta/2} V_{cc_0}(re^{i\theta}) \hat{j}_l(kre^{i\theta}) 
\quad
\left\{
\begin{array}{l}
(c=1, \ldots, n_B) \\ 
(c=n_B+1, \ldots, n)
\end{array}
\right. , \label{SchEq_m_2}
\end{eqnarray}
where $\psi_{l,c}^{(c_0), B,\theta} (r)$ ($\psi_{l,c}^{(c_0), sc,\theta} (r)$) 
means the $\psi_{l,c}^{(c_0), B} (r)$ ($\psi_{l,c}^{(c_0), sc} (r)$) 
complex-scaled as following Eq. (\ref{Eq:CSforWfnc}). 
The complex-scaled Hamiltonian for the channel $c$ is 
\begin{eqnarray}
H_c^{l,\theta} = -e^{-2i\theta}\frac{\hbar^2}{2\mu_c} \frac{d^2}{dr^2} 
+ e^{-2i\theta}\frac{\hbar^2}{2\mu_c} \frac{l(l+1)}{r^2} + V_{D,c}(re^{i\theta}) + M_{T, c} .
\end{eqnarray}

The wave functions for the closed channels, $\{ \psi^{(c_0), B}_{l,c} (r) \}$, 
are square-integrable since the threshold energy of these channels is 
above the incident energy $E$. The complex-scaled ones are 
also square-integrable. As for the open channels whose threshold 
energies are below the $E$,   
the scattered part of wave functions, $\{ \psi^{(c_0), sc}_{l,c} (r) \}$, 
are not square-integrable, since they behave in the asymptotic region as 
\begin{eqnarray}
\psi^{(c_0), sc}_{l,c} (r) \; \rightarrow \; 
k_c f_{l,cc_0}(k_c) \hat{h}^{(+)}_l(k_c r) \; 
\propto \; \exp\{ i (k_c r-l\pi/2)\} \quad
{\rm at} \; r \rightarrow \infty, 
\end{eqnarray}
where $\hat{h}^{\pm}_l (kr)$ is the Riccati-Hankel function. 
But, they are transformed to be square-integrable functions 
due to the complex scaling. The asymptotic behavior of 
the complex-scaled scattered part of wave function is 
\begin{eqnarray}
\psi^{(c_0), sc, \theta}_{l,c} (r) \; \propto \;
 i^{-l} \exp\{ ik_cr\cos\theta -k_c r\sin\theta\} 
\quad {\rm at} \; r \rightarrow \infty. 
\end{eqnarray}
It is easy to understand that the $\psi^{(c_0), sc, \theta}_{l,c}$ 
becomes a square-integrable function for $0 < \theta < \pi$. 
Thus, since both the complex-scaled 
wave functions $\{ \psi_{l,c}^{(c_0), B,\theta} (r)\}$ and 
$\{ \psi_{l,c}^{(c_0), sc,\theta} (r)\}$ are square-integrable, 
they can be expanded with Gaussian base $\{ G_i(r) \}$ as 
\begin{eqnarray}
\left\{
\begin{array}{c}
\psi_{l,c}^{(c_0), B,\theta} (r) \\
\psi_{l,c}^{(c_0), sc,\theta} (r)
\end{array}
\right\}
\; \equiv \; \psi_{l,c}^{(c_0),\theta} (r) 
\; = \; \sum_{j=1}^N t^{(c_0)}_{c,j}(\theta) \, G_j(r) . \label{Gauss_exp_m}
\end{eqnarray}
In the present study, we use a common set of normalized Gaussians 
for all channels: 
\begin{eqnarray}
G_i(r) = N_l(b_i) \, r^{l+1} \, \exp \left[-\frac{ r^2}{2b_i^2}\right], \; 
N_l(b_i) = b_i^{-(2l+3)/2} \left\{ \frac{2^{l+2}}{(2l+1)!! \sqrt{\pi}}\right\}^{1/2}.
\label{Eq:GaussE_CSWF}
\end{eqnarray}

Inserting the Eq. (\ref{Gauss_exp_m}) into the coupled equations 
(\ref{SchEq_m_2}), linear equations for the unknown variables 
$\{t^{(c_0)}_{c,i}(\theta)\}$ are obtained: 
\begin{eqnarray}
\sum_j \left[ \, \left( E O_{ij} - H^{l,\theta}_{c, ij} \right) 
\, t^{(c_0)}_{c,j}(\theta) 
- \sum_{c'=1}^n V_{cc',ij}^\theta \, t^{(c_0)}_{c',j}(\theta) \, \right] 
\; = \; b^\theta_{cc_0, i}, \label{Eq:CSwfnc}
\end{eqnarray}
where each matrix element indicates 
$O_{ij} = \langle G_i | G_j \rangle$, 
$H^{l,\theta}_{c, ij} = \langle G_i | H^{l,\theta}_{c} |G_j \rangle$,  
$V^\theta_{cc', ij} = \langle G_i | V_{cc'}(re^{i\theta}) |G_j \rangle$ and 
\begin{eqnarray}
b^\theta_{cc_0, i} = e^{i\theta/2} \int^\infty_0 dr \; G_i(r) \, 
V_{cc_0}(re^{i\theta}) \, \hat{j}_l(kre^{i\theta}). \label{Eq:def_bcc0i}
\end{eqnarray}

We explain how to calculate scattering amplitudes in the 
remaining part of this section. 
With the scattering wave functions $\{ \Phi_{l,c}^{(c_0)} (r)\}$, 
the scattering amplitude between the initial channel $c_0$ and 
the final channel $c$ is expressed as 
\begin{eqnarray}
f_{l, cc_0} (k_c)= -\frac{2\mu_c}{\hbar^2 k_c k_{c_0}} \;
\sum_{c'=1}^{n} 
 \int^\infty_0 dr \; \hat{j}_l(k_c r) \, 
V_{cc'}(r) \, \Phi_{l,c'}^{(c_0)} (r) . \label{Eq:flcc0}
\end{eqnarray} 
It is obtained with the help of Green function, 
as a detailed explanation is given in~\ref{app_scatt_m}. 
By inserting Eq. (\ref{Eq:Phi_cc0}) into the $\Phi_{l,c'}^{(c_0)} (r)$ of 
the above equation, 
the full scattering amplitude is decomposed to 
the Born term $f_{l, cc_0}^{Born}(k_c)$ attributed to the incoming wave 
$\hat{j}_l(k_{c_0} r)$ and the other part $f_{l, cc_0}^{sc}(k_c)$ 
attributed to the scattered wave $\psi_{l,c}^{(c_0)} (r)$. 
The Born term can be obtained by the numerical integration. 
In the calculation of $f_{l, cc_0}^{sc}(k_c)$, 
the integration path  can be modified 
from $r$-axis to $re^{i\theta}$-line due to the Cauchy's theorem.  
Therefore, the $f_{l, cc_0}^{sc}(k_c)$ is equal to $f_{l, cc_0}^{sc,\theta}(k_c)$ that can be 
calculated with the complex-scaled wave functions 
$\{ \psi_{l,c}^{(c_0),\theta} (r) \}$, as  
\begin{eqnarray}
f_{l,cc_0}^{sc,\theta} (k_c) 
 =  -\frac{2\mu_c}{\hbar^2 k_c k_{c_0}} \; e^{i\theta/2} 
\sum_{c'=1}^{n} 
 \int^\infty_0 dr \; \hat{j}_l(k_c r e^{i\theta}) \, 
V_{cc'}(re^{i\theta}) \, \psi_{l,c}^{(c_0),\theta} (r). 
\end{eqnarray}
Thus, we can obtain the full scattering amplitudes as 
\begin{eqnarray}
f_{l, cc_0} (k_c) & = & f_{l, cc_0}^{Born} (k_c)+ f_{l,cc_0}^{sc} (k_c)\\
f_{l, cc_0}^{Born} (k_c) & = & -\frac{2\mu_c}{\hbar^2 k_c k_{c_0}} \;
 \int^\infty_0 dr \; \hat{j}_l(k_c r) \, 
V_{cc_0}(r) \, \hat{j}_l(k_{c_0} r) \label{Eq:fsc_Born}\\
f_{l,cc_0}^{sc} (k_c) & = & 
f_{l,cc_0}^{sc,\theta} (k_c) \simeq -\frac{2\mu_c}{\hbar^2 k_c k_{c_0}} \;
\sum_{c'=1}^n \sum_{j=1}^N t_{c',j}(\theta) \, b^\theta_{c'c, i}. 
\label{Eq:fsc}
\end{eqnarray}
The $f_{l,cc_0}^{sc,\theta} (k_c)$ can be obtained 
with the matrix elements $\{b^\theta_{c'c, i}\}$ by using the  
Eqs. (\ref{Gauss_exp_m}) and (\ref{Eq:def_bcc0i}). 

We summarize the essential points of the CS-WF 
at the end of this section. 
The first point is that the incoming part $\hat{j}_l(k_{c_0}r)$ 
is separated from the scattered wave functions $\{\psi_{l,c}^{(c_0),sc} (r)\}$ 
as shown in Eq. (\ref{Eq:Phi_cc0}) and 
that only the scattered parts $\{\psi_{l,c}^{(c_0)} (r)\}$ are 
complex-scaled. The complex scaling is used to make a non-square 
integrable function transformed to a square integrable one.  
If the full scattering wave function $\Phi^{(c_0)}_{l,c} (r)$ 
is complex-scaled, it does not become a square-integrable function 
because of Riccati-Bessel function $\hat{j}_l(k_{c_0}r)$ 
that contains both components of $\exp\{\pm i (kr-\frac{l\pi}{2})\}$ 
in the asymptotic region. 
The second point is that we calculate the scattering amplitude 
with the complex-scaled function $\psi_{l,c}^{(c_0),\theta} (r)$, 
instead of the $\psi_{l,c}^{(c_0)} (r)$ which is needed 
in usual calculation of the scattering amplitude. 
By the Cauchy's theorem, 
the amplitude can be obtained with the $\psi_{l,c}^{(c_0),\theta} (r)$ 
which is expressed with a square-integrable base. We note that 
the scattering amplitudes calculated in this way are independent 
of the scaling angle $\theta$.

In case of the semi-relativistic kinematics, 
the formula of scattering amplitudes is obtained by 
replacing the reduced mass $\mu_c$ in Eqs. (\ref{Eq:fsc_Born}) 
and (\ref{Eq:fsc}) with 
the reduced energy $\tilde{\omega}_c$. (See~\ref{app_scatt_SR}.)
The scattering amplitudes are calculated 
in the same way as the non-relativistic 
kinematics as explained above, by replacing 
the matrix elements of kinetic-energy term with 
those for the semi-relativistic kinematics. 
These matrix elements are given in~\ref{Mat_Kin_SR}.  

\subsection{Test of ccCSM for scattering amplitude}

\begin{figure}
\centerline{
  \includegraphics[width=.50\textwidth]{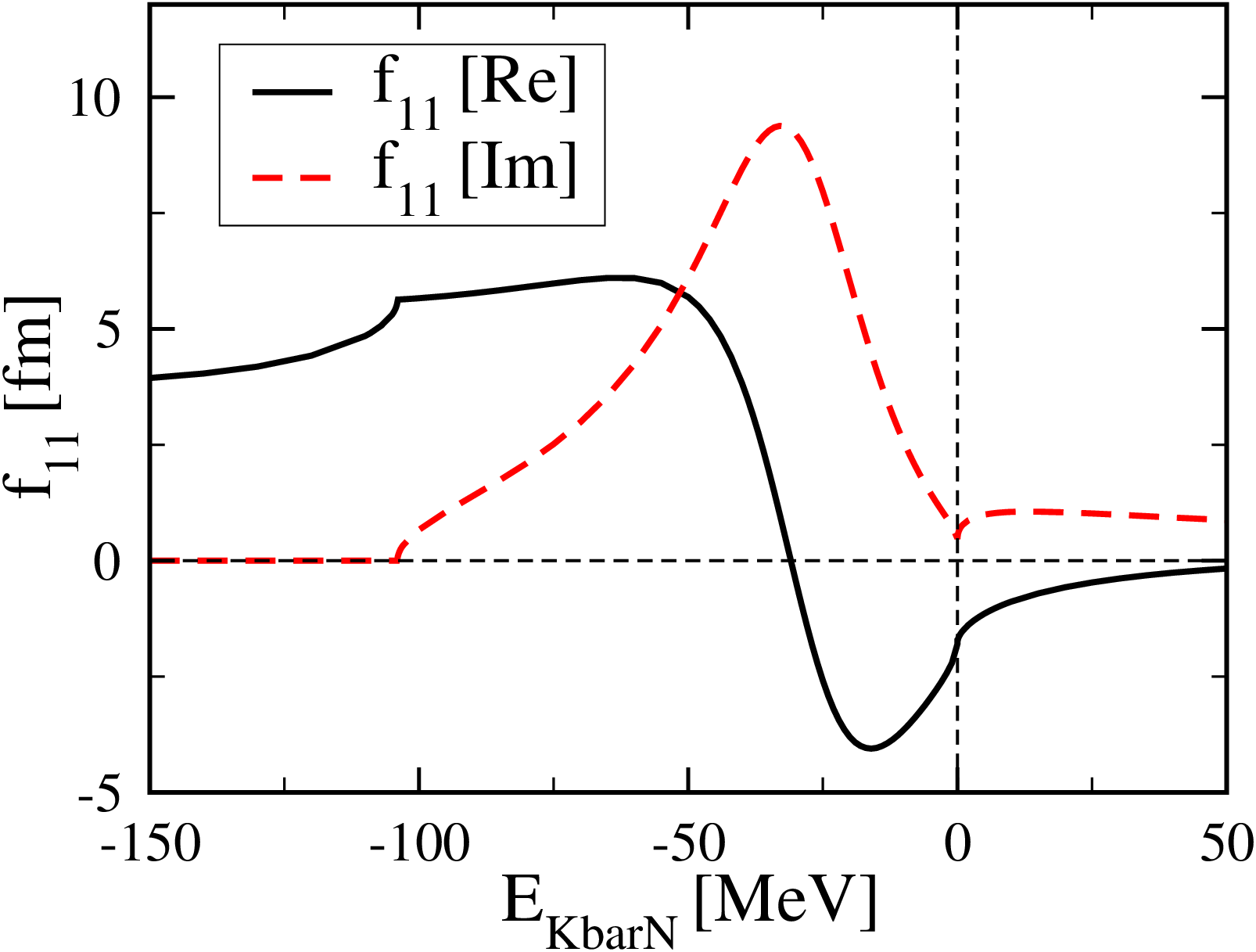}
}
  \caption{The $\bar{K}N \rightarrow\bar{K}N$ scattering amplitude 
($f_{11}$) in the $I=0$ channel calculated with a phenomenological 
potential \cite{AY_2002}. 
``$\rm{E_{KbarN}}$'' means the energy of the system measured from 
the $\bar{K}N$ threshold. 
 \label{f11_AYpot}}
\end{figure}

In this subsection, we test the CS-WF method by applying it to $I=0$ $\bar{K}N$-$\pi\Sigma$ system, 
since it is the first time to apply this method to a meson-baryon system. 

First, we use a phenomenological $\bar{K}N$ potential \cite{AY_2002}. 
Except for its energy independence,
this potential is similar to our KSW-type potential; 
both are for coupled $\bar{K}N$-$\pi\Sigma$ channels,
and given in a single Gaussian form in $r$-space.
Therefore, we employ this potential for the first test. 

Fig.~\ref{f11_AYpot} shows the $\bar{K}N$ scattering amplitude obtained in the CS-WF method. 
In Ref. \cite{AY_2002}, the same Schr\"odinger equation is solved in the usual way
and the resulting $\bar{K}N$ amplitude is shown in Fig. 1 of Ref. \cite{AY_2002}.
By comparing them, one can see that our calculation reproduces the original result well. 
Two scattering lengths also agree well:
$-1.77+i0.47$ fm in our 
calculation and $-1.76+i0.46$ fm in the original work. 


\begin{figure}
  \includegraphics[width=.47\textwidth]{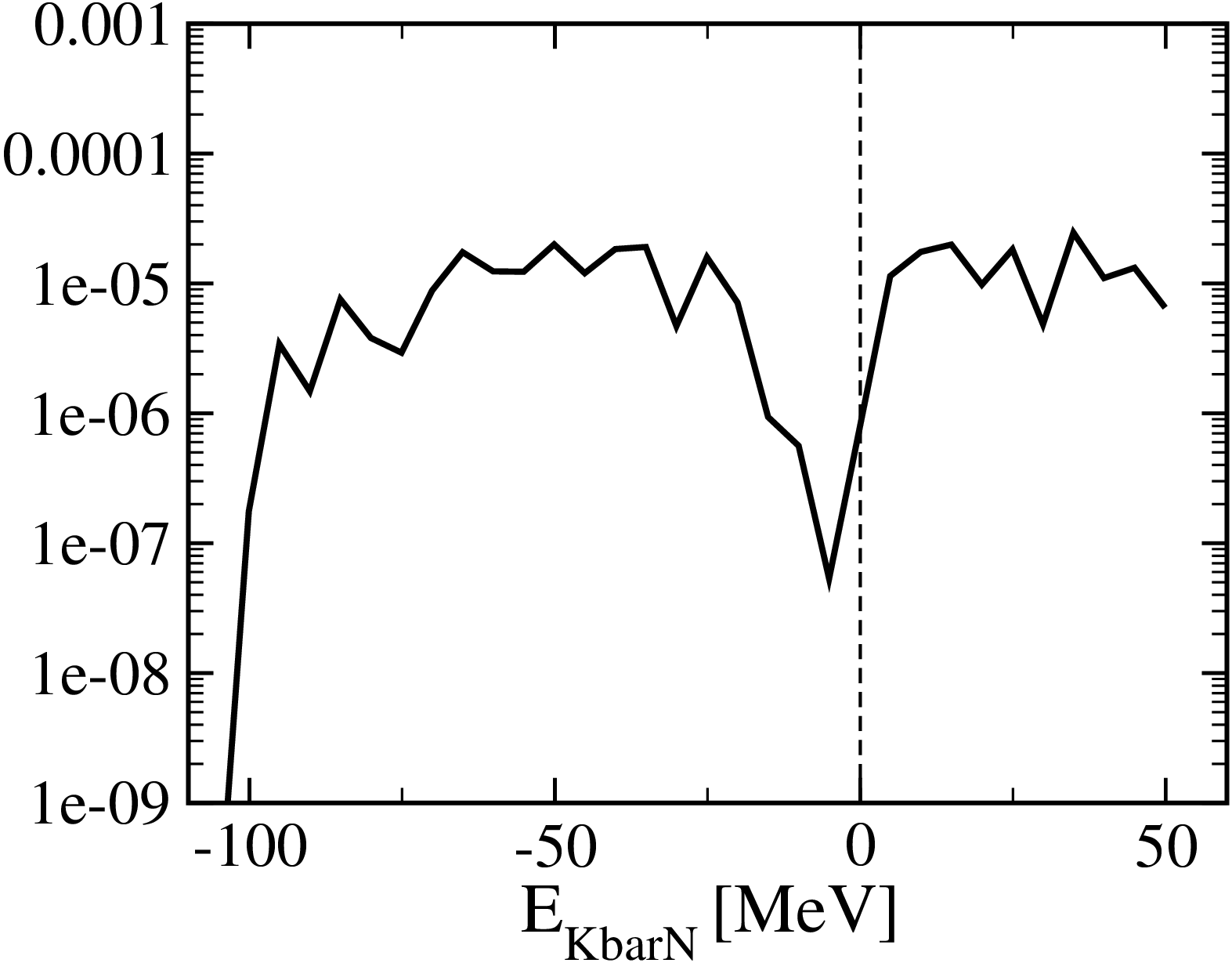}
  \hspace{0.3cm}
  \includegraphics[width=.47\textwidth]{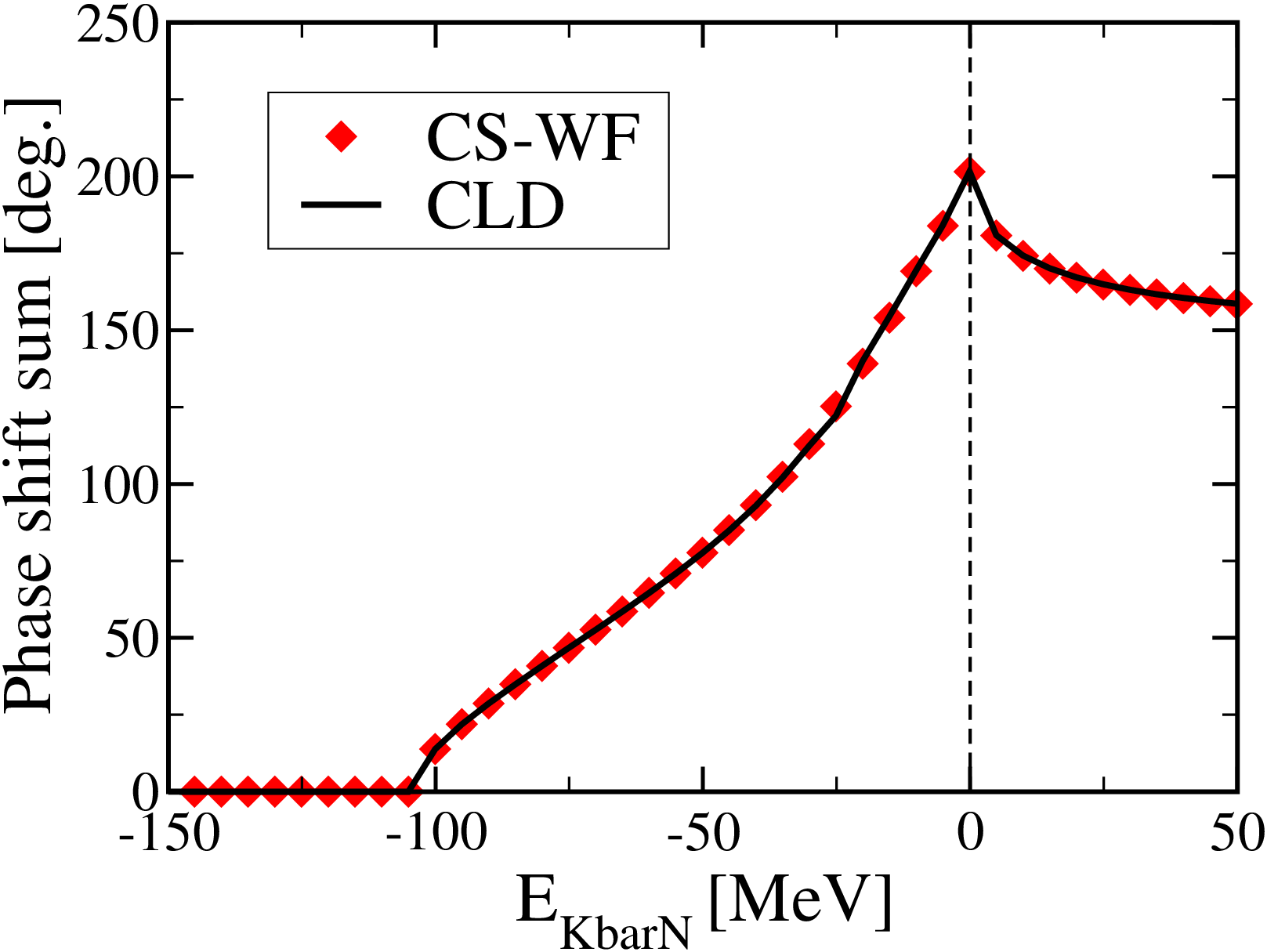}
  \caption{Test calculation of our method for the non-relativistic kinematics. 
(Left) Unitarity violation of the $S$-matrix. (Right) Comparison of 
the phase-shift sum between CS-WF (red diamond) and CLD (black line). 
Here, the $I=0$ channel is considered. 
\label{ccCSMtest_NR}}
\end{figure}

\begin{figure}
  \includegraphics[width=.47\textwidth]{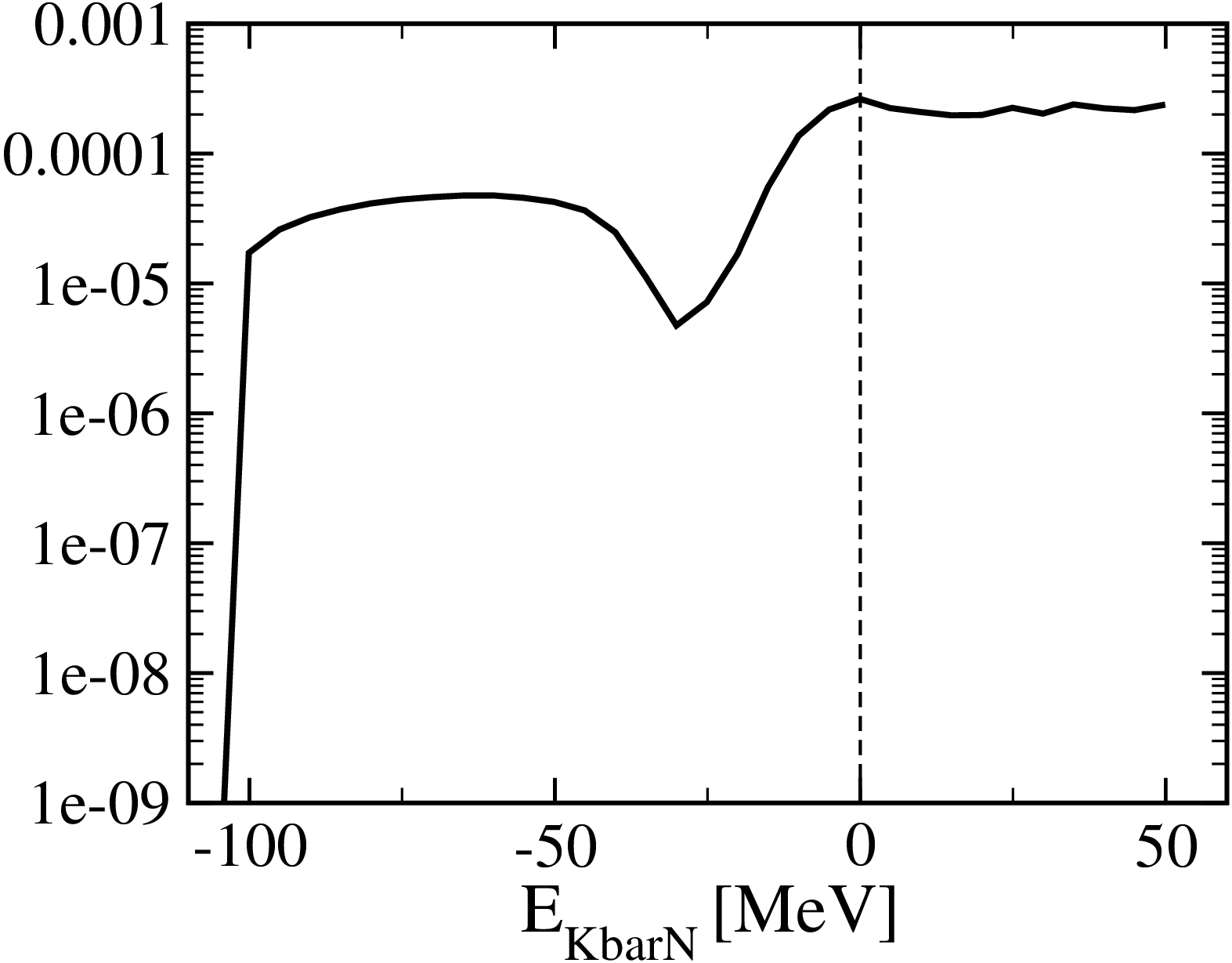}
  \hspace{0.3cm}
  \includegraphics[width=.47\textwidth]{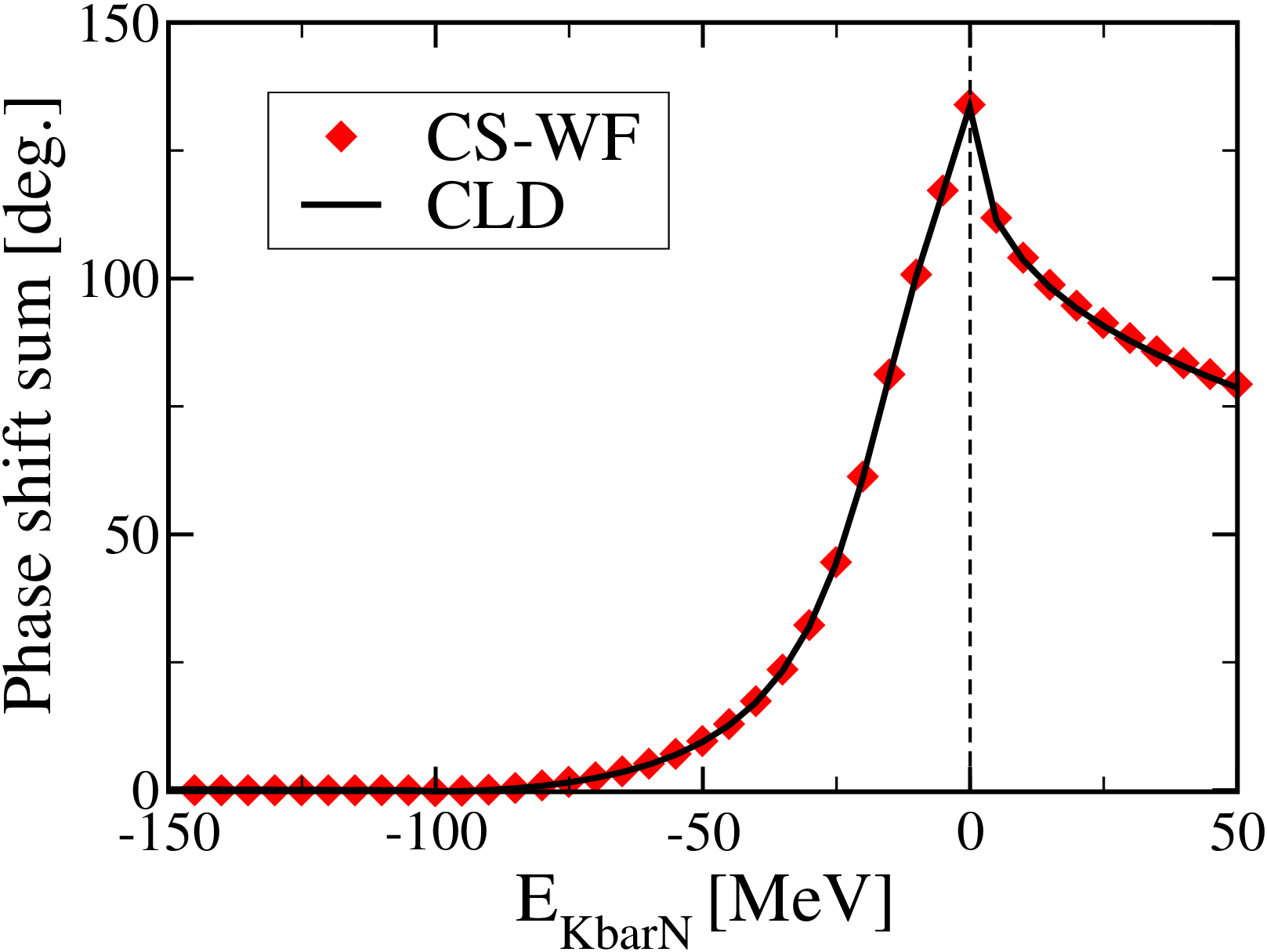}
  \caption{Same as Fig.~\ref{ccCSMtest_NR}, but 
for the semi-relativistic kinematics.  \label{ccCSMtest_SR}}
\end{figure}

Next, we test the CS-WF method with our energy-dependent potential.
We use the NRv2 potential (Eq. (\ref{PotMB_NRv2})) in the NR kinematics case
and the KSW-type potential (Eq. (\ref{PotMB_SR})) in the SR kinematics case.
For both kinematics, we check violation of the unitarity of $S$-matrix; 
$| |\det S| - 1 |$. We calculate the $S$-matrix for the partial wave $l$  
with the scattering amplitude obtained by the CS-WF method. 
The $S$-matrix element for channels $\alpha$ and $\beta$ 
is related to the scattering amplitude as 
\begin{eqnarray}
S^l_{\beta,\alpha} = \delta_{\beta,\alpha} + 2ik_{\alpha} 
\sqrt{\frac{k_\beta/\mu_\beta}{k_\alpha/\mu_\alpha}} f^l_{\beta,\alpha}, 
\label{Smat}
\end{eqnarray} 
where $k_\alpha$ is the wave number in the channel $\alpha$. 
The reduced masses \{$\mu_\alpha$\} in the NR case are replaced with 
the reduced energies \{$\tilde{\omega}_\alpha$\} in the SR case. 
As shown in the left panel of Fig.~\ref{ccCSMtest_NR}, in the NR case 
the magnitude of the unitarity violation 
is confirmed to be significantly small of the order of $10^{-5}$ 
in $\bar{K}N$ energy region of $-100$ to $50$ MeV. 
In the SR case, the violation is slightly larger than that in the NR case 
but still keeps the $10^{-4}$ level as shown in Fig.~\ref{ccCSMtest_SR}, left panel.
We consider that the larger violation 
is attributed to the numerical integration of kinetic-energy term 
in the SR calculation (see~\ref{Mat_Kin_SR}).

We check also the phase-shift sum.
We compare sum of the present phase shifts, $\delta_{\bar{K}N}+\delta_{\pi\Sigma}$, 
with that obtained by the continuum level density (CLD) method \cite{CSM-CLD:Suzuki}.
In the CLD method, the phase-shift sum is given by eigenvalues of the complex-scaled Hamiltonian, 
but the phase shift of each channel is not.
Phase shift is extracted from the $S$-matrix (Eq. (\ref{Smat})) as 
\begin{eqnarray}
S=
\left(
\begin{array}{cc}
S_{11} & S_{12} \\
S_{21} & S_{22}
\end{array}
\right)
=
\left(
\begin{array}{cc}
\cos 2\epsilon \, e^{i \delta_1} & i \sin \epsilon \, e^{i(\delta_1+\delta_2)} \\
i \sin \epsilon \, e^{i(\delta_1+\delta_2)} & \cos 2\epsilon \, e^{i \delta_2} 
\end{array}
\right), 
\end{eqnarray}
where $\delta_i$ is phase shift of channel $i$ and 
$\epsilon$ is a mixing parameter. 
As seen in the right panel of Figs.~\ref{ccCSMtest_NR} and~\ref{ccCSMtest_SR}, 
the phase-shift sums calculated in the two methods 
agree with each other quite well in both the kinematics. 

\section{Results}
\label{}

We show results of our calculation for $\bar{K}N$-$\pi Y$ system with the ccCSM. 
%
As mentioned in the section~\ref{Sec:Kin_Int}, 
we use our KSW-type potential and two kinds of its non-relativistic approximation.
We use both the non-relativistic kinematics and the semi-relativistic kinematics. 
Hereafter we abbreviate non-relativistic and semi-relativistic as ``NR'' and ``SR'', respectively. 
Natural combinations of potential types and kinematics are following: 

\noindent
\begin{tabular}{cll}
-&Case NRv1& One of non-relativistic approximated KSW-type potentials 
\\
&&(Eq. (\ref{PotMB_NRv1})), NRv1, with 
the non-relativistic kinematics. \\ 
-&Case NRv2 &The other non-relativistic approximated KSW-type potentials 
\\
&&(Eq. (\ref{PotMB_NRv2})), NRv2, with 
the non-relativistic kinematics. \\
-&Case SR &The KSW-type potential (Eq. (\ref{PotMB_SR})) with the semi-relativistic\\
&& kinematics. 
\end{tabular}

We consider $f_\pi$ in the KSW-type potential as a parameter in our model. 
In the present study, it is varied around the physical values 
of $f_\pi\simeq93$ MeV and $f_K\simeq110$ MeV. 
We examine four cases of $f_\pi=90$, 100, 110 and 120 MeV. 
The results shown in the following sections are obtained 
with $f_\pi=110$ MeV if the $f_\pi$ value is not specified.

\subsection{Scattering amplitude of $I=0$ $\bar{K}N$-$\pi\Sigma$ system}

\begin{table}
\caption{Range parameters for $I=0$ $\bar{K}N$-$\pi\Sigma$ system. 
$f_\pi=110$ MeV case. 
$d_{\bar{K}N,\, \bar{K}N}$ ($d_{\pi\Sigma,\, \pi\Sigma}$) means the 
range parameter of Gaussian potential in $\bar{K}N$-$\bar{K}N$ 
($\pi\Sigma$-$\pi\Sigma$) channel. $a_{\bar{K}N\,(I=0)}$ is the 
$I=0$ $\bar{K}N$ scattering length calculated with given range parameters. 
``Martin'' means that obtained by Martin's analysis \cite{Exp:ADMartin}. 
All quantities are in unit of fm. \label{RangePara}}
\begin{tabular*}{\textwidth}{l@{\hspace{0.5cm}}|@{\hspace{0.7cm}}r@{\hspace{0.5cm}}rcr@{\hspace{0.5cm}}r@{\hspace{0.7cm}}|r}
\hline
Case       & NRv1 & NRv2                  & \hspace{0.3cm} & SR-A & SR-B  & \hspace{0.1cm}Martin\\
\hline
Kinematics & \multicolumn{2}{c}{Non-rela.}& & \multicolumn{2}{c|}{Semi-rela.} & --- \\
Potential  & NRv1 & NRv2                  & & \multicolumn{2}{c|}{KSW-type}   & --- \\
\hline
&&&&&&\\
$d_{\bar{K}N,\, \bar{K}N}$    & 0.440 & 0.438 && 0.499 & 0.369 & ---\\
$d_{\pi\Sigma,\, \pi\Sigma}$  & 0.605 & 0.636 && 0.712 & 0.348 & ---\\ 
&&&&&&\\
Re $a_{\bar{K}N\,(I=0)}$ & $-1.701$ & $-1.700$ && $-1.700$ & $-1.696$ &$-1.70$\\
Im $a_{\bar{K}N\,(I=0)}$ & 0.681    & 0.681    && 0.681    & 0.681 & 0.68\\
\hline
\end{tabular*}
\end{table}

First, we determine the range parameters of Gaussian functions, 
$\{d^{I=0}_{\alpha \beta}\}$, in the meson-baryon potentials defined in 
Eqs. (\ref{PotMB_SR}), (\ref{PotMB_NRv1}) and (\ref{PotMB_NRv2}). 
In the present study, with an assumption of $d_{\bar{K}N,\pi\Sigma} 
= d_{\pi\Sigma,\bar{K}N} = (d_{\bar{K}N,\bar{K}N}+d_{\pi\Sigma,\pi\Sigma})/2$, 
we search for the two of the range parameters, 
$d_{\bar{K}N,\bar{K}N}$ and $d_{\pi\Sigma,\pi\Sigma}$, so as 
to reproduce 
the complex value of $\bar{K}N$ scattering length with $I=0$ which 
was obtained by Martin's analysis; 
$a_{\bar{K}N(I=0)}=-1.70+i0.68$ fm \cite{Exp:ADMartin}. 
For all combinations of kinematics and meson-baryon potential, 
we can find the range parameters which reproduce the Martin's value. 
The determined range parameters and the resulting $\bar{K}N$ scattering length 
for $f_\pi=110$ MeV case are listed in Table~\ref{RangePara}. 
It should be noted that in the semi-relativistic case we find two sets 
of the range parameters as shown on the right two columns in the table. 
Hereafter, we denote them as ``SR-A'' 
and ``SR-B'', respectively. Also in cases of other $f_\pi$ values, 
we determine the range parameters as given in Table 
\ref{RangePara_100-120}. 

\begin{figure}
  \includegraphics[width=.47\textwidth]{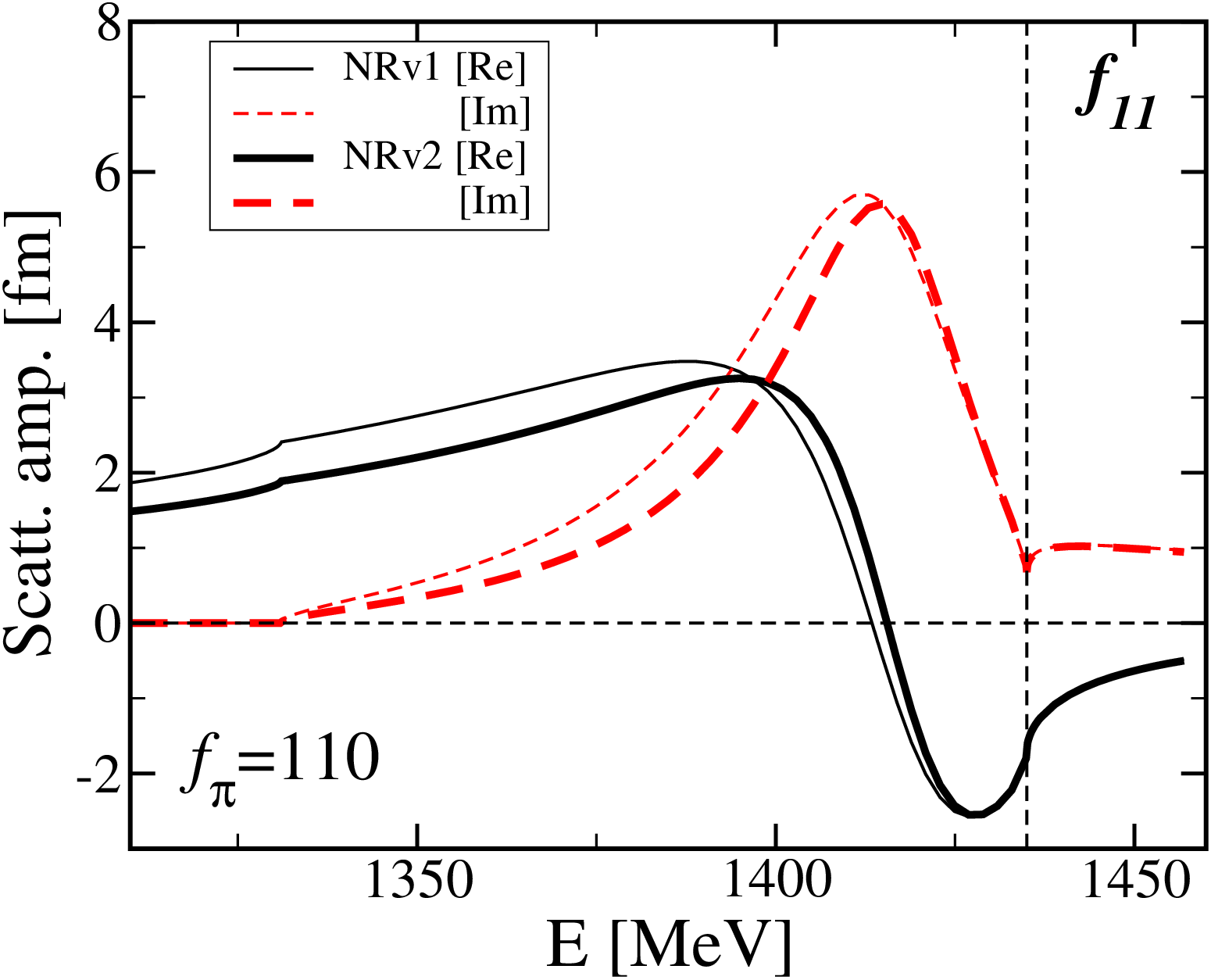}
  \hspace{0.3cm}
  \includegraphics[width=.47\textwidth]{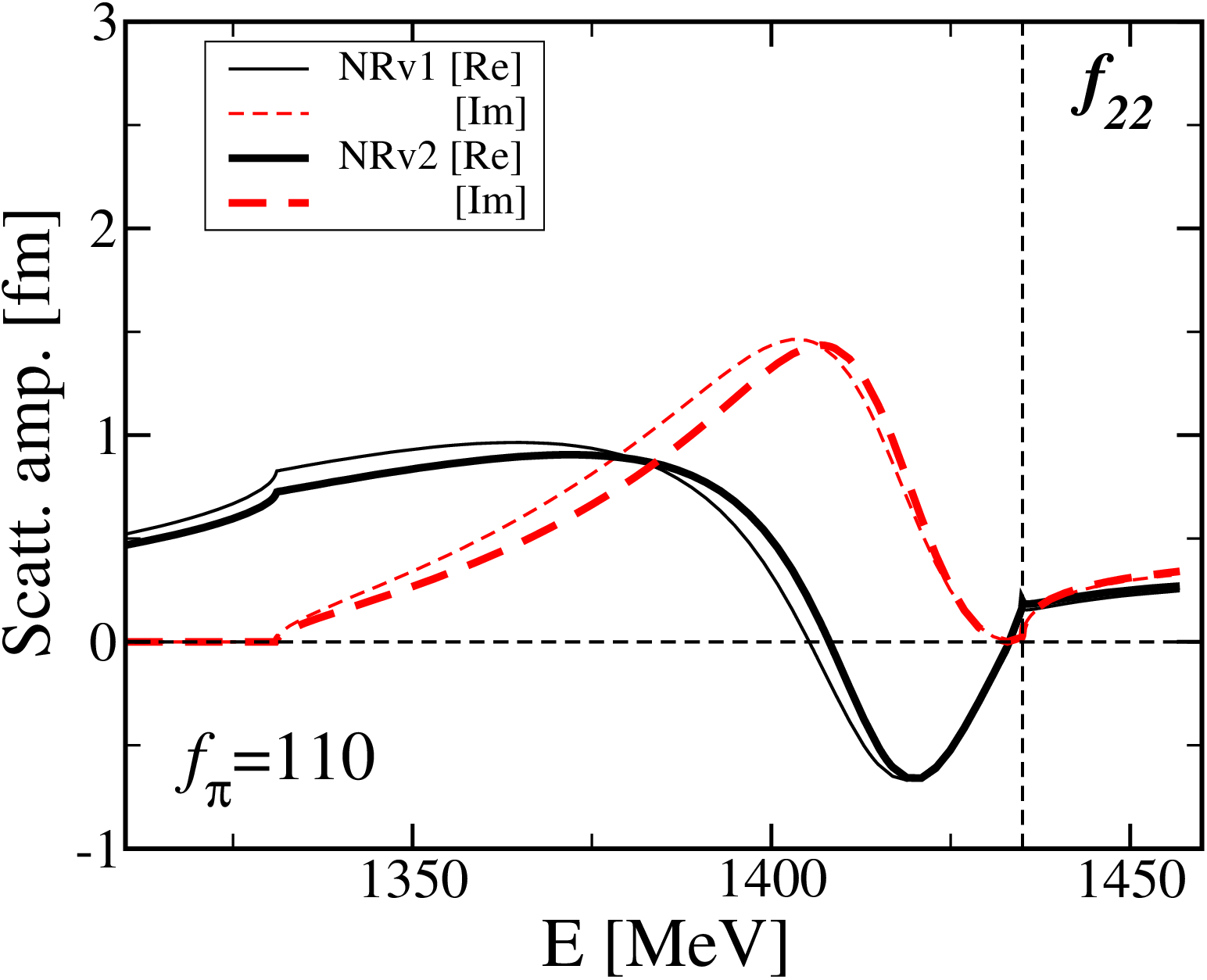}
  \caption{$I=0$ scattering amplitudes for NRv1 and NRv2 cases with 
$f_\pi=110$ MeV. 
The scattering amplitudes of NRv1 (NRv2) are shown with thin (bold) line. 
The real (imaginary) part of scattering amplitude is drawn with 
a black-solid (red-dashed) line. 
Left (right) panel shows $\bar{K}N$ ($\pi\Sigma$) scattering amplitude. 
Vertical dashed line means the $\bar{K}N$ threshold. 
\label{KSW_I=0_caseNR}}
\end{figure}

\begin{figure}
  \includegraphics[width=.47\textwidth]{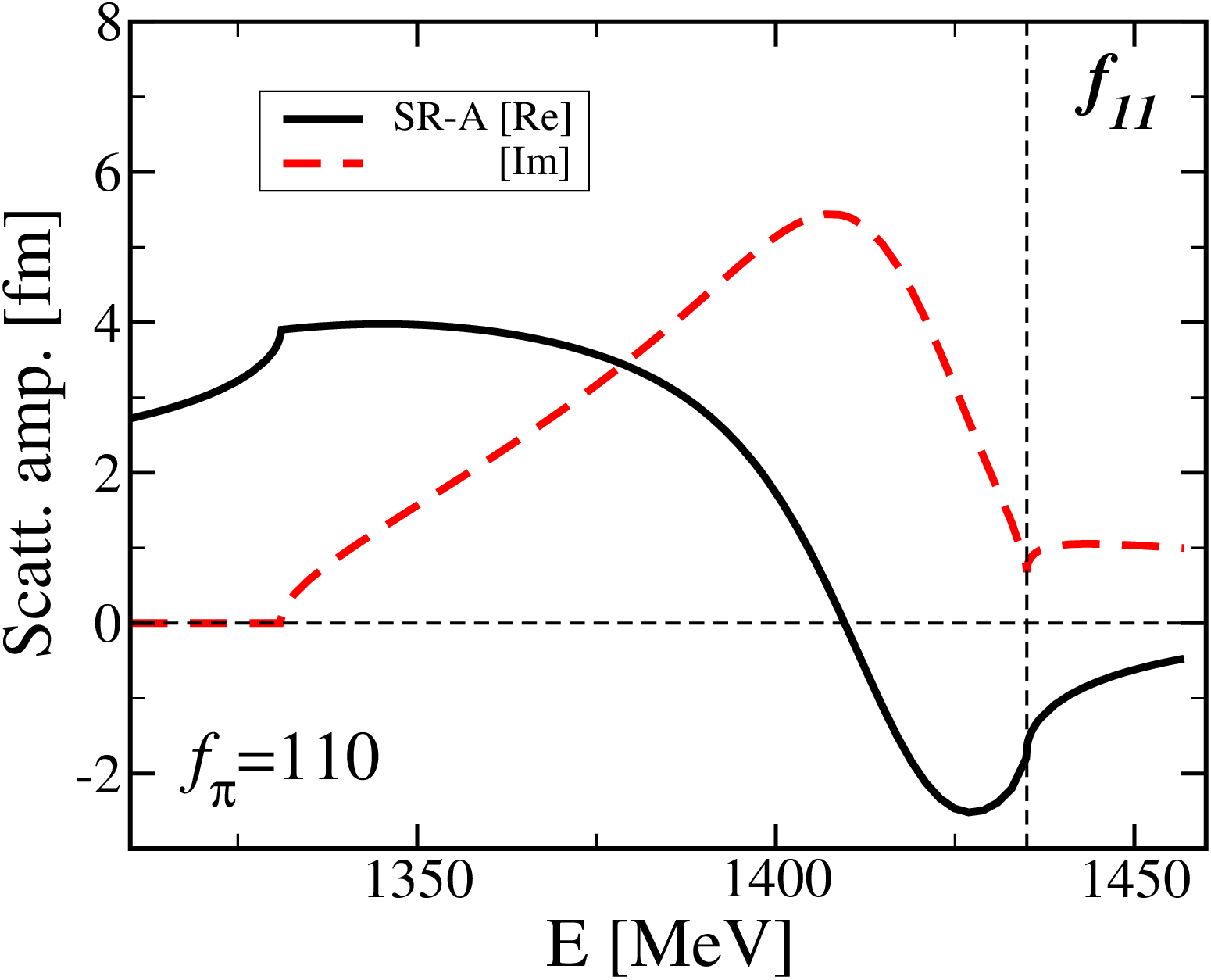}
  \hspace{0.3cm}
  \includegraphics[width=.47\textwidth]{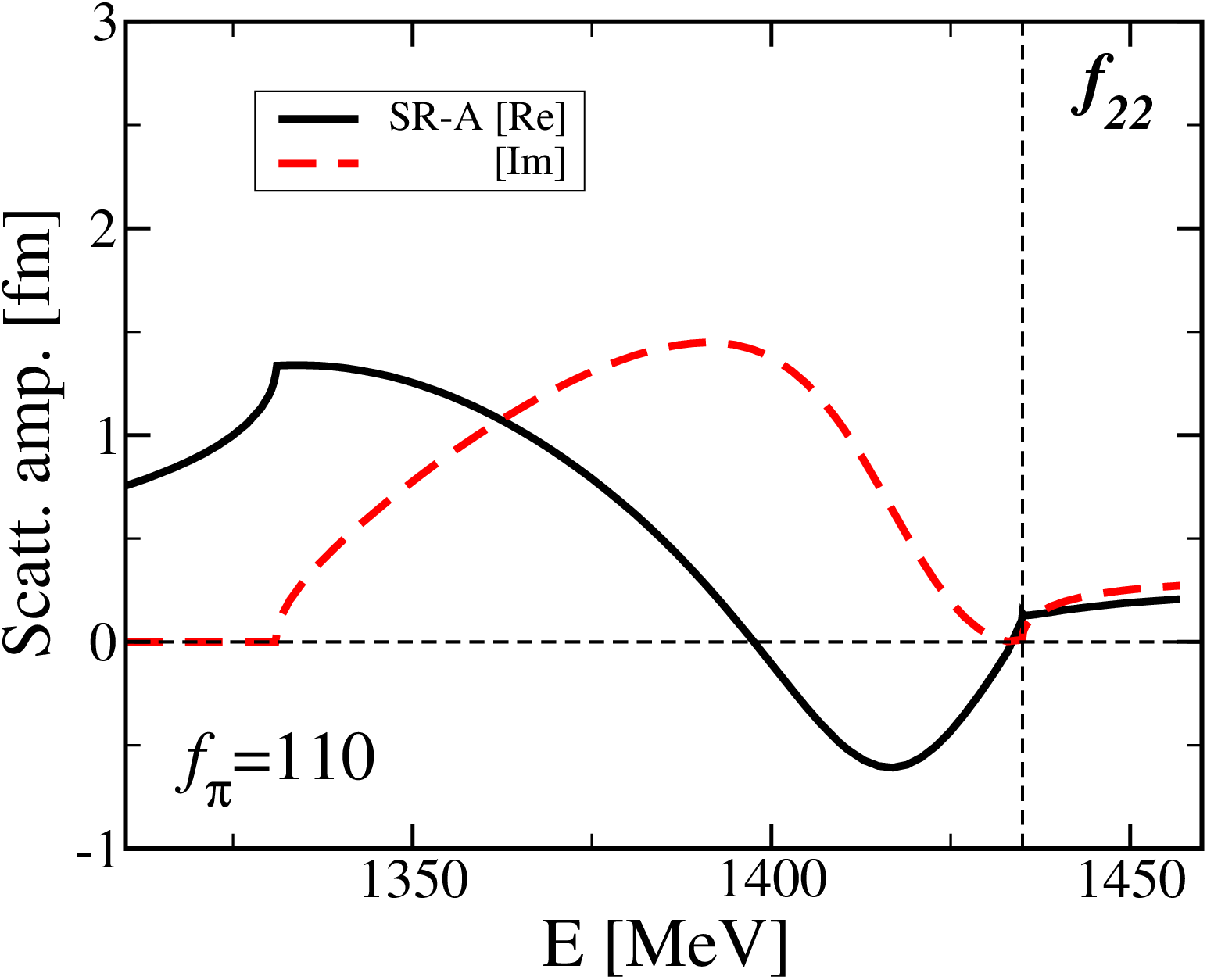}
  \caption{$I=0$ scattering amplitudes for SR-A case with $f_\pi=110$ MeV. 
Similar to Fig.~\ref{KSW_I=0_caseNR}.
\label{KSW_I=0_caseSR-A}}
\end{figure}

\begin{figure}
  \includegraphics[width=.47\textwidth]{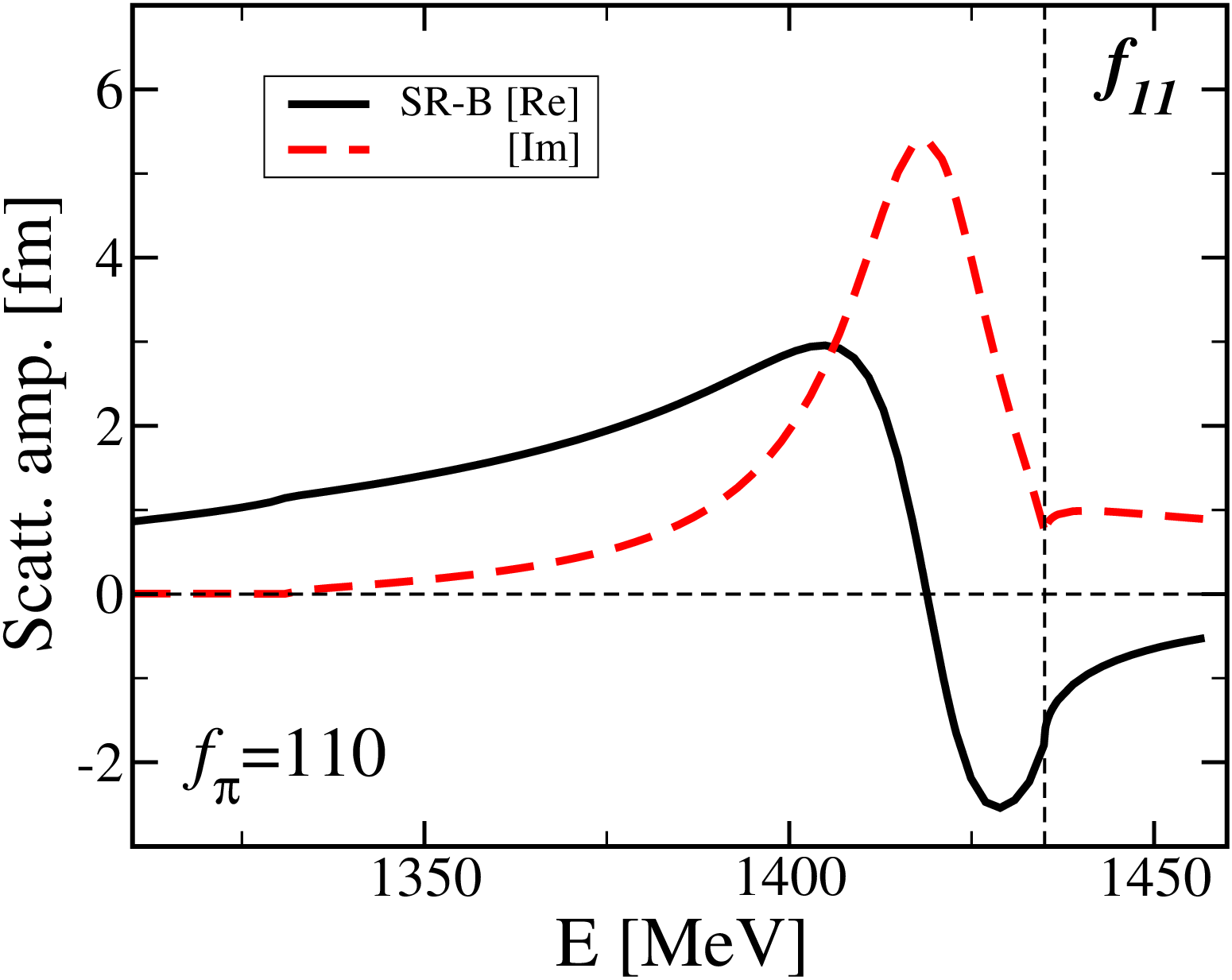}
  \hspace{0.3cm}
  \includegraphics[width=.47\textwidth]{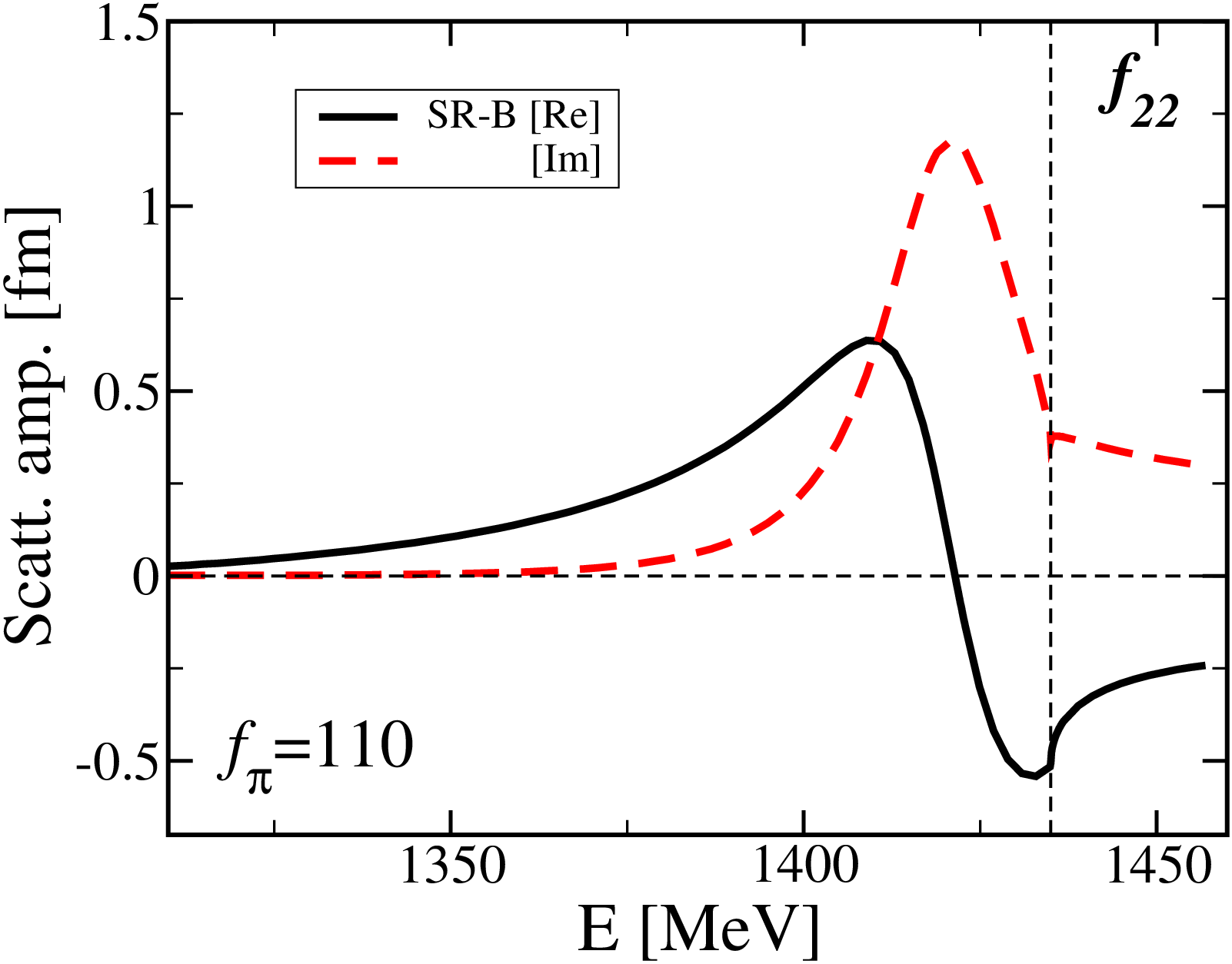}
  \caption{$I=0$ scattering amplitudes for SR-B case with $f_\pi=110$ MeV. 
Similar to Fig.~\ref{KSW_I=0_caseNR}.
\label{KSW_I=0_caseSR-B}}
\end{figure}

Using the meson-baryon potentials with these range parameters, 
we calculate the scattering amplitude in the $I=0$ channel. 
Figs.~\ref{KSW_I=0_caseNR} - \ref{KSW_I=0_caseSR-B} show 
the $\bar{K}N$ and $\pi\Sigma$ scattering amplitudes 
for $f_\pi=110$ MeV. 
As seen in Figs.~\ref{KSW_I=0_caseNR} and~\ref{KSW_I=0_caseSR-A}, 
the NR and SR-A give quantitatively the same scattering amplitudes. 
Let us see the $I=0$ scattering amplitudes in more detail.  
In two non-relativistic cases NRv1 and NRv2, there is almost no difference 
between their scattering amplitudes.  
Compared with these NR cases, magnitude of scattering amplitudes 
becomes larger near the $\pi\Sigma$ threshold in a semi-relativistic 
case SR-A, though near the $\bar{K}N$ threshold 
the scattering amplitudes of both cases are the same. 
Fig.~\ref{KSW_I=0_caseSR-B} is a result of the other semi-relativistic case 
SR-B. The global distribution of $\bar{K}N$ scattering amplitude is quite 
similar 
to the NR result. However, the $\pi\Sigma$ scattering amplitude 
is very different from that of NR. 

\begin{figure}
  \includegraphics[width=.47\textwidth]{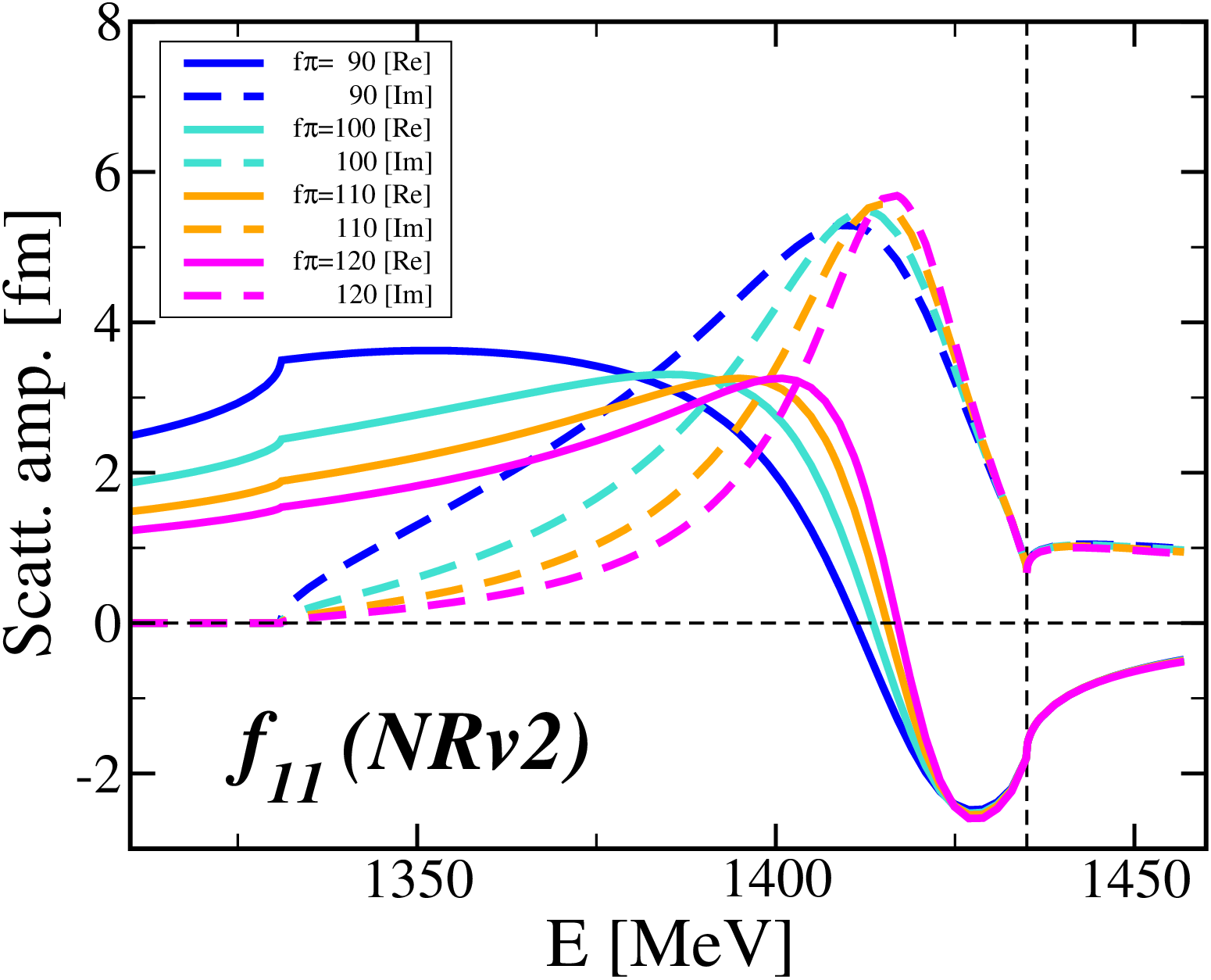}
  \hspace{0.3cm}
  \includegraphics[width=.47\textwidth]{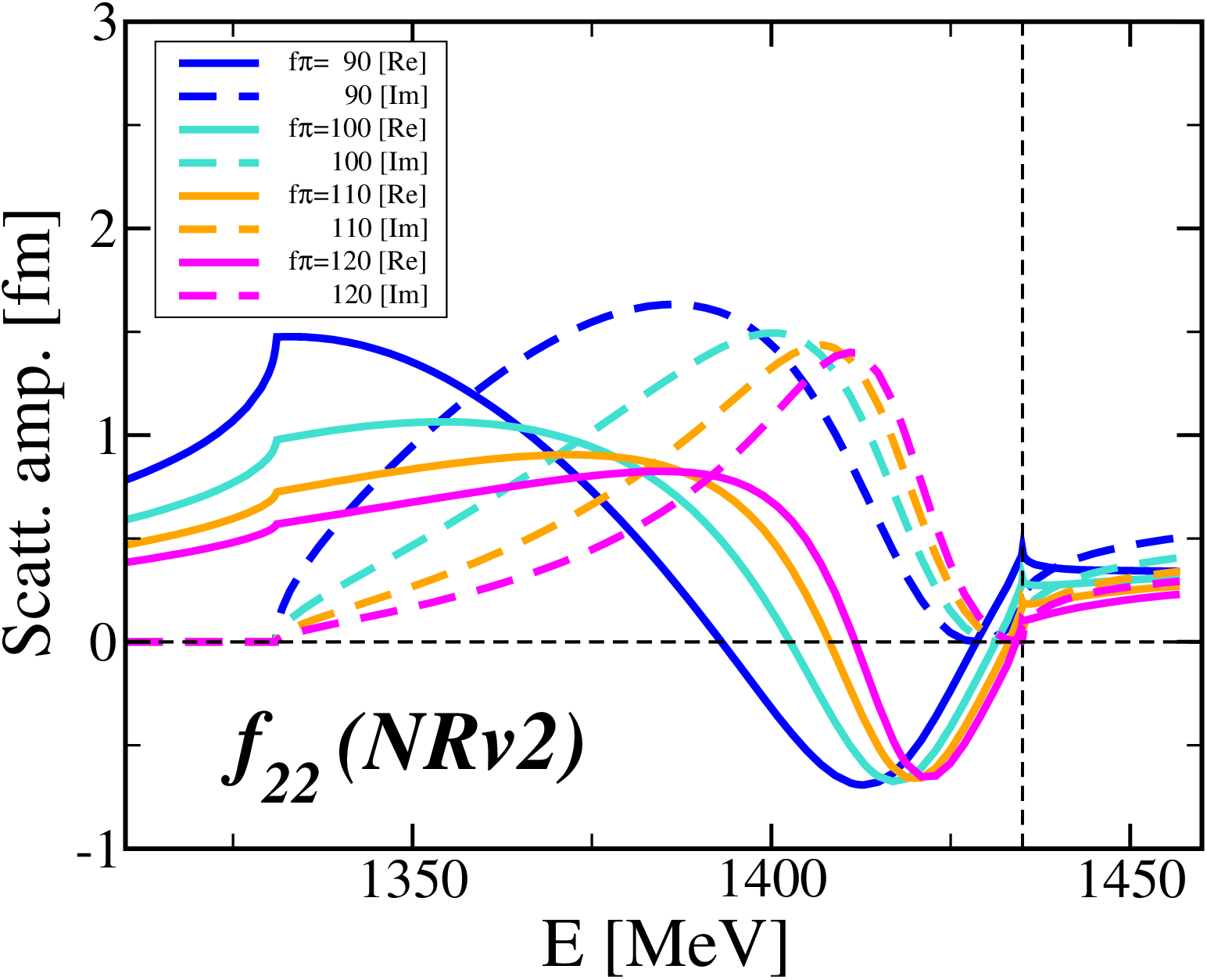}
  \caption{$I=0$ scattering amplitudes for NRv2 case, calculated with 
various $f_\pi$ values. Blue, light blue, orange and magenta lines 
correspond to $f_\pi$=90, 100, 110 and 120 MeV, respectively.   
Solid (dashed) line indicates the real (imaginary) part of 
scattering amplitude. 
Left (right) panel shows $\bar{K}N$ ($\pi\Sigma$) scattering amplitude.
\label{KSW_I=0_mfpi_NRv2}}
\end{figure}

\begin{figure}
  \includegraphics[width=.47\textwidth]{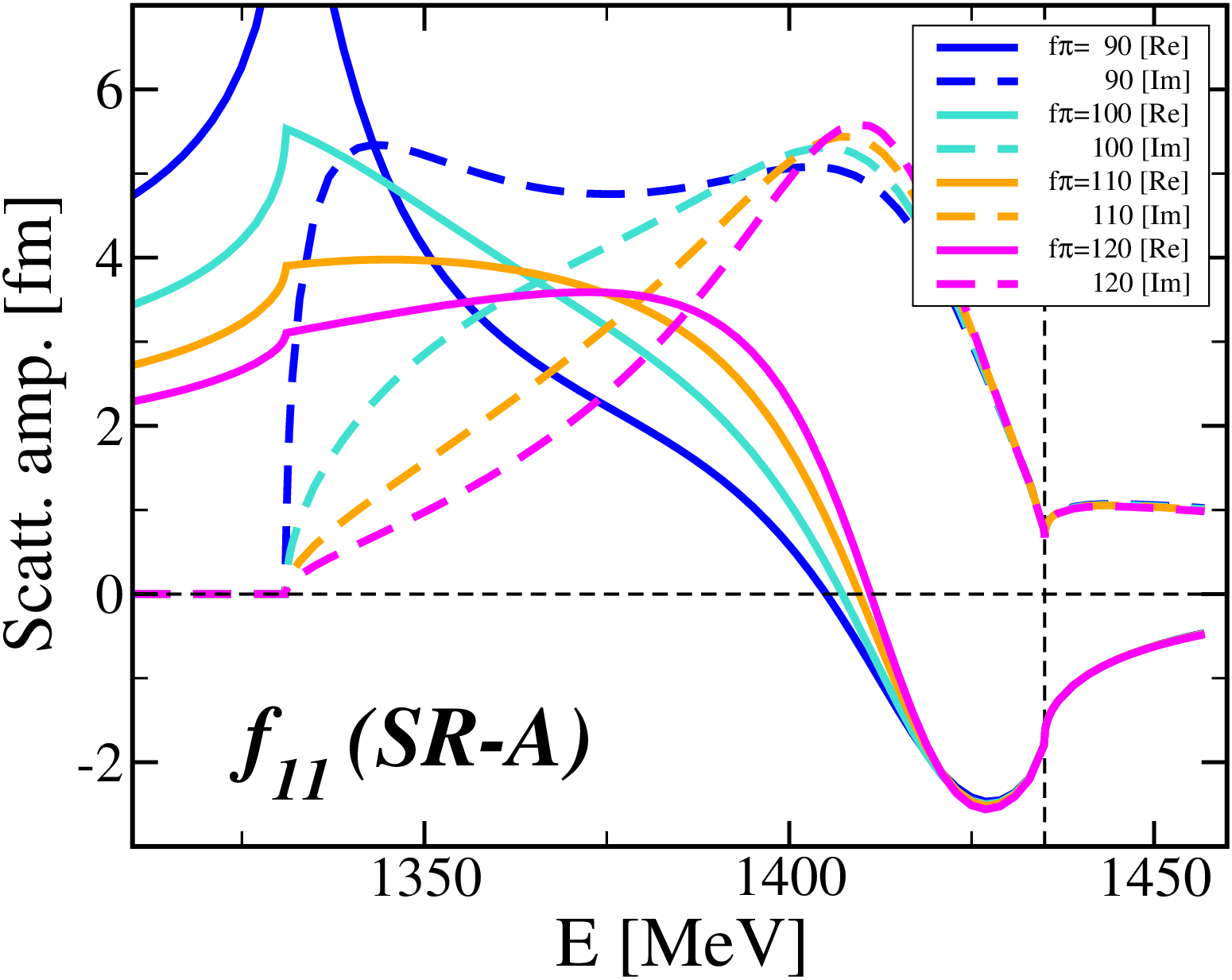}
  \hspace{0.3cm}
  \includegraphics[width=.47\textwidth]{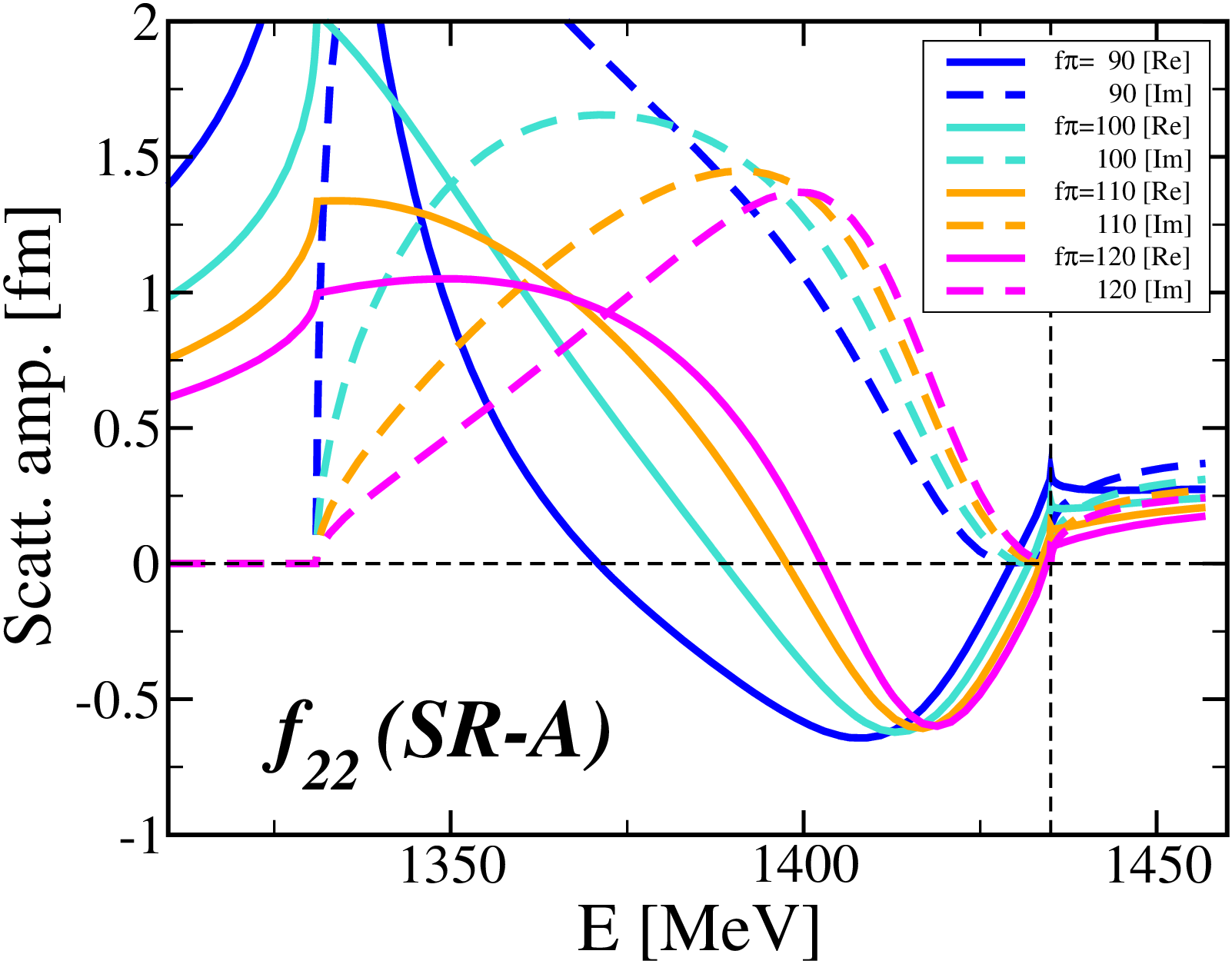}
  \caption{$I=0$ scattering amplitudes for SR-A case, 
calculated with various $f_\pi$ values. 
Similar to Fig.~\ref{KSW_I=0_mfpi_NRv2}.
\label{KSW_I=0_mfpi_SR-A}}
\end{figure}

\begin{figure}
  \includegraphics[width=.47\textwidth]{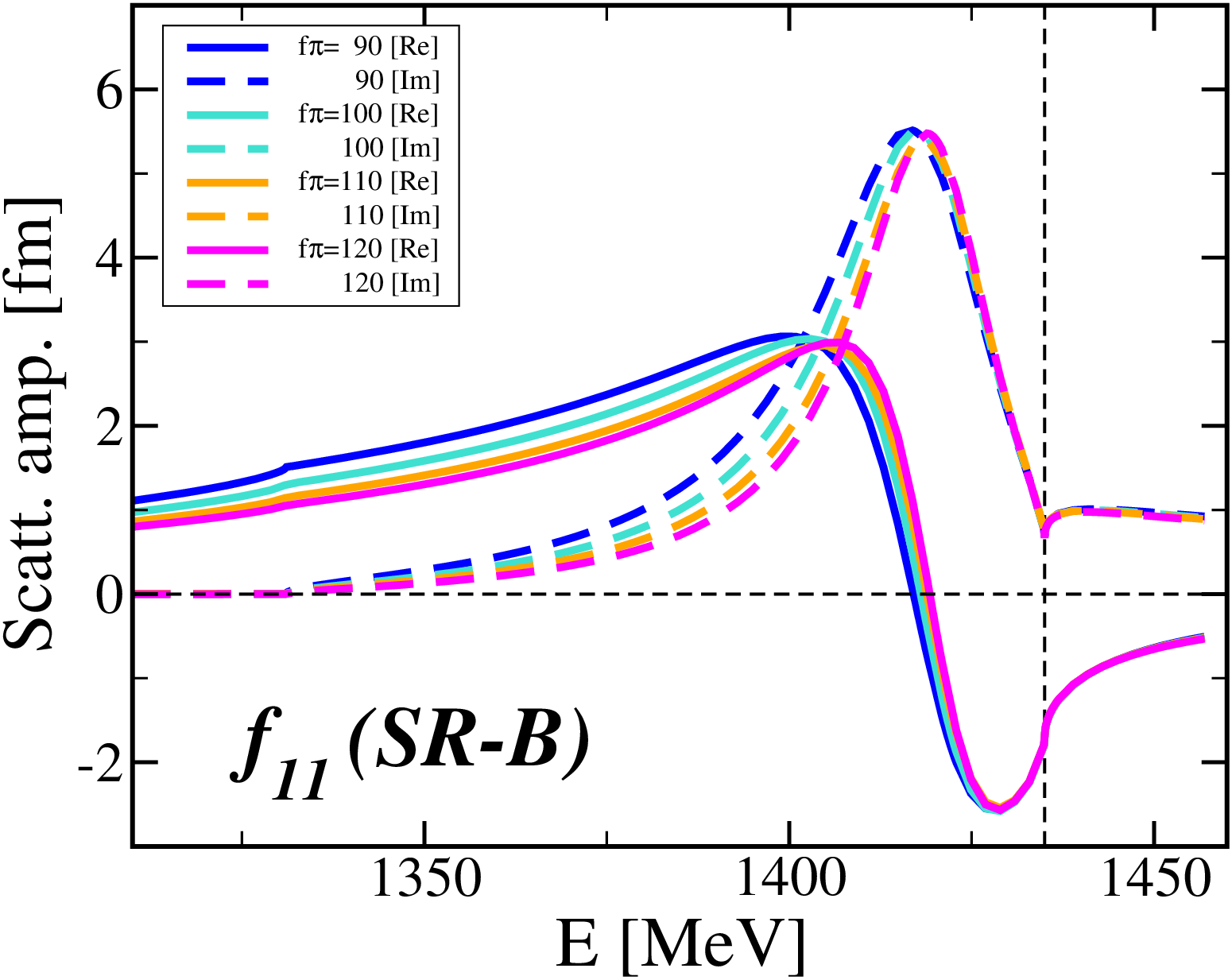}
  \hspace{0.3cm}
  \includegraphics[width=.47\textwidth]{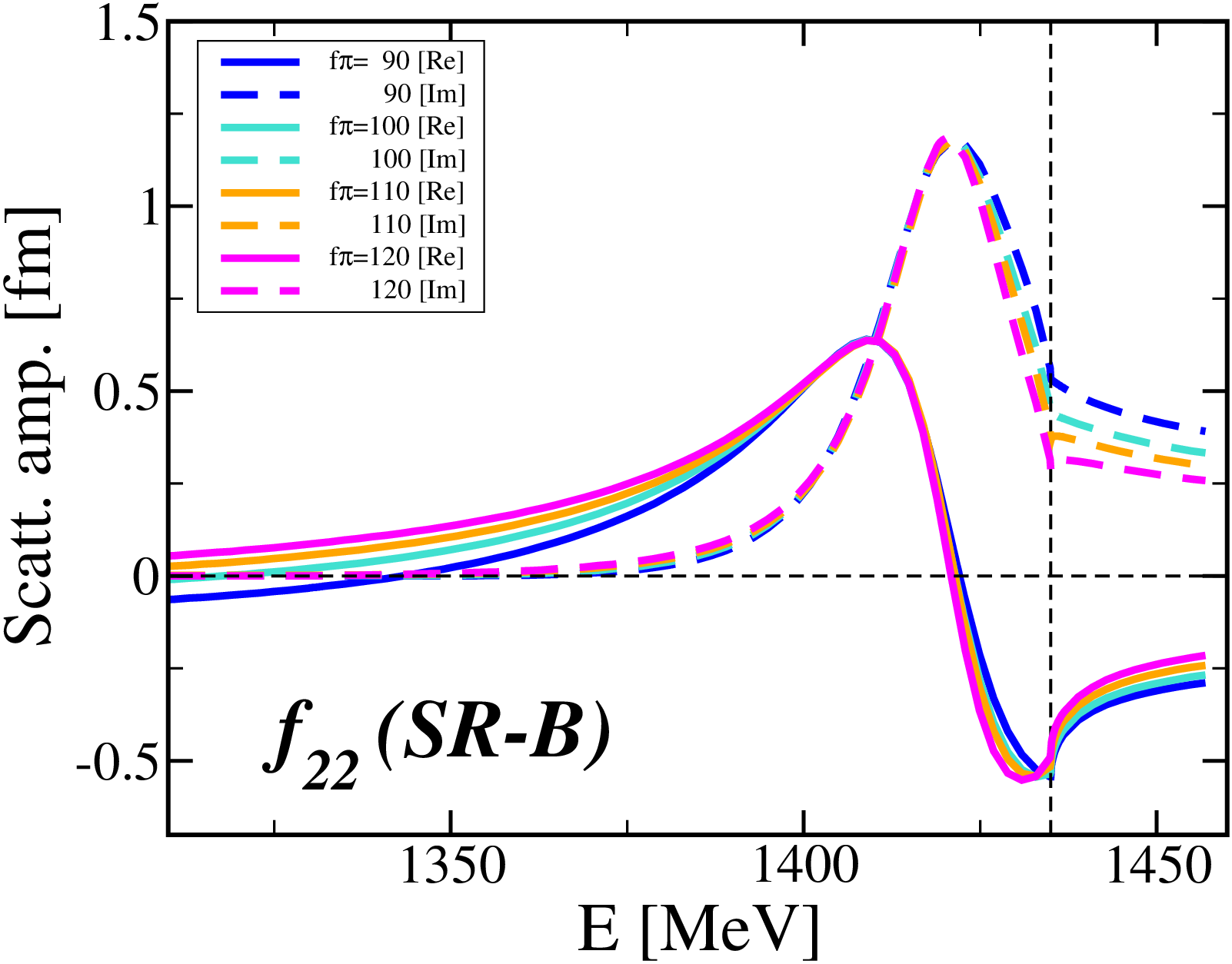}
  \caption{$I=0$ scattering amplitudes for SR-B case, 
calculated with various $f_\pi$ values. 
Similar to Fig.~\ref{KSW_I=0_mfpi_NRv2}. 
\label{KSW_I=0_mfpi_SR-B}}
\end{figure}

\begin{figure}
  \includegraphics[width=.47\textwidth]{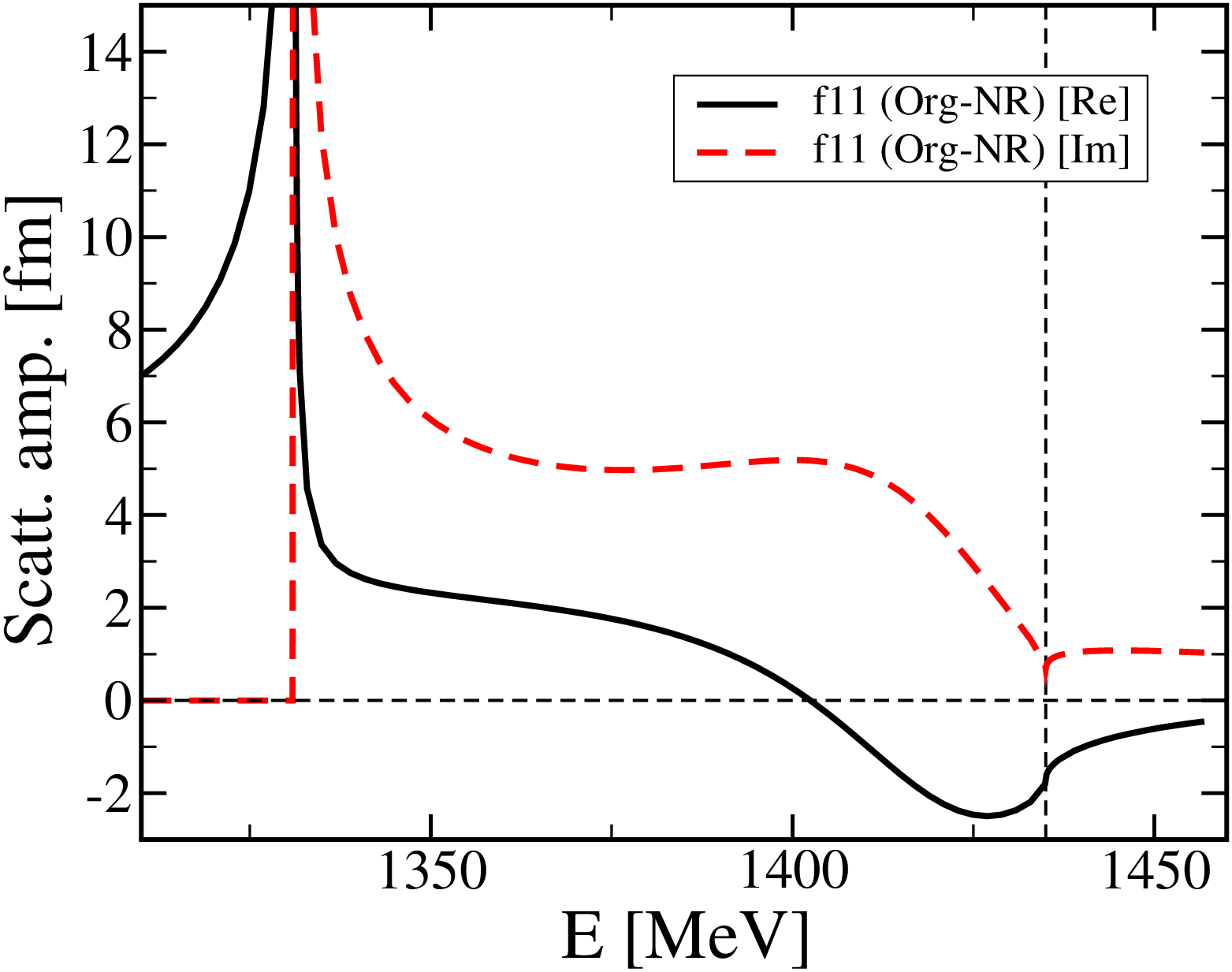}
  \hspace{0.3cm}
  \includegraphics[width=.47\textwidth]{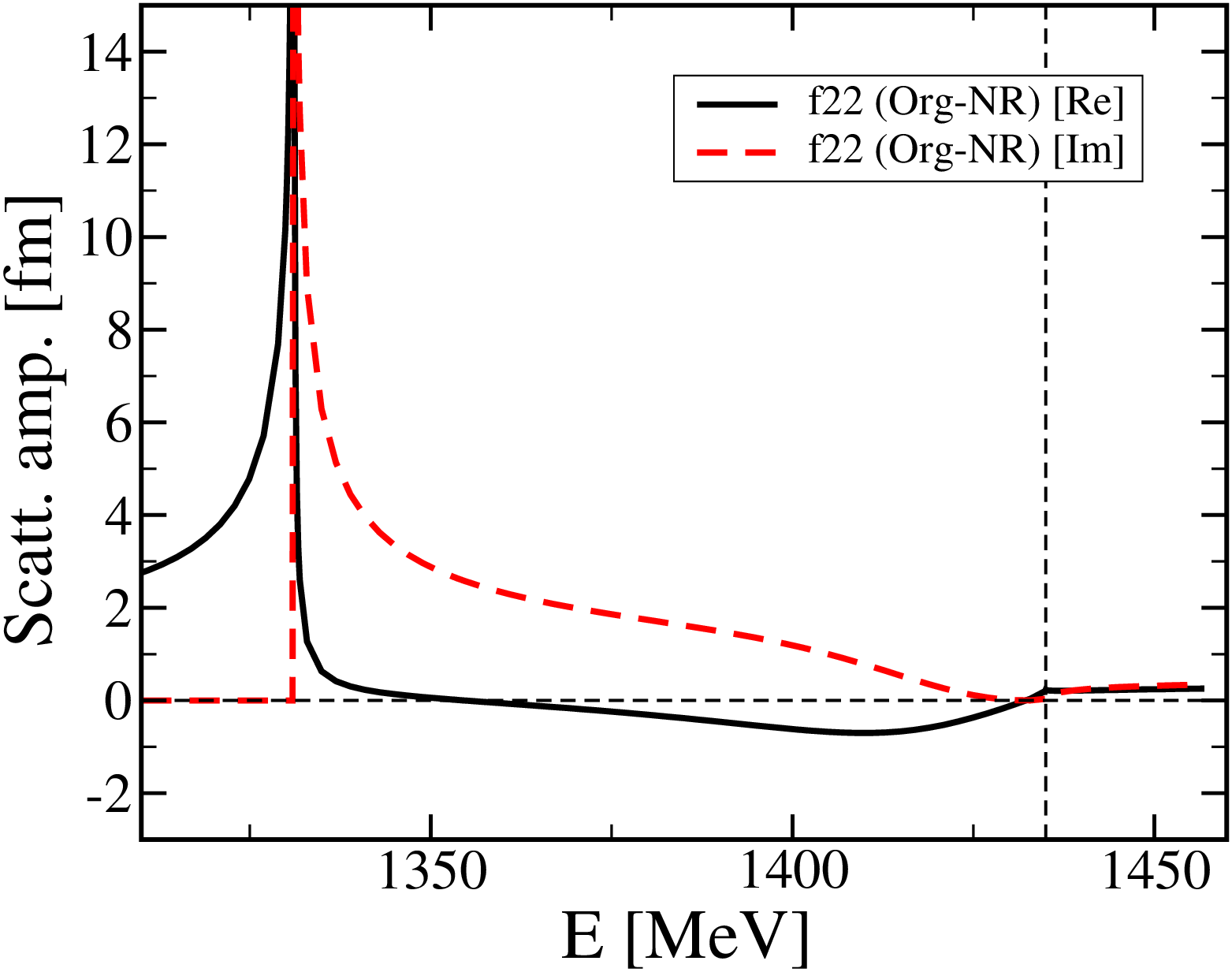}
  \caption{$I=0$ scattering amplitudes calculated with the KSW-type potential 
used in NR kinematics. $f_\pi=90$ MeV. Similar to Fig.~\ref{KSW_I=0_caseNR}. 
The $\bar{K}N$ scattering length calculated is 
$a_{\bar{K}N (I=0)}=-1.709+i0.679$ fm when the range parameters are 
$(d_{\bar{K}N,\, \bar{K}N}, d_{\pi\Sigma,\, \pi\Sigma}) = (0.593, 0.541)$ fm. 
\label{KSW_I=0_case0}}
\end{figure}

We investigate the $f_\pi$ dependence of the scattering amplitude. 
Figs.~\ref{KSW_I=0_mfpi_NRv2} - \ref{KSW_I=0_mfpi_SR-B} show 
the scattering amplitudes of the NRv2, SR-A and SR-B cases, respectively, 
in which $f_\pi$ is varied from 90 to 120 MeV. 
In the NR case, it is found that the scattering amplitude depends strongly 
on the $f_\pi$ value, especially near the $\pi\Sigma$ threshold as shown 
in Fig.~\ref{KSW_I=0_mfpi_NRv2}. 
The SR-A case has a similar tendency as shown 
in Fig.~\ref{KSW_I=0_mfpi_SR-A}. 
On the other hand, in the SR-B case which is the other semi-relativistic case, 
the amplitudes don't depend on the $f_\pi$ value so much. 
(See Fig.~\ref{KSW_I=0_mfpi_SR-B}.)  
In all cases, the amplitudes far below $\bar{K}N$ threshold 
tend to be less attractive at larger value of $f_\pi$.

For instruction, we have examined a case where kinematics and potential 
are mismatched.   
When the KSW-type potential is used under the non-relativistic kinematics,  
the scattering amplitude behaves singularly as shown 
in Fig.~\ref{KSW_I=0_case0}, which is the most typical case ($f_\pi=90$ MeV). 
Both the $\bar{K}N$ and $\pi\Sigma$ 
scattering amplitudes are singular at the $\pi\Sigma$ threshold. 
By investigating the relation between scattering length $a_{\pi\Sigma}$ 
and effective range $r_{e}$ in the $\pi\Sigma$ channel, 
it is found that a virtual state is generated in this case. 
Obtained values of $(a_{\pi\Sigma}, r_{e})$ are $(61,-6.3)$ fm. 
According to Appendix A in Ref. \cite{KN-pS:IHJ}, 
$(a_{\pi\Sigma}, r_{e})$ satisfying the condition $-a_{\pi\Sigma}/2 < r_{e}$
is an indication of the existence of a virtual state without decay width. 
This virtual state causes such a singularity in scattering amplitudes. 
We consider that the KSW-type potential should be used under 
the relativistic kinematics (semi-relativistic, at least) 
since the flux factor involved in this potential 
is based on relativistic kinematics.

\begin{figure}
\centerline{
  \includegraphics[width=.47\textwidth]{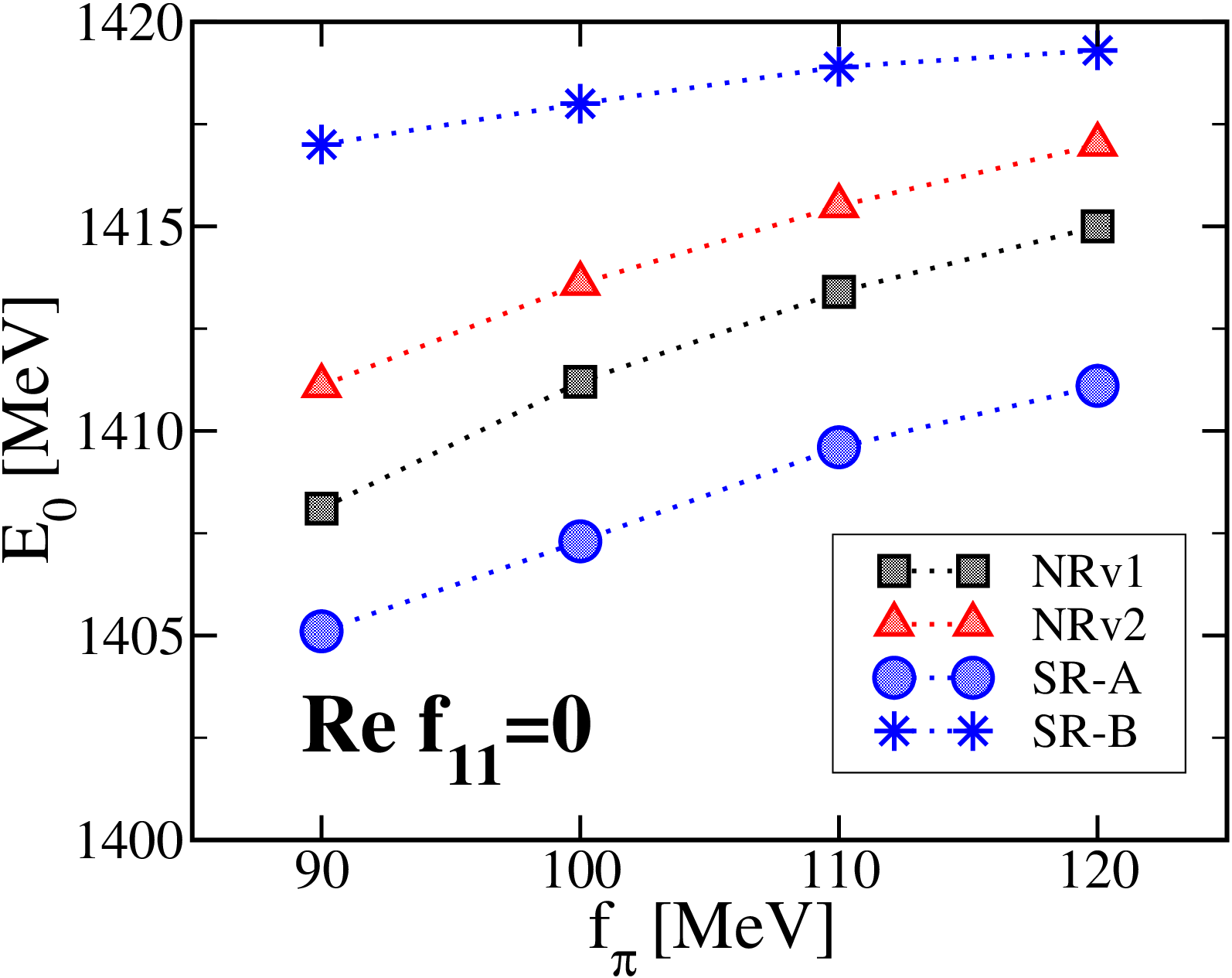}
\hspace{0.3cm}
  \includegraphics[width=.47\textwidth]{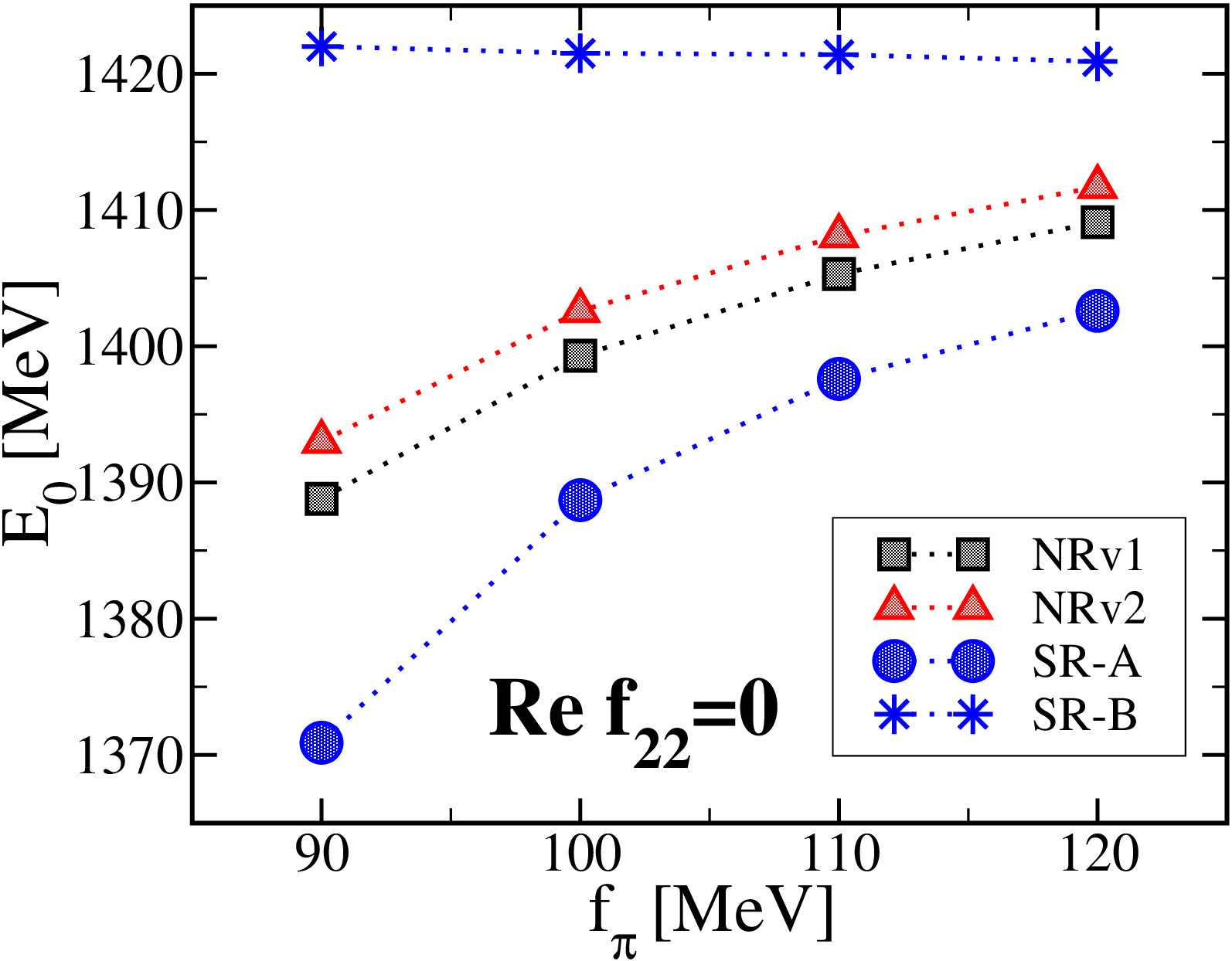}
}
  \caption{
Resonance energies estimated from scattering amplitudes. $f_\pi=90 \sim 120$ MeV
(Left) $E_{0}(\bar{K}N)$ obtained from the $\bar{K}N$ scattering amplitude. 
(Right) $E_{0}(\pi\Sigma)$ obtained from the $\pi\Sigma$ one. 
Black square and red triangle mean NRv1 and NRv2, respectively. 
Blue circle and asterisk mean SR-A and SR-B, respectively. 
\label{KSW_I=0_f=0}}
\end{figure}

In all cases, the resonance structure is found 
in the $\bar{K}N$ and $\pi\Sigma$ scattering amplitudes 
below the $\bar{K}N$ threshold. 
We estimate the resonance energies $E_{0}(\bar{K}N)$ and $E_{0}(\pi\Sigma)$ 
from the scattering amplitudes $f_{\bar{K}N}$ and $f_{\pi\Sigma}$ by 
putting the conditions on 
${\rm Re} \, f_{\bar{K}N}(E_{0}(\bar{K}N)) = 0$ and 
${\rm Re} \, f_{\pi\Sigma}(E_{0}(\pi\Sigma)) = 0$, respectively. 
For various $f_\pi$, $E_{0}(\bar{K}N)$ and $E_{0}(\pi\Sigma)$ 
are shown in Fig.~\ref{KSW_I=0_f=0}. 
In the non-relativistic kinematics 
the estimated energies are not so different between NRv1 and NRv2, 
since the scattering amplitudes are almost same in these cases as shown 
in Fig.~\ref{KSW_I=0_caseNR}. 
In the semi-relativistic kinematics, there are two sets of the parameters, 
SR-A and SR-B, as explained before. Since these two sets 
give quite different scattering amplitudes as shown in Figs. 
\ref{KSW_I=0_mfpi_SR-A} and~\ref{KSW_I=0_mfpi_SR-B}, the resonance energies 
are also different between them. 
Compared with the NR cases, 
$E_{0}(\bar{K}N)$ and $E_{0}(\pi\Sigma)$ in SR-A are smaller, 
while those in SR-B are larger. In particular, the $f_\pi$ dependence 
of $E_{0}(\pi\Sigma)$ in SR-B is quite different from the NR cases, 
while that in SR-A is rather strong but qualitatively similar to them. 
In SR-B, the resonance energies (especially $E_{0}(\pi\Sigma)$) are 
quite stable for $f_\pi$ value and remain to be around 1420 MeV.


\subsection{Property of the $I=0$ $\bar{K}N$-$\pi\Sigma$ resonant state}
\label{Sec:Resonace_I=0}

\begin{table}
\caption{Pole position of the $I=0$ $\bar{K}N$-$\pi\Sigma$ system 
and the meson-baryon distance in the pole state. 
$(E_R, -\Gamma/2)$ indicates the complex energy of the resonance pole. 
$B_{\bar{K}N}$ means the binding energy that is the $E_R$ measured from the $\bar{K}N$ threshold. The unit of these energies is MeV. 
$\sqrt{\langle r^2 \rangle}_{\bar{K}N}$, $\sqrt{\langle r^2 \rangle}_{\pi\Sigma}$ and $\sqrt{\langle r^2 \rangle}_{\bar{K}N+\pi\Sigma}$ indicate the meson-baryon mean distance of $\bar{K}N$, $\pi\Sigma$ and total components, respectively. 
These values are given in unit of fm. 
$f_\pi=110$ MeV case. \label{Poles_I=0}}
\begin{tabular*}{\textwidth}{l@{\hspace{0.3cm}}|@{\hspace{0.3cm}} r@{\hspace{0.5cm}}r@{\hspace{0.8cm}}r@{\hspace{0.5cm}}r}
\hline
Case  & NRv1 & NRv2 & SR-A & SR-B  \\
\hline
Kinematics& \multicolumn{2}{c}{Non-rela.} & \multicolumn{2}{c}{Semi-rela.} \\
Potential & NRv1 & NRv2 & \multicolumn{2}{c}{KSW-type}   \\
\hline
&&&&\\
$E_R$ & 1416.6 & 1417.8 & 1419.5 &1420.0 \\
$\Gamma/2$ & 19.5 & 16.6 & 25.0 &12.8 \\
($B_{\bar{K}N}$) & (18.4) & (17.2) & (15.5) &(15.0) \\
&&&&\\
$\sqrt{\langle r^2 \rangle}_{\bar{K}N}$ &$1.31-0.37i$&$1.37-0.37i$& 
$1.22-0.47i$ &$1.18-0.49i$\\
$\sqrt{\langle r^2 \rangle}_{\pi\Sigma}$&$0.39+0.05i$&$0.37+0.04i$& 
$0.13+0.05i$ &$0.11-0.06i$\\
$\sqrt{\langle r^2 \rangle}_{\bar{K}N+\pi\Sigma}$ &$1.36-0.34i$&$1.42-0.34i$& 
$1.22-0.47i$ &$1.18-0.49i$\\
\hline
\end{tabular*} 
\end{table}

Using the meson-baryon potential determined in the previous section, 
we investigate the resonance in the $I=0$ $\bar{K}N$-$\pi\Sigma$ system. 
In practice, we search poles on the complex-energy plane with the usual 
complex scaling method as explained in the section~\ref{CSM:usual}. 
In all NR and SR cases, one pole is clearly found.\footnote{
Preceding studies based on the chiral SU(3) theory reported 
that there exist two poles in the $I=0$ channel 
\cite{ChU:Jido, ChU:HW} We will discuss a signature of another pole later.} 
We denote the complex energy of a resonance pole as $(E_R, -\Gamma/2)$. 
The found poles for the case $f_\pi=110$ MeV are shown in Table~\ref{Poles_I=0}. The real part of energy $E_R$ is well determined 
to be about 1420 MeV, independently of kinematics 
and potential types. The imaginary part of energy $\Gamma/2$ depends 
on the kinematics. The $\Gamma/2$ is about 18 MeV in the NR. 
In the SR, it should be noted that the two potentials give quite 
different values; $\Gamma/2 \simeq 25$ MeV in the SR-A and 
$\Gamma/2 \simeq 13$ MeV in the SR-B.  

\begin{figure}
\centerline{
  \includegraphics[width=.60\textwidth]{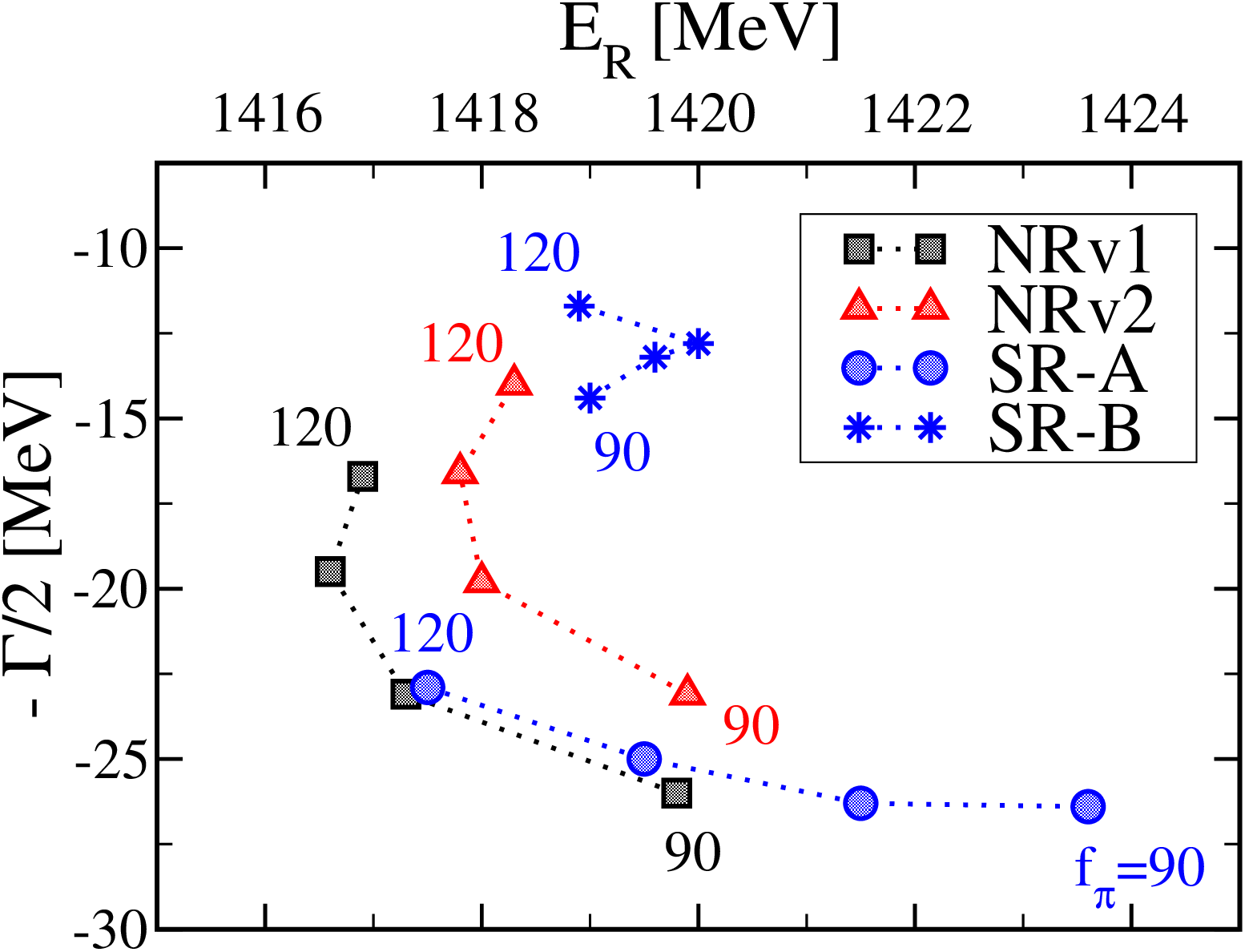}
}
  \caption{$I=0$ pole in the complex-energy plane, calculated with 
NRv1, NRv2, and SR for various $f_\pi$ values. 
Black square and red triangle mean NRv1 and NRv2, respectively. 
Blue circle and asterisk mean SR-A and SR-B, respectively. 
The number shown in the panel indicates the $f_\pi$ value which increases 
continuously from 90 to 120 along the connected dashed line 
in each case. 
\label{KSW_I=0_poles_mfpi}}
\end{figure}

Fig.~\ref{KSW_I=0_poles_mfpi} shows the pole positions of 
all cases when the $f_\pi$ value is varied from 90 MeV to 120 MeV. 
In the NR cases, 
$E_R$ is stable for $f_\pi$ and is 1417-1420 MeV, 
but $\Gamma/2$ is ranging from 14 MeV to 26 MeV.
In the SR cases, the poles of SR-A and those of SR-B show  
completely different behavior for the variation of $f_\pi$. 
In the SR-A case, the pole moves widely in the $E_R$ direction 
from 1417 MeV to 1424 MeV, while keeping $\Gamma/2$ to be about 25 MeV. 
On the other hand, in the SR-B case the poles are found to distribute 
in compact region. The pole position $(E_R, -\Gamma/2)$  is determined 
with small deviation; $(1419.5\pm1, \, -13\pm2)$ MeV. 
From these results, it seems that the real energy of the pole is 
rather determined well, compared to the imaginary energy.  
The imaginary energy of the pole indicates a half width 
decaying to $\pi\Sigma$. In our study we have a constraint for $\bar{K}N$ 
channel but no constraint for $\pi\Sigma$ channel. 
We consider that the large deviation of $\Gamma/2$ is 
due to the lack of constraint condition for $\pi\Sigma$ channel.  

Here, we mention the difference between the SR-A and SR-B which are two 
solutions of the semi-relativistic case. 
Observing the pole behavior for the $f_\pi$ variation in Fig. 
\ref{KSW_I=0_poles_mfpi}, 
we notice that the pole behavior of the SR-A is qualitatively 
the same  as that of the NR cases, while the SR-B shows completely 
different behavior from the NR cases. 
In the SR-A case, when the $f_\pi$ decreases, 
the pole moves with $E_R$ and $\Gamma/2$ decreasing 
in the same way as the NR cases. 
However, in the SR-B case the pole moves quite in a different way.
Taking into account also results of the scattering amplitude as 
given in the previous section, 
we consider that {\it a semi-relativistic solution, SR-A, can be regarded as 
a kind of semi-relativistic version of the non-relativistic solutions 
(NRv1 and NRv2)}, because it has qualitatively the similar properties 
to the NR's. The other one, the SR-B, is a unique solution 
to the semi-relativistic kinematics, which has completely different 
properties from the NR's.

We have investigated another pole which is expected to exist, 
because many studies of the $I=0$ $\bar{K}N$-$\pi\Sigma$ system reported 
that this system has a double pole structure 
when an energy-dependent chiral SU(3) potential is used as we use 
\cite{ChU:Jido, ChU:HW}; 
The higher pole state is slightly below the $\bar{K}N$ threshold 
and with small width, 
while the lower one is far below the $\bar{K}N$ threshold and 
with large width. (For instance, the former is around (1432, $-$17) MeV
and  the latter is around (1400, $-$80) MeV \cite{ChU:HW}.)
In our study, certainly we have found self-consistent solutions 
which seem to indicate a lower pole of the two poles; 
In the non-relativistic kinematics, it is found at the complex energy 
$(\sim 1360, -90\sim-40)$ MeV in case of NRv2, and in the semi-relativistic kinematics it is around $(1350\sim1390, -100\sim-30)$ in case of SR-A, 
though no self-consistent solutions are found in the deep binding region 
in case of SR-B.  
In the previous subsection scattering amplitudes of 
the NRv2 and SR-A are shown in Figs.~\ref{KSW_I=0_caseNR} and 
\ref{KSW_I=0_caseSR-A}, respectively. 
These scattering amplitudes are found to have qualitatively a similar 
feature to the amplitudes obtained in the former studies 
\cite{ChU:HW, KN-pS:IHJ} which involve the double pole structure. 
With this fact, 
the self-consistent solution of the NRv2 and SR-A is expected 
to be a lower pole. 
However, the position of these poles, in particular the imaginary energy, 
rather depends on the scaling angle $\theta$. In the NRv2 case, the separation 
of the pole from continuum states indicated by the $2\theta$-line on 
the complex-energy plane seems insufficient. 
We consider that these difficulties to specify the pole position are 
due to the limitation of the numerical accuracy. 
When poles have large decay width compared to excitation energy, 
it is known empirically that such poles are difficult 
to be found by the CSM with Gaussian base, 
since the spatial oscillation of the complex-scaled wave function 
is not well described with them. 
Therefore, 
we cannot conclude yet that the double pole structure is confirmed 
in the present our analysis. We need more investigation of the lower pole. 

\begin{figure}
  \includegraphics[width=.32\textwidth]{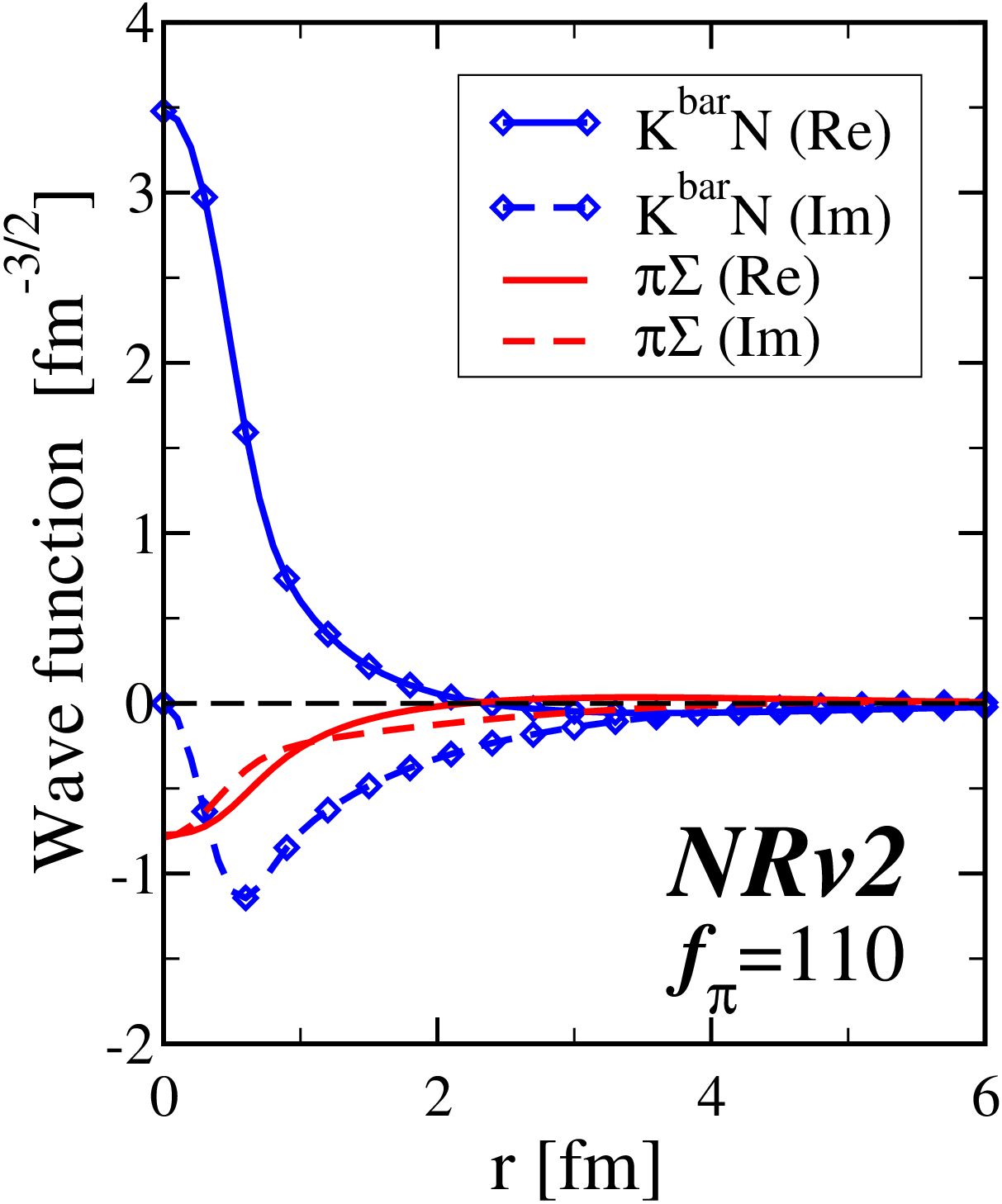}
  \hspace{0.0cm}
  \includegraphics[width=.32\textwidth]{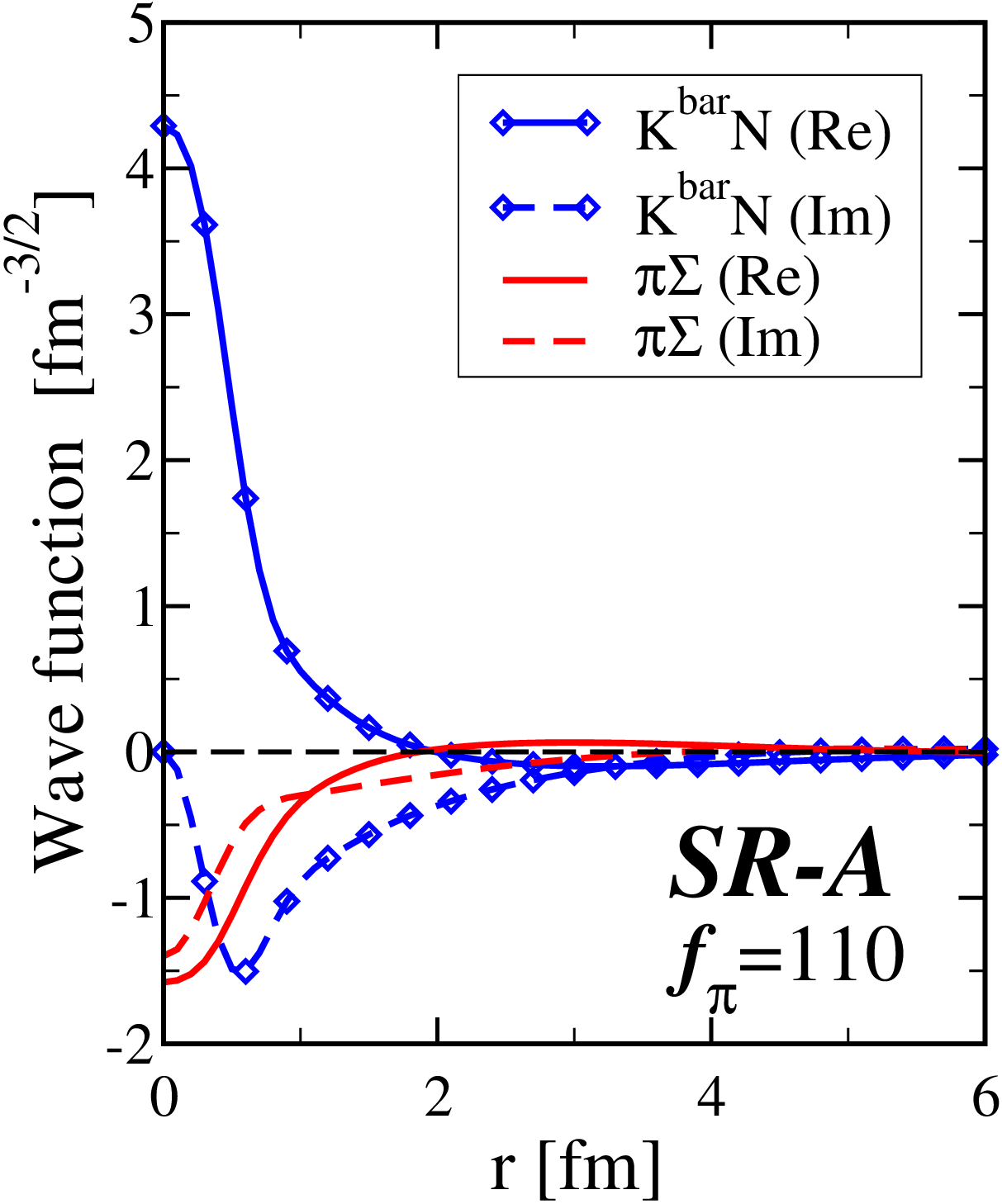}
  \hspace{0.0cm}
  \includegraphics[width=.32\textwidth]{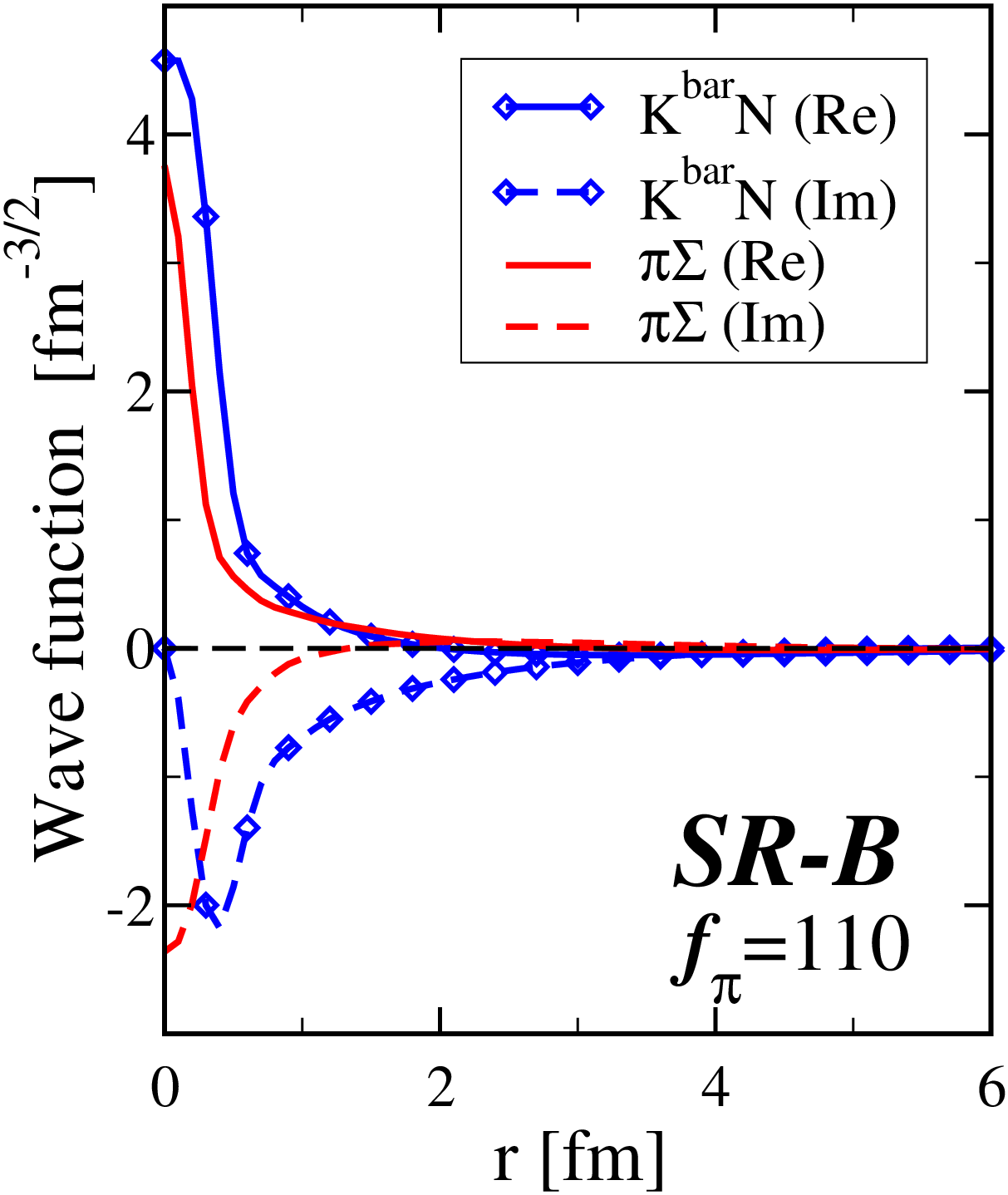}
  \caption{Complex-scaled wave function of each component 
in the $I=0$ pole state at $\theta=30^\circ$. 
We choose a phase such that the $\bar{K}N$ 
wave function becomes real at $r=0$. 
Blue (red) lines show $\bar{K}N$ ($\pi\Sigma$) wave function, whose 
real (imaginary) part is drawn with solid (dashed) lines. 
Left: NRv2, middle: SR-A, right: SR-B. 
$f_\pi=110$ MeV case. \label{KSW_I=0_wfnc}}
\end{figure}

We are interested in the inertial structure of $\Lambda(1405)$, 
which has been investigated theoretically in various ways 
\cite{ChU_L1405str:Yamagata, ChU_L1405str:Sekihara, L1405str:Takeuchi}. 
We check the wave function of the pole state obtained in 
our ccCSM calculation. 
Fig.~\ref{KSW_I=0_wfnc} shows complex-scaled wave functions 
of each component with the scaling angle $\theta=30^\circ$. 
Here, the wave functions are multiplied by an appropriate phase factor 
so that the $\bar{K}N$ wave function becomes real at $r=0$. 
Without this phase factor, the complex-scaled wave function of 
$\bar{K}N$-$\pi\Sigma$ is normalized using Eq. (\ref{Eq:Norm_CSwfnc}). 
Both of $\bar{K}N$ and $\pi\Sigma$ wave functions are confirmed to be 
well localized. It is noted that localization of the $\pi\Sigma$ 
component is due to the complex scaling, in spite that the state is 
above $\pi\Sigma$ threshold.  
In these wave functions, it can be confirmed also that 
a semi-relativistic case SR-B shows especially different nature 
from other cases. 
The SR-B wave function is much compact and its real part of $\pi\Sigma$ 
component has the opposite phase to other cases. 
The wave function of the other semi-relativistic case SR-A 
is similar to that of a non-relativistic one NRv2. 

The mean distance between meson and baryon in 
the resonant state is calculated as 
\begin{eqnarray}
\langle \, \hat{r}_{MB}^2 \, \rangle= 
\langle \tilde{\Phi}_\theta | \, 
\hat{r}^2_{MB, \, \theta} \, | \Phi_\theta \rangle, \label{Eq:Rrms}
\end{eqnarray}
where $\hat{\bm r}_{MB} = \hat{\bm r}_{\rm Meson} -  \hat{\bm r}_{\rm Baryon}$
and $\Phi_\theta$ means the complex-scaled wave function of the resonance pole. 
It should be noted that 
the matrix elements of resonant state are obtained independently of $\theta$
because the properties of the resonant wave functions are uniquely determined 
as the Gamow states \cite{CSM_Rrms:Homma}. 
The expectation value calculated with Eq. (\ref{Eq:Rrms}) is inevitably  
a complex number because the resonant state is treated as
 a Gamow state in our framework. Therefore, the root-mean square distance, 
$\sqrt{\langle \, \hat{r}_{MB}^2 \, \rangle}$, is also a complex number. 
Certainly its physical meaning is still unclear, but we show this quantity 
as a reference for the size. We believe that it is useful 
for us to get a feeling of the size of the system. 
Indeed, since the imaginary part of the obtained complex mean distance 
is smaller than its real part as will be shown later, we consider that 
the real part can be regarded as a mean distance 
with a physical meaning \cite{CSM_Rrms:Homma}. 

If such an interpretation for the complex-valued distance 
calculated within the ccCSM is accepted, 
the mean distance between meson and baryon 
is $\sim$1.4 fm in NR kinematics and $\sim$1.2 fm in SR kinematics 
when $f_\pi=110$ MeV,  as shown in Table~\ref{Poles_I=0}. 
Those for other $f_\pi$ values are given in Table~\ref{ResPole_100-120}. 
In both semi-relativistic cases of SR-A and SR-B, 
the mean distance has small $f_\pi$ dependence, remaining about 1.2 fm. 
In the NR case it depends on the $f_\pi$ value similarly 
to the pole position, but it increases slightly from 1.2 fm to 1.4 fm 
when $f_\pi$ increases from 90 MeV to 120 MeV. Thus, it is found that 
the mean distance is almost the same in both two kinematics. 
As for the imaginary part of the complex-valued distance, 
it is certainly small value of 0.5 fm at most. 

Compared to results of other studies,  
the size obtained in our calculation seems rather small, 
even if the modulus of $\sqrt{\langle \, \hat{r}_{MB}^2 \, \rangle}$ 
is regarded as a mean distance between meson and baryon in the $I=0$ system. 
For example, according to the study using a chiral SU(3)-based potential, 
the size is obtained as 1.9 fm with 12 MeV binding energy of $\bar{K}N$ 
($M=1423$ MeV) \cite{Kpp:DHW}. 
It is considered that the difference of sizes 
between two calculations is mainly caused by different definition of 
the resonant state. In the previous study, the state is 
treated as a $\bar{K}N$ bound state, as a result of elimination of 
$\pi\Sigma$ channel and perturbative treatment of the imaginary part of 
potential. On the other hand, the resonant state is a Gamow state 
in the current study since the complex scaling method imposes 
the correct outgoing boundary condition on a solution implicitly. 

\subsection{Comparison with other studies of $I=0$ $\bar{K}N$-$\pi\Sigma$ scattering amplitude}

There are many studies of the $I=0$ $\bar{K}N$-$\pi\Sigma$ scattering 
amplitude. Here, we compare our result mainly with that of 
Ref. \cite{KN-pS:IHJ},  because some of their models are constructed 
under the same condition as our study:  
Their models ``A1'' and ``B E-dep'' employ the Weinberg-Tomozawa term 
as an interaction kernel and are constrained with the $I=0$ $\bar{K}N$ 
scattering length. 
Their amplitudes are calculated under the relativistic kinematics. 
Therefore, their results of models A1 and B E-dep can be directly 
compared with those of the SR case of our study. 

\begin{figure}
  \includegraphics[width=.47\textwidth]{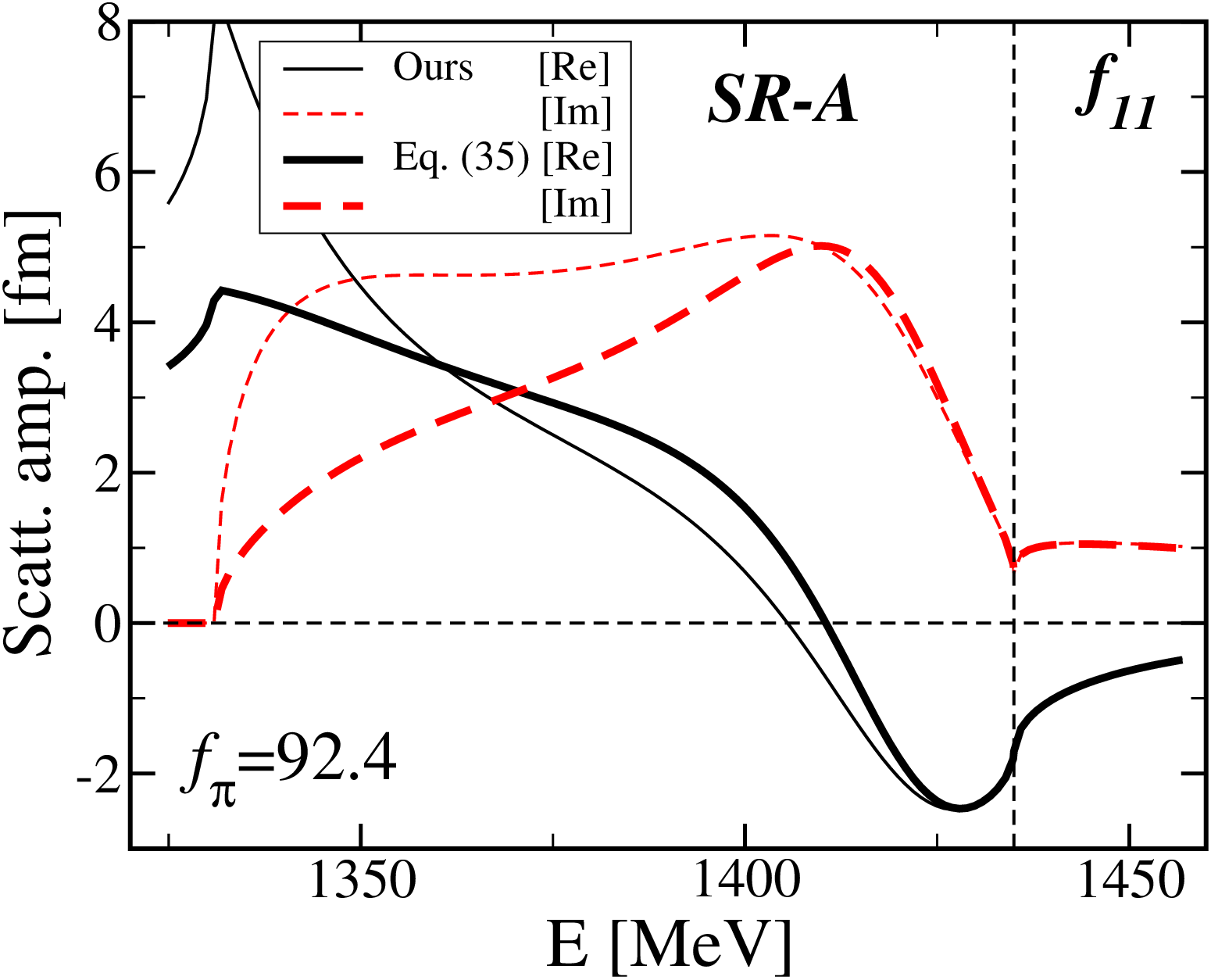}
  \hspace{0.3cm}
  \includegraphics[width=.47\textwidth]{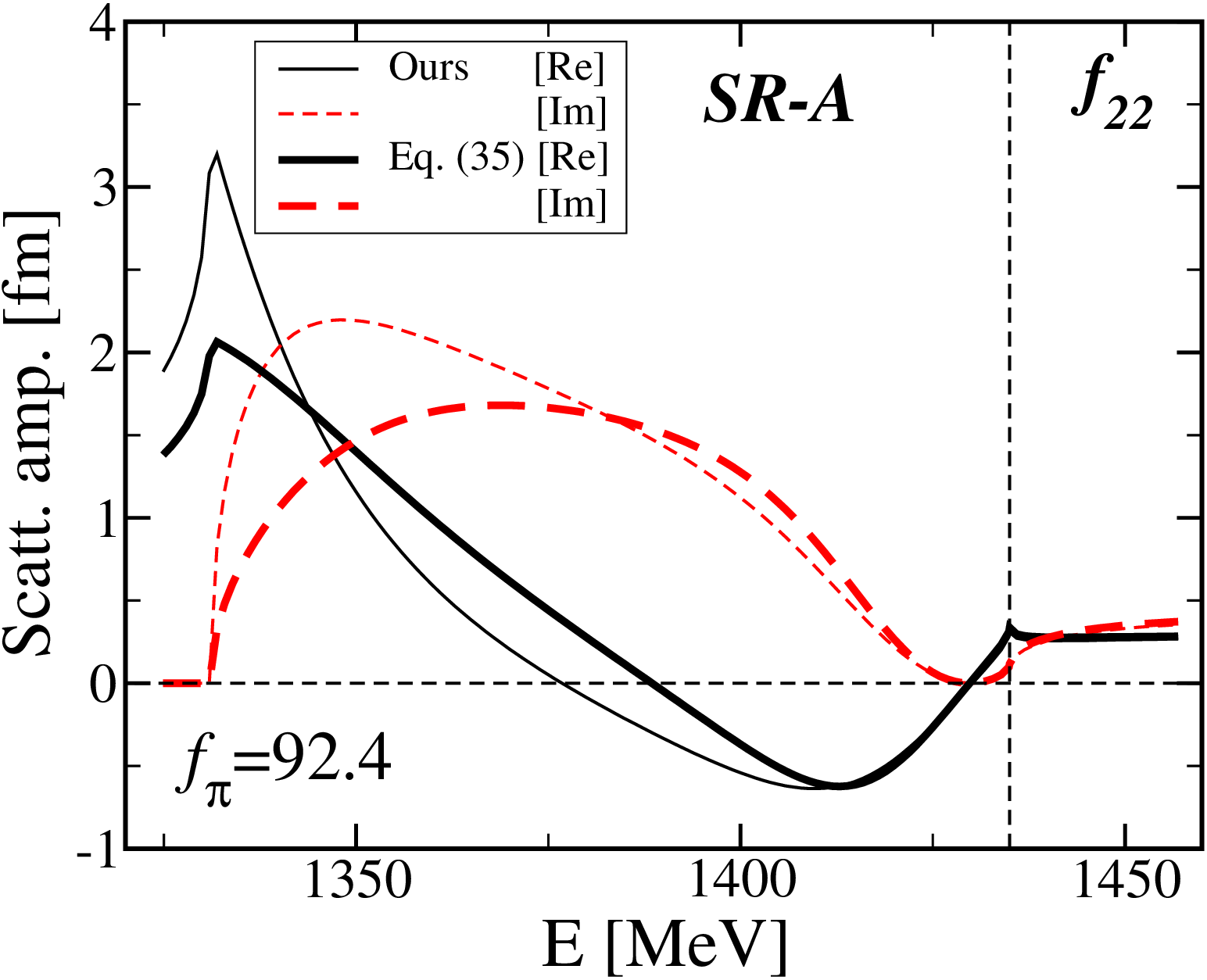}
\\
\\
  \includegraphics[width=.47\textwidth]{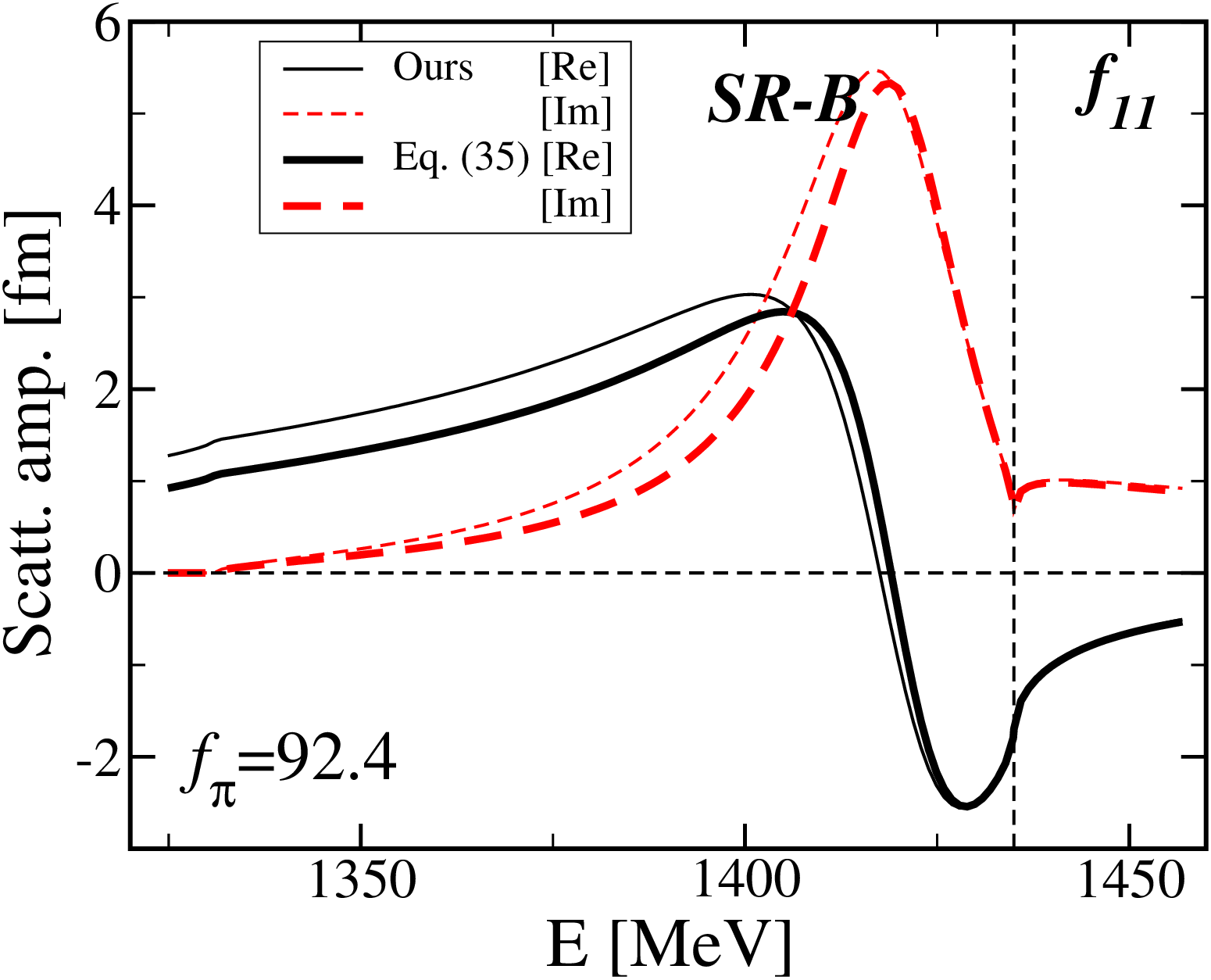}
  \hspace{0.3cm}
  \includegraphics[width=.47\textwidth]{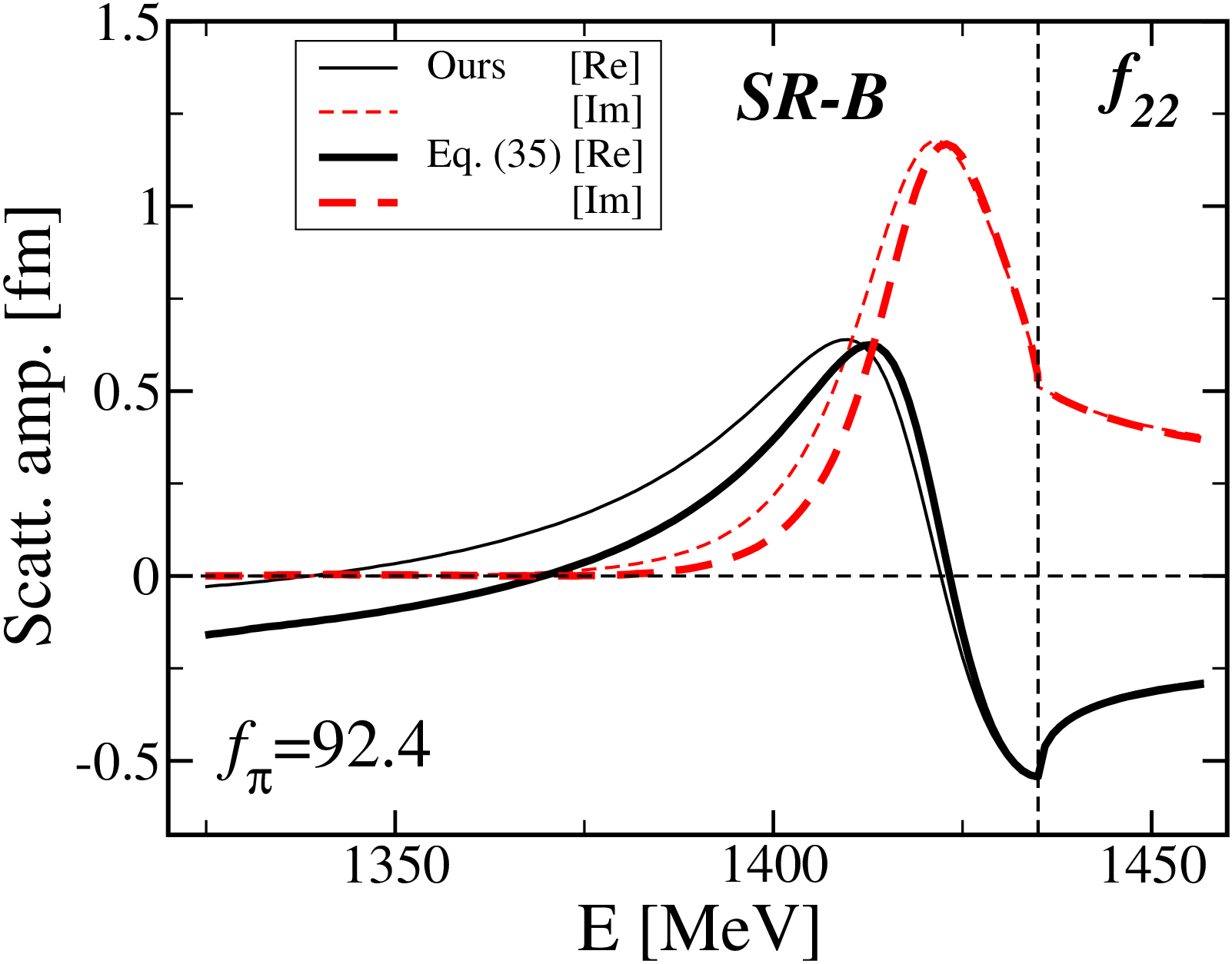}
  \caption{$I=0$ scattering amplitudes calculated with 
different interaction kernels. The thick (thin) line indicates the amplitude 
obtained with the pseudo-potential Eq. (\ref{Eq:Pseudo-pot_IHJ}) 
(Eq. (\ref{PotMB_SR})). 
Left (right) panels show $\bar{K}N$ ($\pi\Sigma$) scattering amplitude. 
Two types of scattering amplitudes, SR-A and SR-B, are given 
in upper and lower panels, respectively.  $f_\pi=92.4$ MeV. \label{Comp-KSW-IHJ}}
\end{figure}

Similarly to Ref. \cite{KN-pS:IHJ}, we set the value of $f_\pi$ to be 
92.4 MeV in our SR calculation. Also with this $f_\pi$ value, 
we find two parameter sets which correspond to SR-A and SR-B, 
while in Ref. \cite{KN-pS:IHJ} a single solution is reported 
for each model. 
Fig.~\ref{Comp-KSW-IHJ} shows the scattering amplitudes of both SR-A and SR-B. 
Compared with the scattering amplitudes of model A1 shown in Fig. 1 
in their paper, both of SR-A and SR-B are found to give different 
amplitudes from them. In a case of SR-A, 
both scattering amplitudes of $\bar{K}N$ and $\pi\Sigma$ behave 
similarly to those of model A1 near the $\bar{K}N$ threshold. 
However, far from the $\bar{K}N$ threshold they are quite different 
from the amplitudes of model A1. In the other case of SR-B, 
the $\bar{K}N$ amplitude is almost the same as that of model A1, whereas 
the $\pi\Sigma$ one is completely different from that of model A1. 
As for resonance poles, we find a pole at (1423.1, $-$26.4) MeV and 
(1419.4, $-$14.1) MeV in SR-A and SR-B, respectively. 
It is known that the models in Ref. \cite{KN-pS:IHJ} give double pole structure. 
The higher pole of them are at (1422, $-$16) MeV in the model A1 
and (1422, $-$22) MeV in the model B E-dep. 
They are rather close to the poles found in the SR-B and SR-A, respectively. 

We notice that the interaction kernel is slightly different between 
our study and Ref. \cite{KN-pS:IHJ}. The energy dependence 
of the interaction kernel is just the sum of meson energies such 
as $\omega_\alpha+\omega_\beta$ which is involved in the pseudo-potential 
$\hat{V}_{MB}^{SR}$ as shown in Eq. (\ref{PotMB_SR}). 
On the other hand, it is given as 
$(2\sqrt{s}-M_\alpha-M_\beta) \sqrt{\frac{E_\alpha+M_\alpha}{2M_\alpha}}\sqrt{\frac{E_\beta+M_\beta}{2M_\beta}}$ in their study. 
Compared with our interaction kernel, the relativistic $q^2$ correction 
and the non-static effect of baryons are taken into account by the first term 
and the additional square-root terms, respectively. 
We investigate the influence of interaction kernels with 
different energy dependence. 
Here, we make a pseudo-potential from their interaction kernel, following 
our ansatz that a relativistic flux factor is used and a Gaussian 
form is assumed: 
\begin{eqnarray}
\hat{V}_{MB}^{Ref. \,[33]} & = & \sum_{\alpha, \beta} 
\; - \frac{C^I_{\alpha \beta}}{8 f_\pi^2} (2\sqrt{s}-M_\alpha-M_\beta) 
\sqrt{\frac{E_\alpha+M_\alpha}{2M_\alpha}}
\sqrt{\frac{E_\beta+M_\beta}{2M_\beta}} \nonumber\\
&& \quad \quad  \quad \quad  \quad \quad  \quad \quad 
\times \sqrt{\frac{M_\alpha M_\beta}{s\, \tilde{\omega}_\alpha \tilde{\omega}_\beta}}
\; g^I_{\alpha \beta} (r) 
\; |\alpha\rangle \langle \beta |. \label{Eq:Pseudo-pot_IHJ}
\end{eqnarray} 

Scattering amplitudes calculated with this pseudo-potential 
are drawn with thick line in Fig.~\ref{Comp-KSW-IHJ}.  
Two sets of range parameters are found also for this potential. 
It is confirmed that the scattering amplitudes differ slightly 
from those obtained with the pseudo-potential $\hat{V}_{MB}^{SR}$. 
In the SR-A, in particular, the amplitudes near the $\pi\Sigma$ 
threshold are suppressed, compared with those with $\hat{V}_{MB}^{SR}$. 
However, they are still larger than the scattering amplitudes 
of the model A1 in Ref. \cite{KN-pS:IHJ}. 
Therefore, the difference between our result and Ref. \cite{KN-pS:IHJ} 
is partially attributed to the difference of the interaction kernel. 
We note that in the SR-B the scattering amplitudes are not so different 
between two interaction kernels, since the magnitude of amplitudes near 
$\pi\Sigma$ threshold is rather small compared with the SR-A model. 
The small scattering amplitudes may be related to the fact that 
the SR-B has rather small range parameters compared with other models, 
as listed in Tables \ref{RangePara} and \ref{RangePara_100-120}.

In addition, we consider that the ansatz of a relativistic flux factor and/or 
the assumption of Gaussian form may also contribute to such a difference. 
We comment on the latter ingredient. The Gaussian form possesses 
such a nature potentially that it enhances the magnitude 
of a scattering amplitude far below a threshold. 
It is easily confirmed by Born approximation 
that the $s$-wave scattering amplitude for single-range 
Gaussian potentials diverges as $|E|^{-1} \exp \, [\,c |E|\,]$ 
when $E\rightarrow -\infty$ \cite{GaussDiv}. 
(The number $c$ is a positive constant.)
Indeed, as mentioned before, the scattering amplitudes of SR-A 
with our potential $\hat{V}_{MB}^{SR}$ has large magnitude near 
the $\pi\Sigma$ threshold (namely far below the $\bar{K}N$ threshold) 
compared with the model A1 of Ref. \cite{KN-pS:IHJ}. 
The similar tendency can be seen in the comparison with another study. 
In Ref. \cite{ChU:HW} only the WT term is used as well but the 
constraint condition for the scattering amplitude is different from ours. 
Anyway, when the amplitudes shown in Fig. 4 in their paper are 
compared with our ones\footnote{This scattering amplitude has a similar shape with that of $f_\pi=120$ MeV case shown in Fig.~\ref{KSW_I=0_mfpi_SR-A}.} 
 of SR-A calculated with the above 
$\hat{V}_{MB}^{Ref. \,[33]}$ using the same $f_\pi$ as 
that of Ref. \cite{ChU:HW} (106.95 MeV), 
it is confirmed that the magnitude of their scattering amplitudes 
are significantly smaller near the $\pi\Sigma$ threshold.
In both the model A1 of Refs. \cite{KN-pS:IHJ} and 
the model of Ref. \cite{ChU:HW}, the dimensional regularization 
is used to obtain a finite result. 
On the other hand, it can be said in the present study that 
such a regularization is realized by using the Gaussian-form potential. 
The different regularization scheme causes the different extrapolation 
of the scattering amplitudes to the subthreshold region. This 
is considered to be a possible reason of the difference between 
our result and results of other studies.

\subsection{$I=1$ $\bar{K}N$-$\pi\Sigma$-$\pi\Lambda$ system}

We make the same investigation on the isospin $I=1$ sector 
which has three channels of  $\bar{K}N$, $\pi\Sigma$ 
and $\pi\Lambda$. 
In our model of the Gaussian-form potential, 
there are six range parameters in this sector. 
However, two of them, $d_{\pi\Lambda, \pi\Sigma}$ 
and $d_{\pi\Lambda, \pi\Lambda}$, give no contribution to the result, 
since the potential strength of these channels are forced to be zero 
due to the SU(3) algebra which is involved in the Weinberg-Tomozawa term 
of effective chiral SU(3) Lagrangian. 
(See $C^{(I=1)}$ in Eq. (\ref{Eq:SU(3)CG}).) 
Since we assume $d_{\bar{K}N, \pi\Sigma}=(d_{\bar{K}N, \bar{K}N}+d_{\pi\Sigma, \pi\Sigma})/2$ similarly to the  $I=0$ case, 
three range parameters, $d_{\bar{K}N, \bar{K}N}$, 
$d_{\pi\Sigma, \pi\Sigma}$ and $d_{\bar{K}N, \pi\Lambda}$, are unknown parameters to be determined. 

Similarly to the $I=0$ case, we constrain the range parameters in 
the potential by the Martin's value of the $I=1$ $\bar{K}N$ scattering 
length; $a_{\bar{K}N(I=1)} = 0.37 + i 0.60$ fm \cite{Exp:ADMartin}. 
However, the three unknown parameters can't be determined by only 
the complex value of $a_{\bar{K}N(I=1)}$. 
Here, we reduce the number of unknown parameters by referring the following 
two facts: 
1. In studies with chiral unitary model, isospin symmetric subtraction 
constants have been often assumed and succeeded to reproduce 
various physical quantities \cite{CUM:Hyodo}. 
2. In a separable potential used in Faddeev-AGS calculation 
of $\bar{K}NN$-$\pi YN$ \cite{Faddeev:Ikeda}, the cut-off parameter for 
the $\bar{K}N$ channel 
is not so different between $I=0$ and $I=1$ sector. 
Based on these facts, we examine three conditions as follows: 

\noindent
\begin{tabular}{cll}
-&Cond. (a)& $d_{\bar{K}N, \bar{K}N}$ is fixed to that of the $I=0$ case. \\
&&$d_{\pi\Sigma, \pi\Sigma}$ and $d_{\bar{K}N, \pi\Lambda}$ are 
searched to reproduce the complex \\
&&value of $a_{\bar{K}N(I=1)}$.
\\ 
-&Cond. (b) &$d_{\bar{K}N, \bar{K}N}$ and $d_{\pi\Sigma, \pi\Sigma}$ are 
fixed to those of the $I=0$ case. \\
&&$d_{\bar{K}N, \pi\Lambda}$ is searched 
to reproduce the real part of $a_{\bar{K}N(I=1)}$. \\
-&Cond. (c) &Similar to the condition (b), but $d_{\bar{K}N, \pi\Lambda}$ 
is searched to\\
&& reproduce the imaginary part of $a_{\bar{K}N(I=1)}$. 
\end{tabular}

\begin{table}
\caption{Range parameters for the $I=1$ $\bar{K}N$-$\pi\Sigma$-$\pi\Lambda$ 
system with non-relativistic and semi-relativistic kinematics (NRv2 and SR-A). 
$f_\pi=110$ MeV. Same as Table~\ref{RangePara}. 
``Condition'' is explained in the text. \label{RangePara_I=1_NR_SR-A}}
\begin{tabular*}{\textwidth}{l@{\hspace{0.5cm}}|@{\hspace{0.5cm}} r@{\hspace{0.7cm}}r@{\hspace{0.7cm}}r@{\hspace{1.2cm}}r@{\hspace{0.7cm}}r@{\hspace{0.7cm}}r}
\hline
Case  & \multicolumn{3}{c}{NRv2} & \multicolumn{3}{c}{SR-A}  \\
\hline
Kinematics & \multicolumn{3}{c}{Non-rela.} & \multicolumn{3}{c}{Semi-rela.} \\
Potential  & \multicolumn{3}{c}{NRv2} & \multicolumn{3}{c}{KSW-type}  \\
\hline
Condition  & (a) & (b) & (c) & (a) & (b) & (c) \\
\hline
\\
$d_{\bar{K}N,\, \bar{K}N}$  & 0.438 & 0.438 & 0.438 & 0.499 & 0.499& 0.499\\
$d_{\pi\Sigma,\, \pi\Sigma}$  & 0.159 & 0.636 & 0.636 & 0.261 & 0.712& 0.712\\ 
$d_{\bar{K}N,\, \pi\Lambda}$  & 0.221 & 0.301 & 0.445 & 0.282 & 0.354& 0.467\\ 
\\
Re $a_{\bar{K}N\,(I=1)}$ & 0.376 & 0.372 & 0.657 & 0.375 & 0.371 & 0.659\\
Im $a_{\bar{K}N\,(I=1)}$ & 0.606 & 1.504 & 0.599 & 0.605 & 1.493 & 0.600\\
\hline
\end{tabular*}
\end{table}

\begin{figure}
  \includegraphics[width=.47\textwidth]{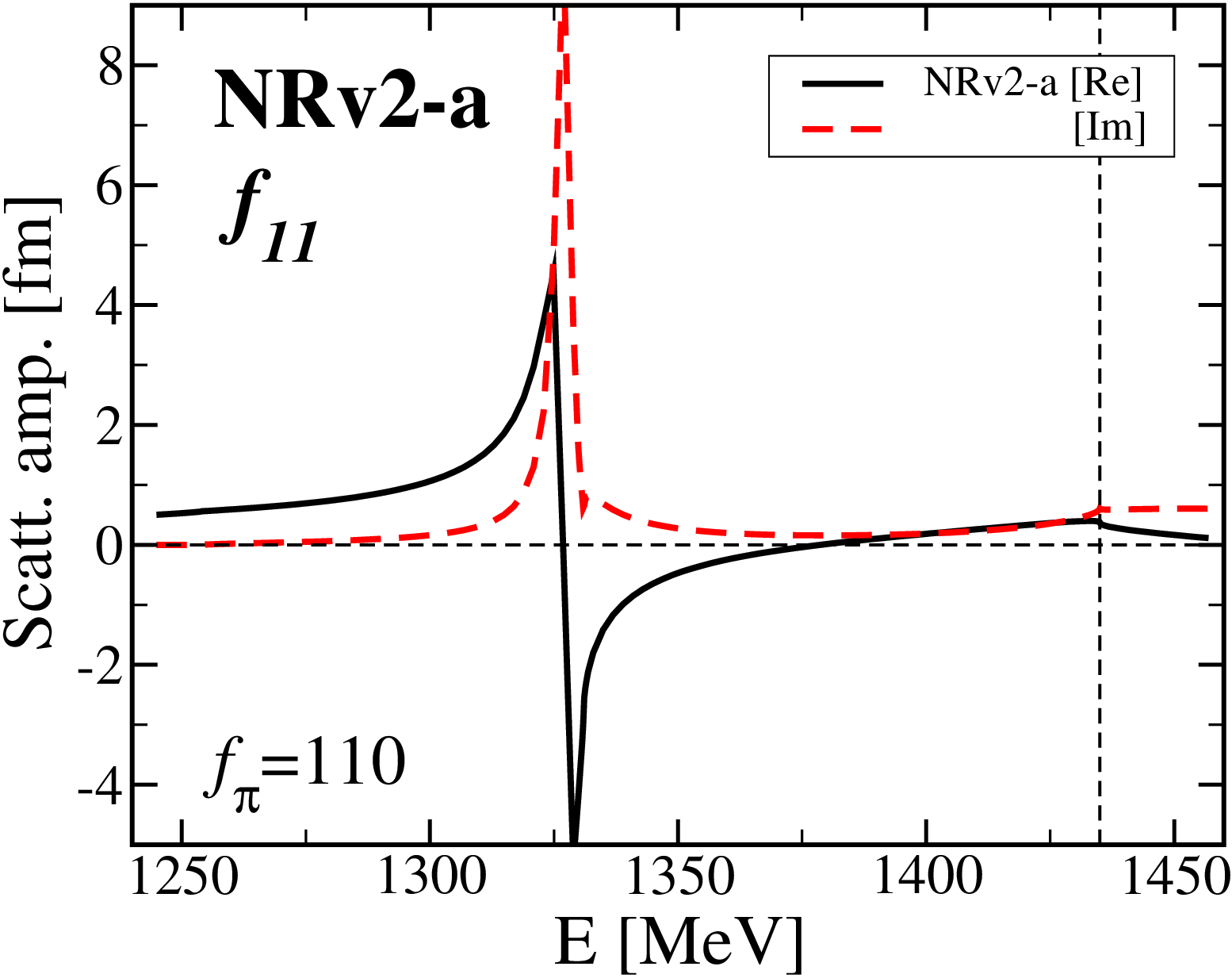}
  \hspace{0.3cm}
  \includegraphics[width=.47\textwidth]{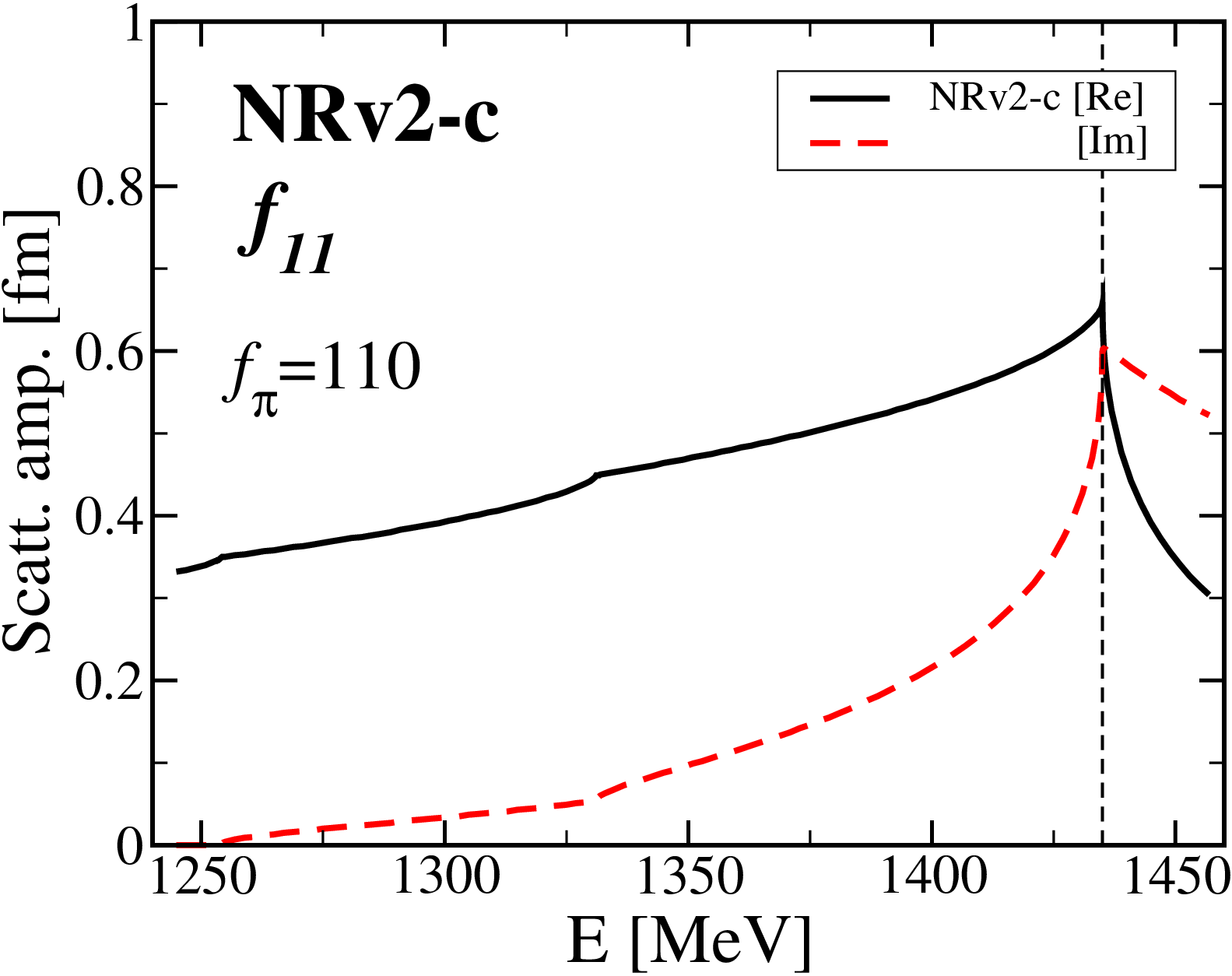}
  \includegraphics[width=.47\textwidth]{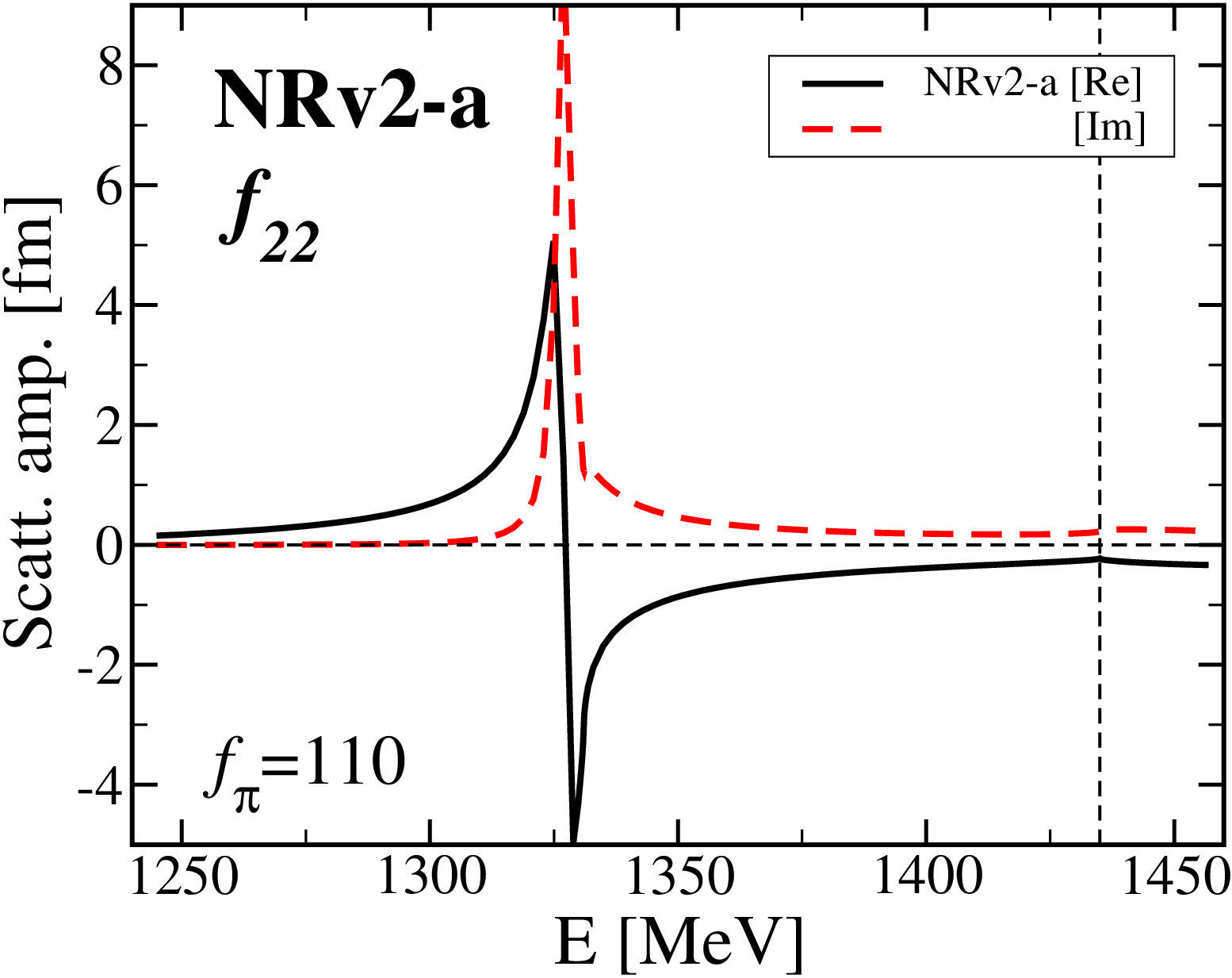}
  \hspace{0.3cm}
  \includegraphics[width=.47\textwidth]{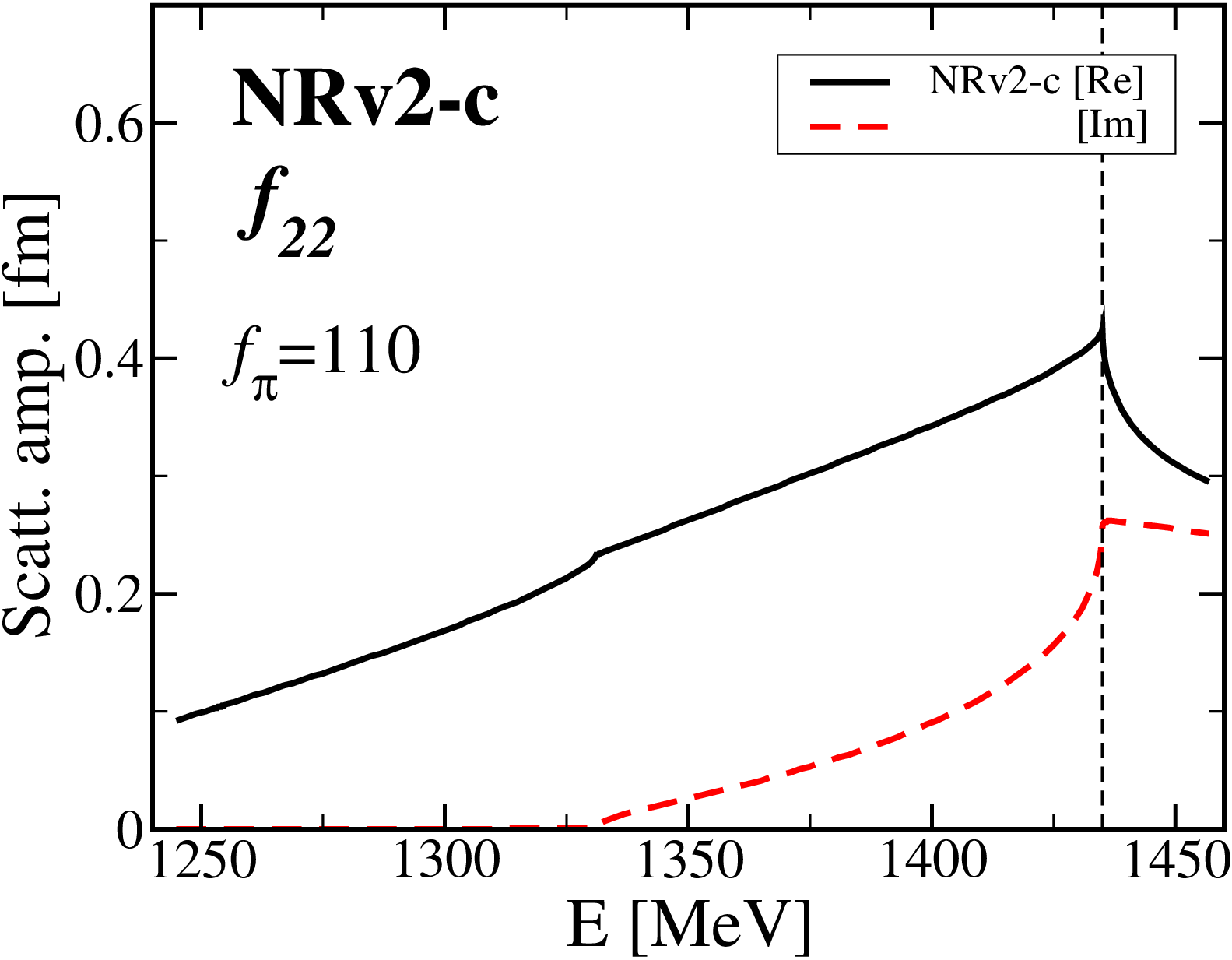}
  \includegraphics[width=.47\textwidth]{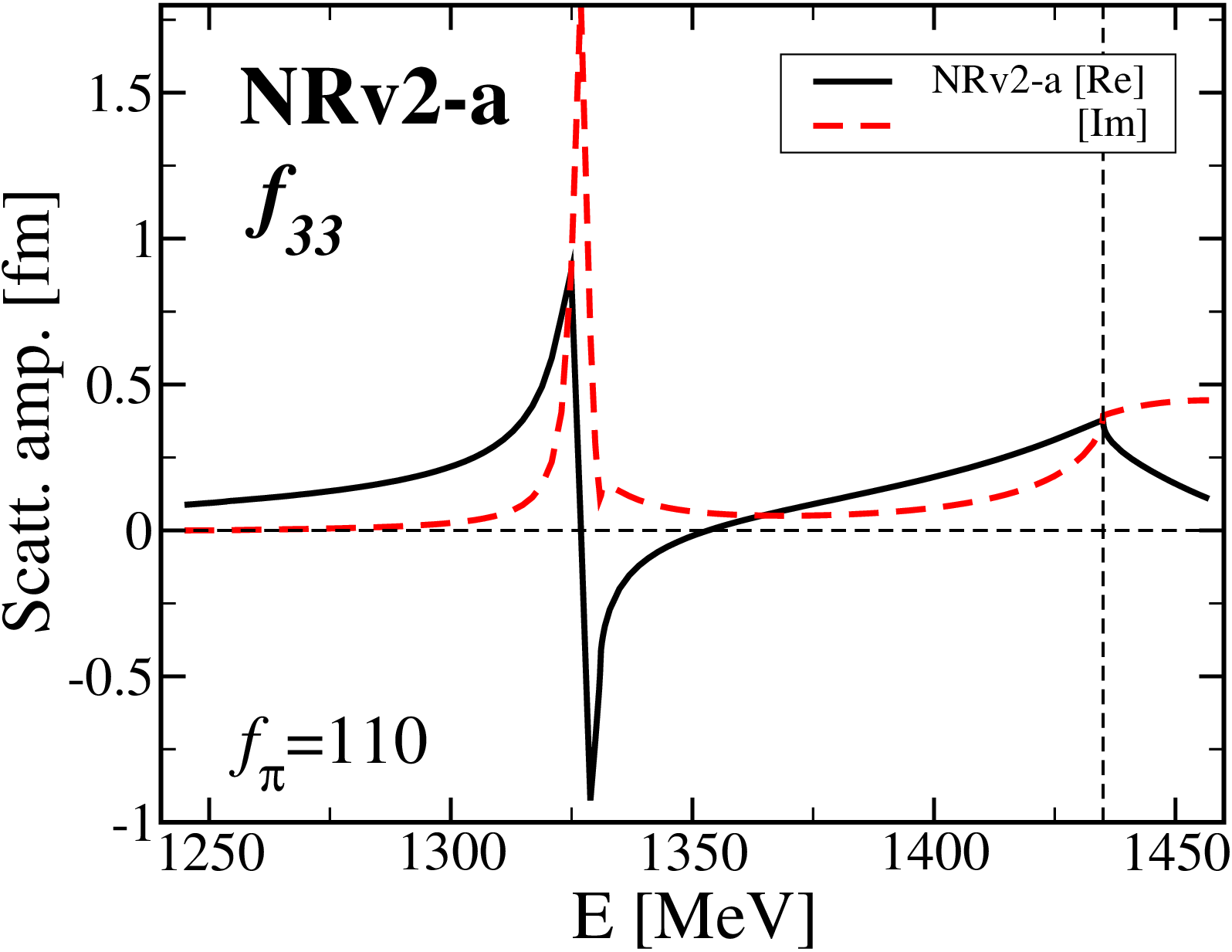}
  \hspace{0.3cm}
  \includegraphics[width=.47\textwidth]{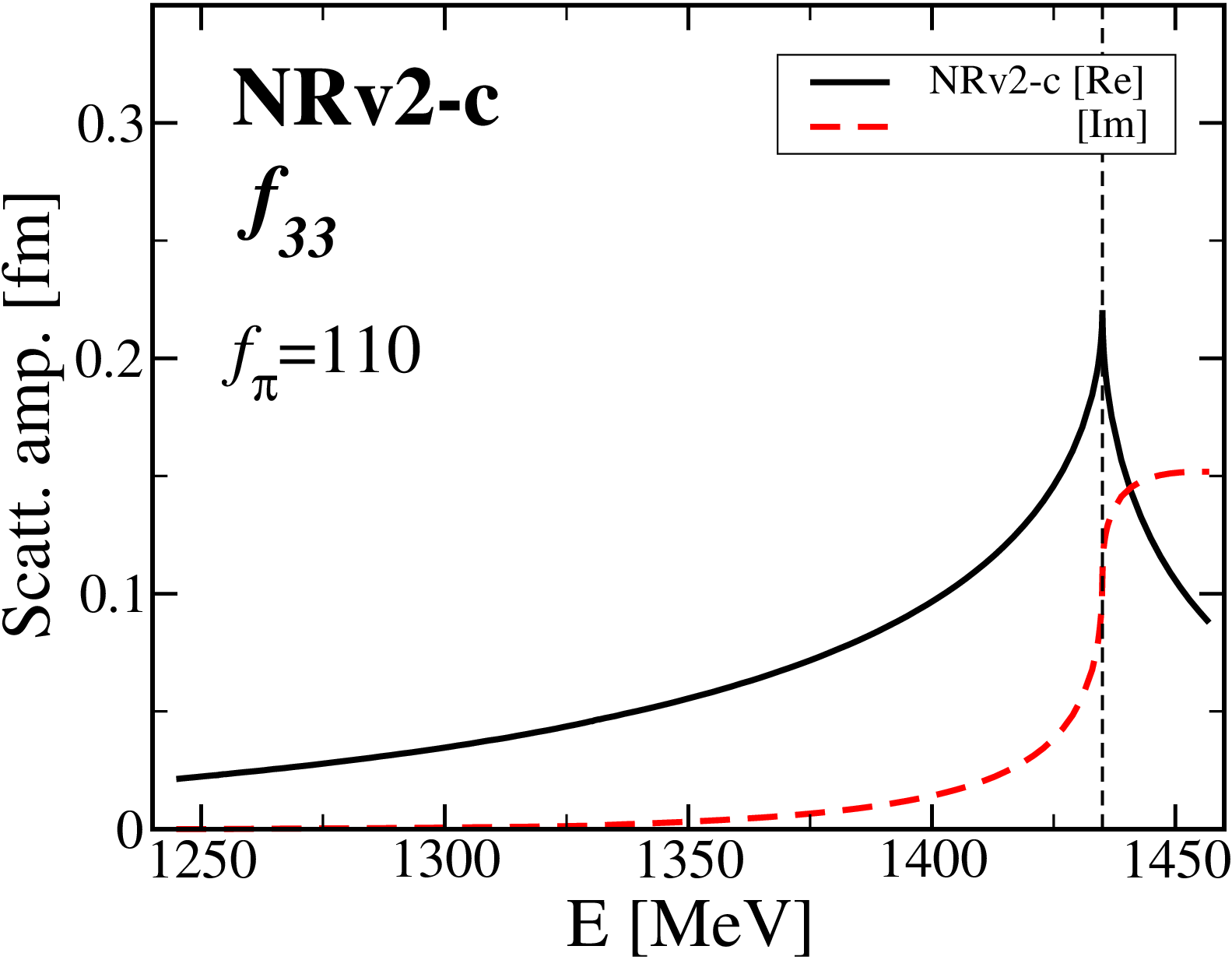}
  \caption{$I=1$ scattering amplitudes for NRv2 case with $f_\pi=110$ MeV. 
The real (imaginary) part of scattering amplitude is drawn with 
a black-solid (red-dashed) line. 
Left (right) panels are calculated with the condition (a) (condition (c)). 
Top: $\bar{K}N \rightarrow \bar{K}N$, middle: $\pi\Sigma \rightarrow \pi\Sigma$, bottom: $\pi\Lambda \rightarrow \pi\Lambda$. 
 \label{KSW_I=1_NR}}
\end{figure}

We describe mainly the result obtained with $f_{\pi}=110$ MeV. 
 In the NR case, we can find a set of range parameters 
which satisfy each condition (a)-(c), as shown in Table 
\ref{RangePara_I=1_NR_SR-A}.
However, it is found that in the condition (a) the $I=1$ scattering amplitude 
has a sharp resonance structure slightly 
below $\pi\Sigma$ threshold (Fig.~\ref{KSW_I=1_NR}, left column),
\footnote{In this case we have found a pole on the complex-energy plane 
by the ccCSM at $(E_R, -\Gamma/2)=(1327.2, -1.8)$ MeV. Certainly, 
this pole exists only by 4 MeV below $\pi\Sigma$ threshold (=1331 MeV).}
 although no narrow resonant states in $I=1$ sector 
have been confirmed theoretically and experimentally 
in such energy region. 
In the conditions (b) and (c), such a resonance 
structure does not appear in all the scattering amplitudes (Fig.~\ref{KSW_I=1_NR}, right column).

\begin{figure}
  \includegraphics[width=.47\textwidth]{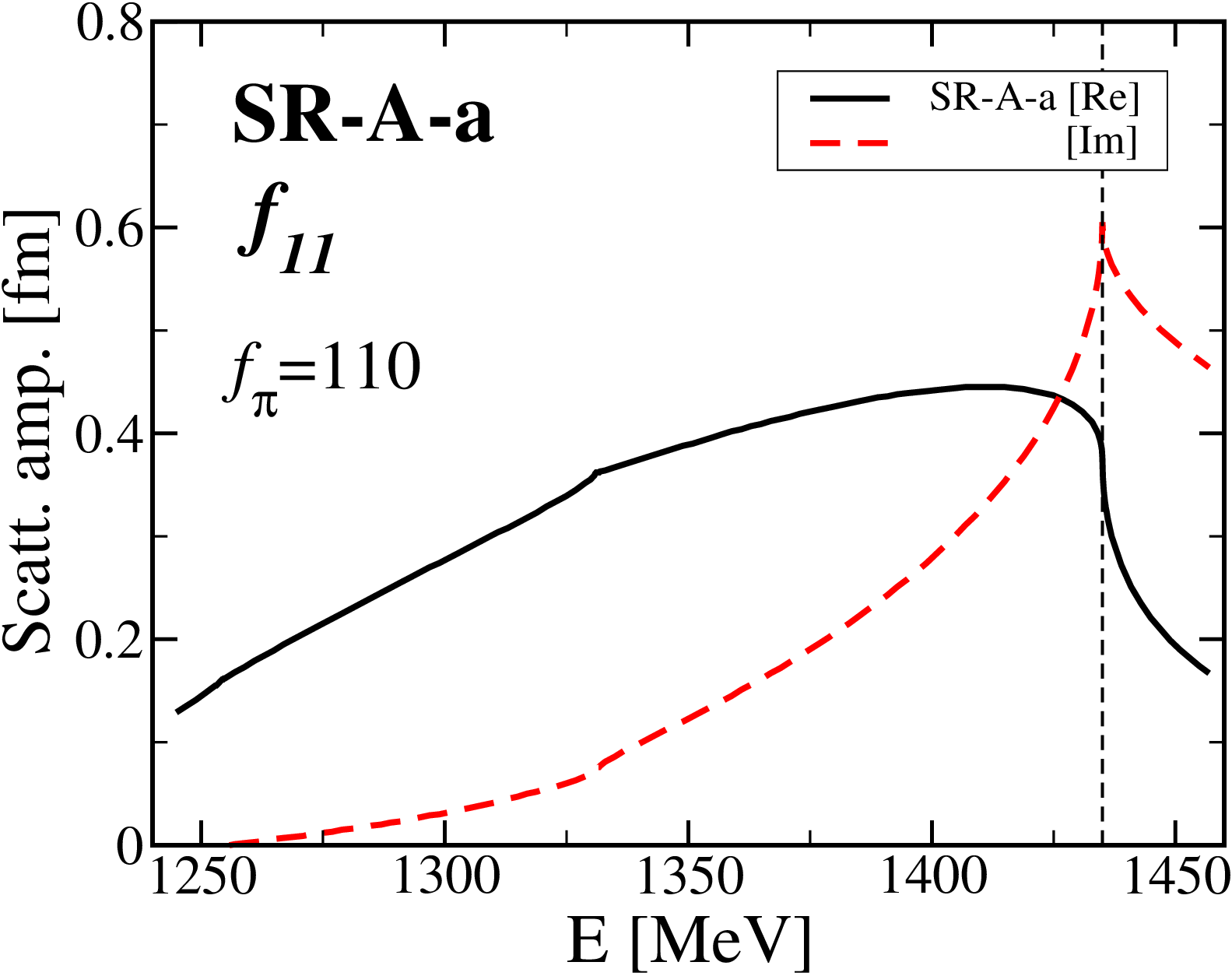}
  \hspace{0.3cm}
  \includegraphics[width=.47\textwidth]{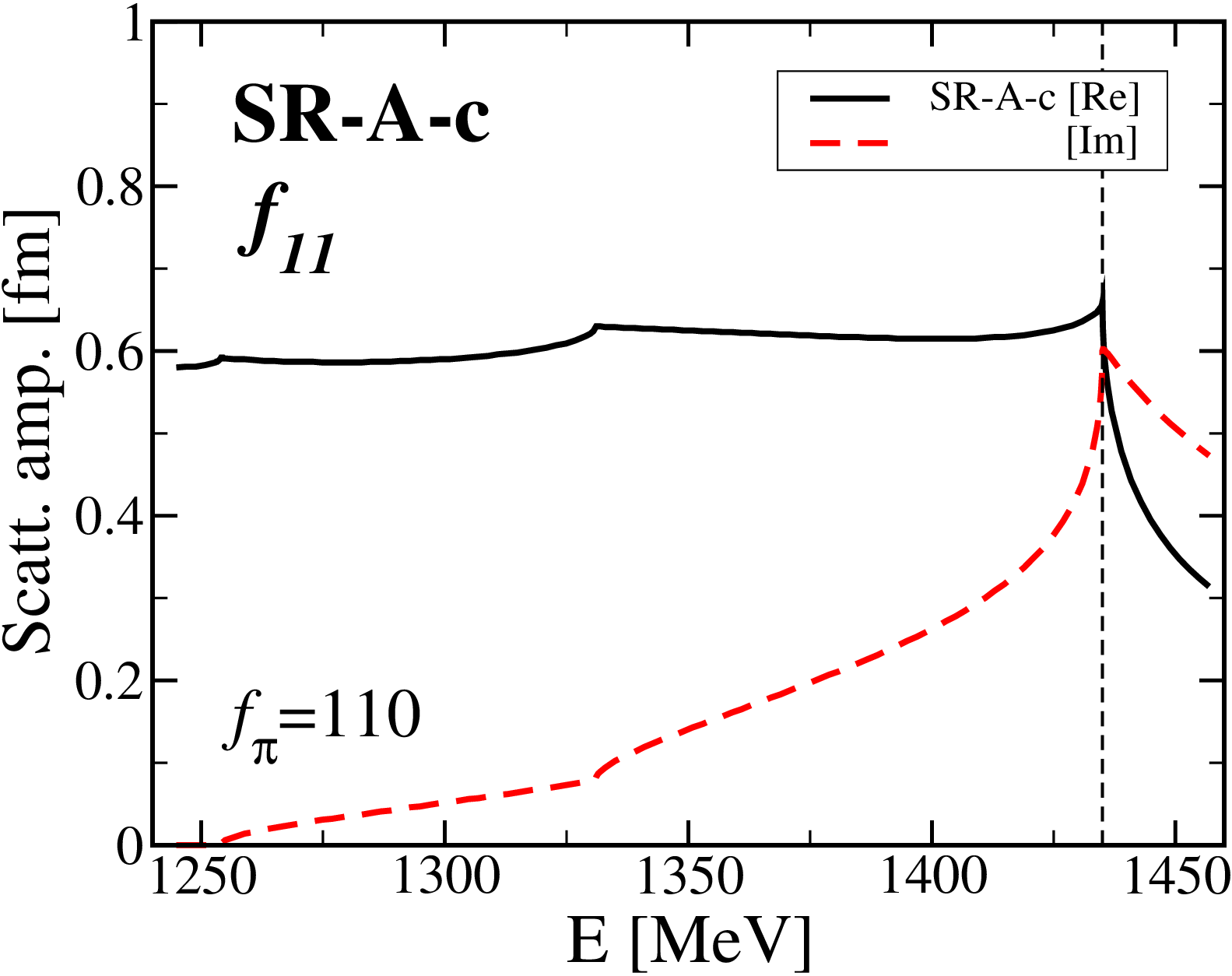}
  \includegraphics[width=.47\textwidth]{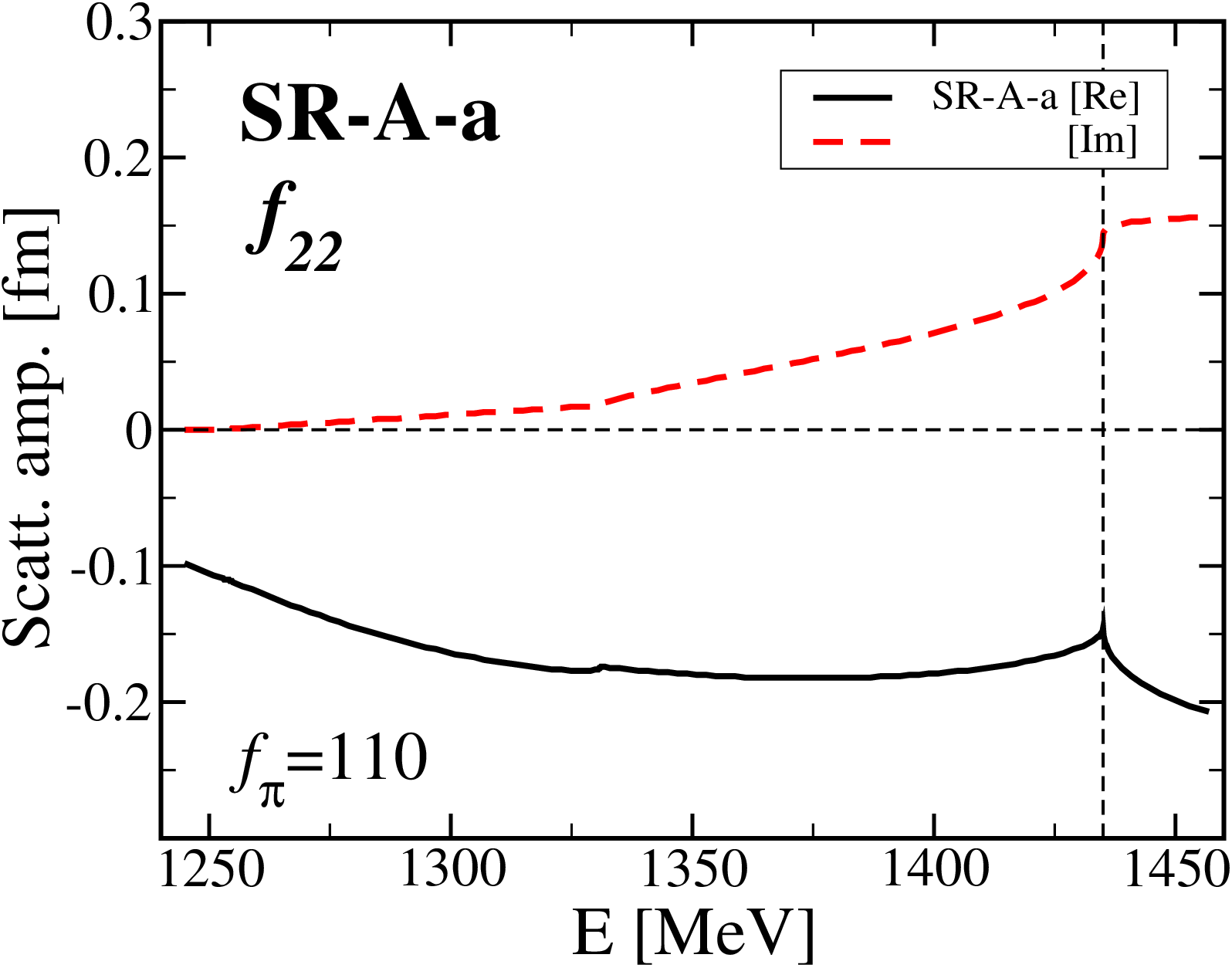}
  \hspace{0.3cm}
  \includegraphics[width=.47\textwidth]{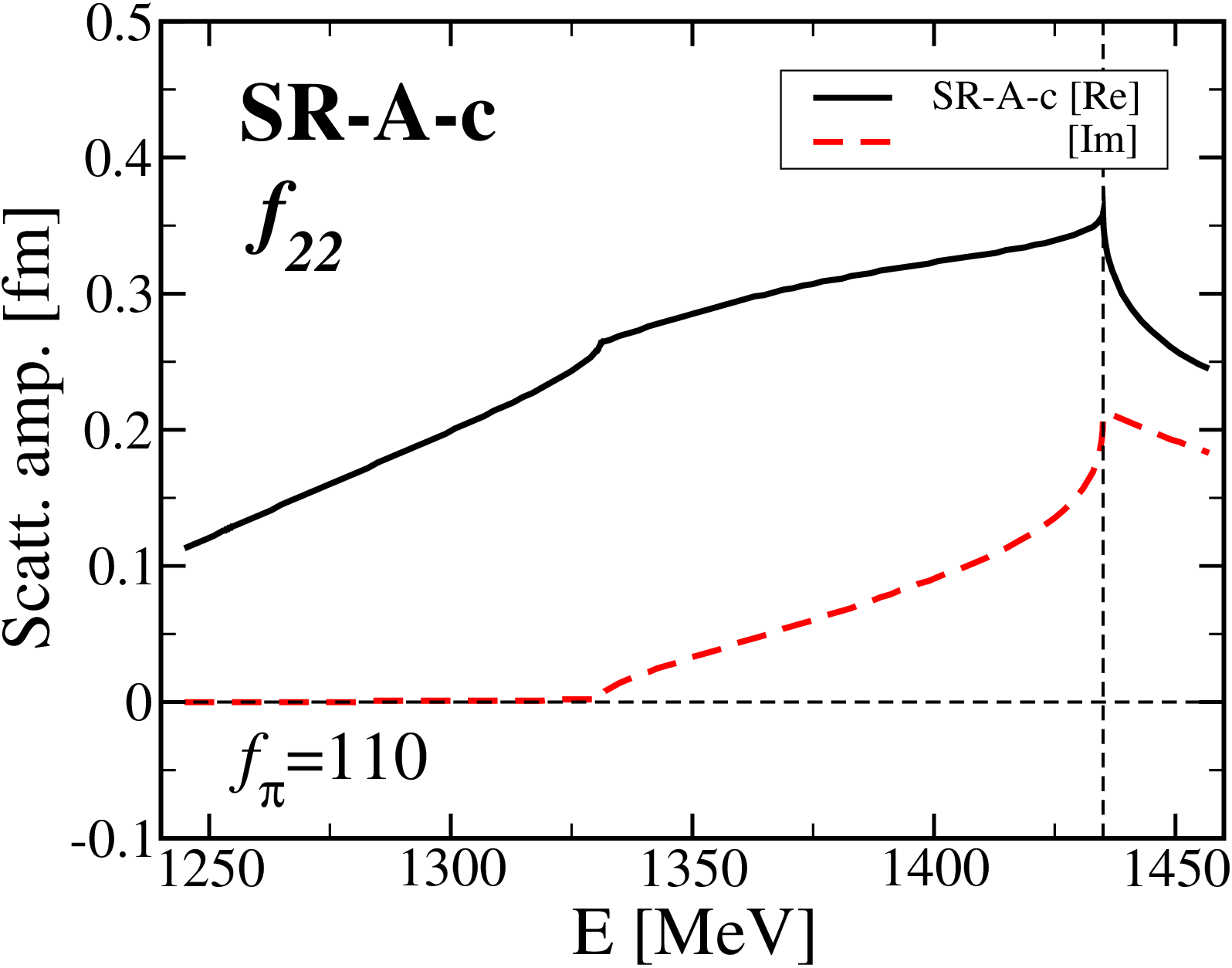}
  \includegraphics[width=.47\textwidth]{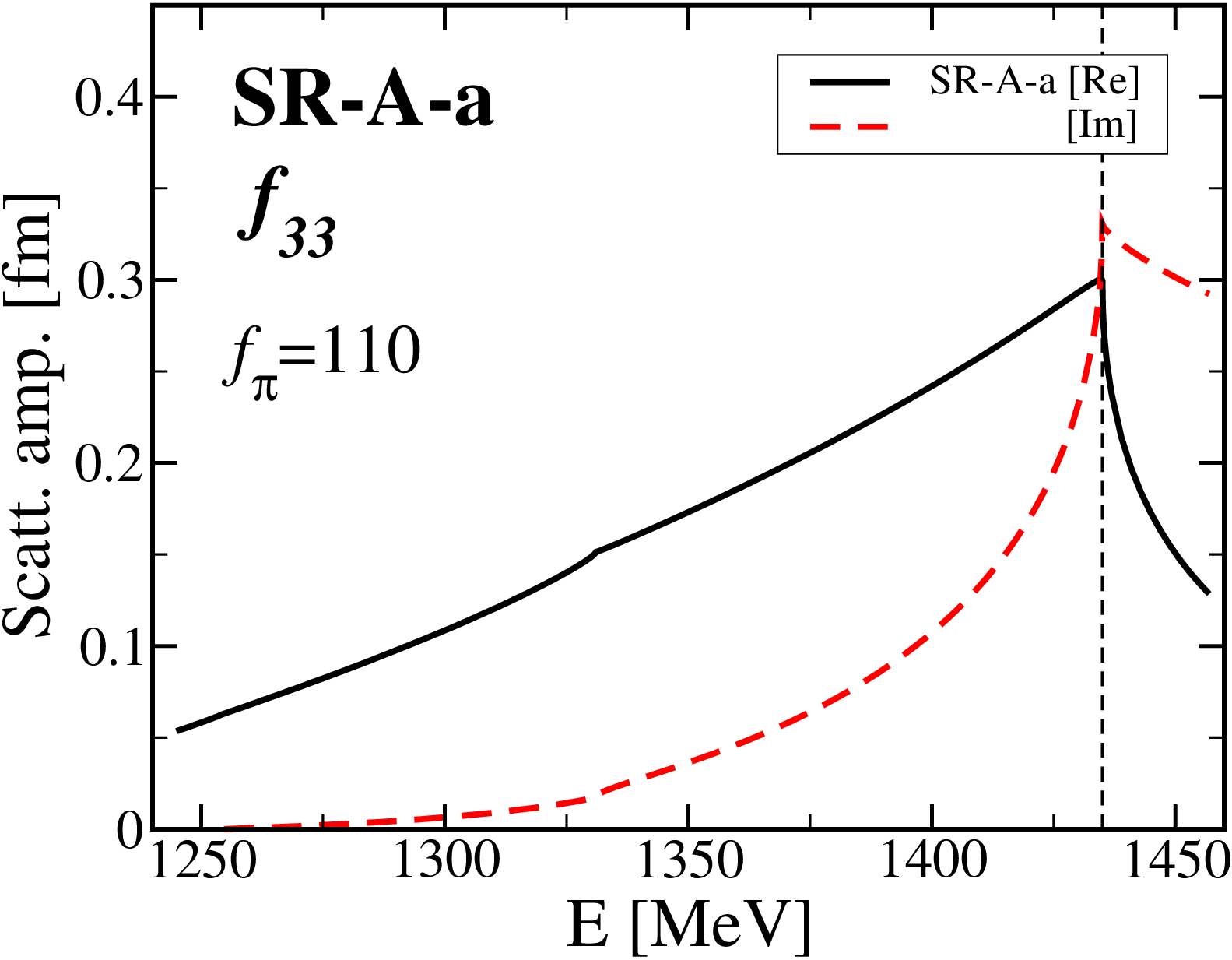}
  \hspace{0.3cm}
  \includegraphics[width=.47\textwidth]{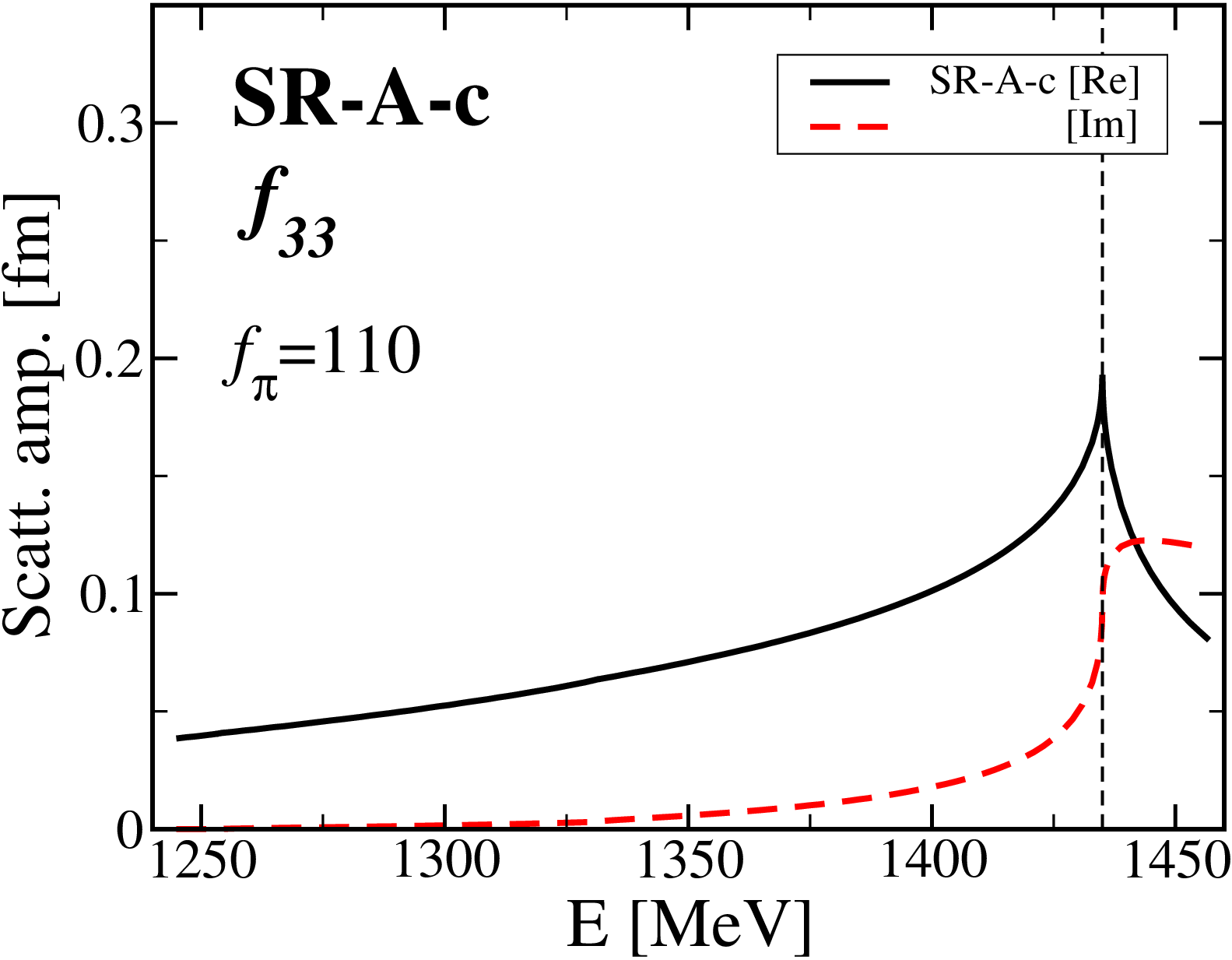}
  \caption{$I=1$ scattering amplitudes for SR-A case with $f_\pi=110$ MeV. 
Similar to Fig.~\ref{KSW_I=1_NR}. \label{KSW_I=1_SR-A}}
\end{figure}

In a semi-relativistic case, SR-A, we can find a range parameter set 
for each condition (a) to (c), in the same way as NRv2. 
(given in three right columns of Table~\ref{RangePara_I=1_NR_SR-A})
The scattering amplitudes calculated with the condition (a) are shown 
in the left column of Fig.~\ref{KSW_I=1_SR-A}. 
As seen in this figure, the $\pi\Sigma$ scattering amplitude 
indicates repulsive nature of $\pi\Sigma$-$\pi\Sigma$ channel, 
in spite that the direct $\pi\Sigma$ potential is originally attractive. 
This is considered to be a consequence of the coupled-channel effect. 
Calculated with the condition (c), all scattering amplitudes of SR-A 
are quantitatively the same as those of NRv2 with the condition (c) 
(see the right column of Figs.~\ref{KSW_I=1_NR} and~\ref{KSW_I=1_SR-A}). 

\begin{figure}
  \includegraphics[width=.47\textwidth]{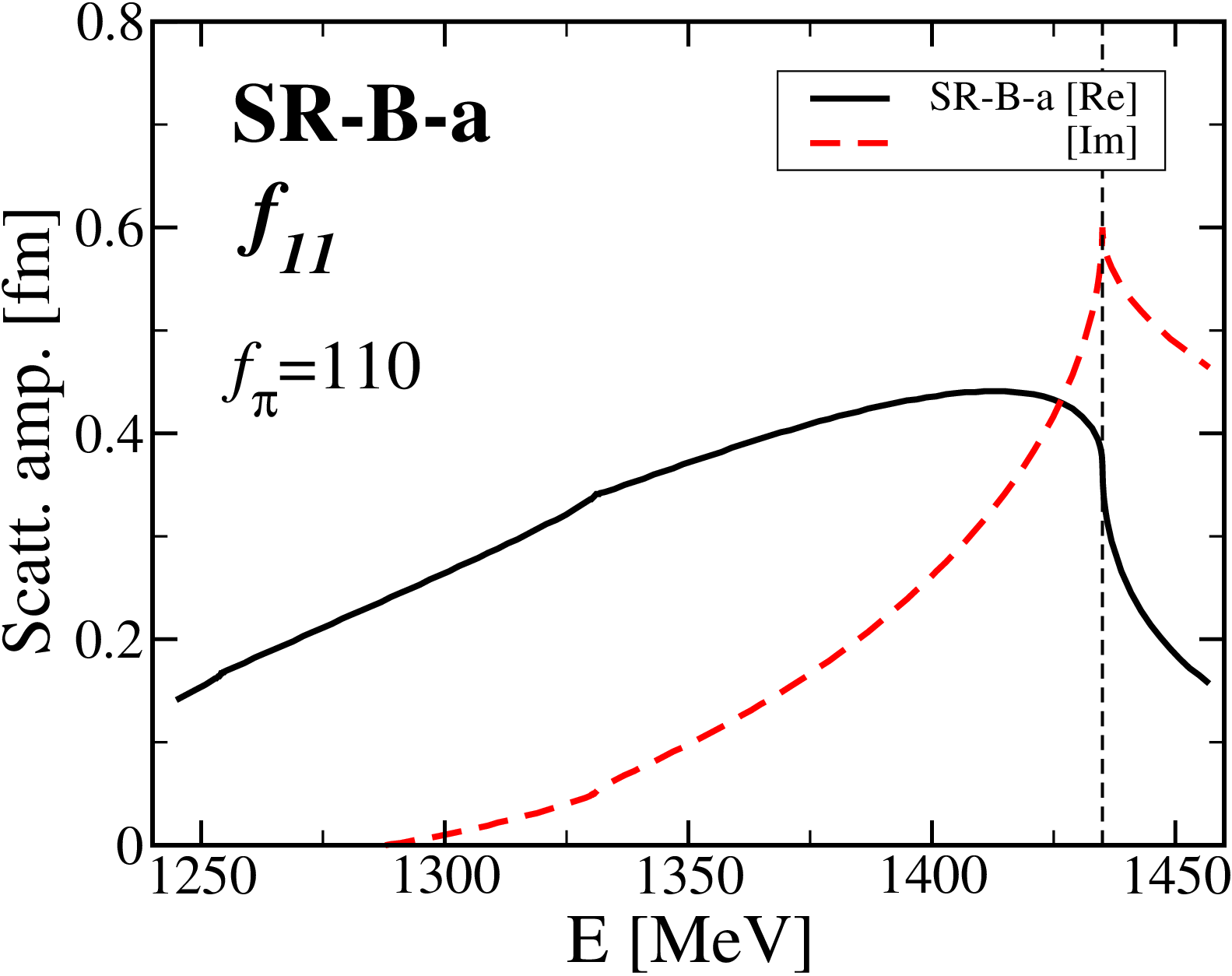}
  \hspace{0.3cm}
  \includegraphics[width=.47\textwidth]{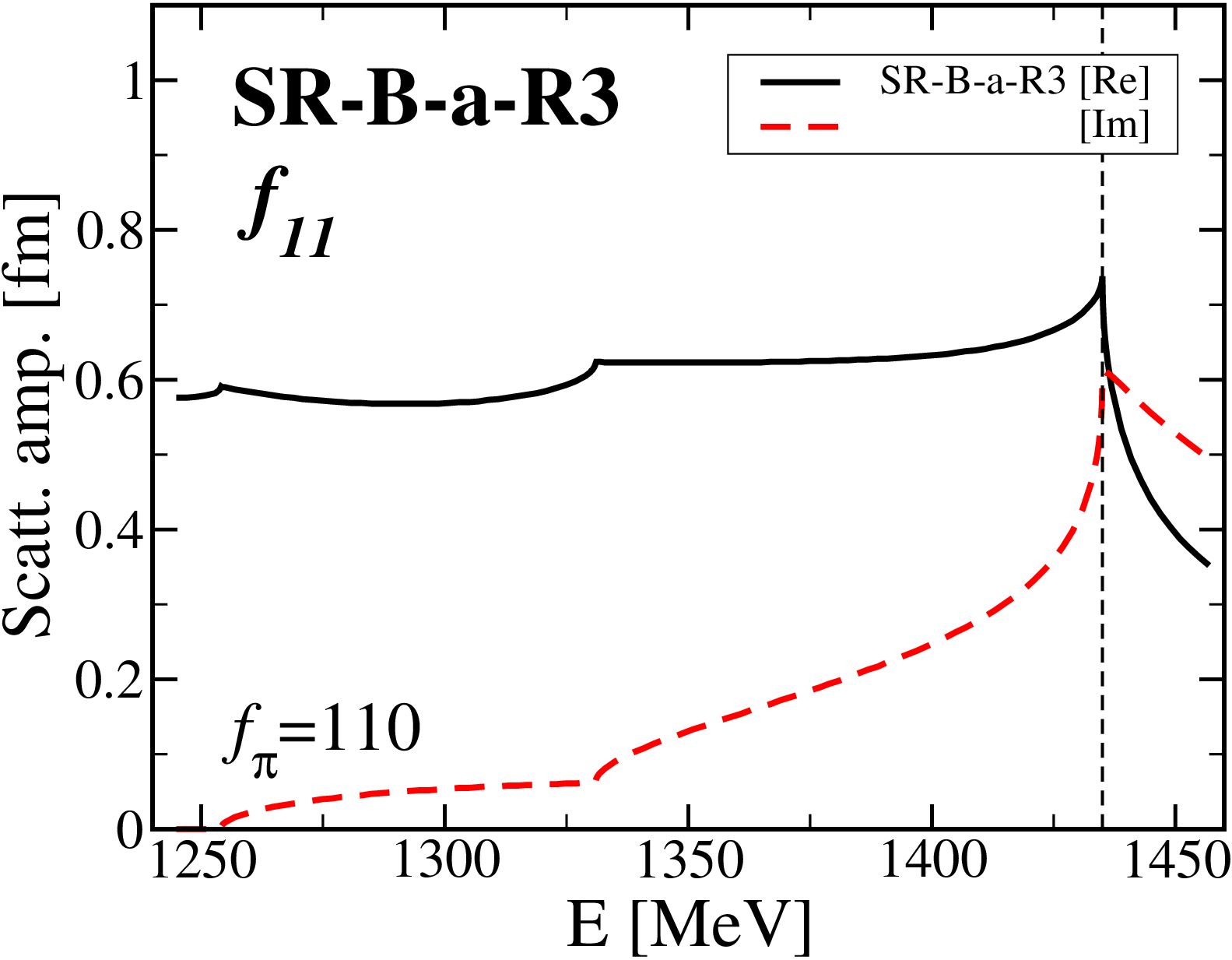}
  \includegraphics[width=.47\textwidth]{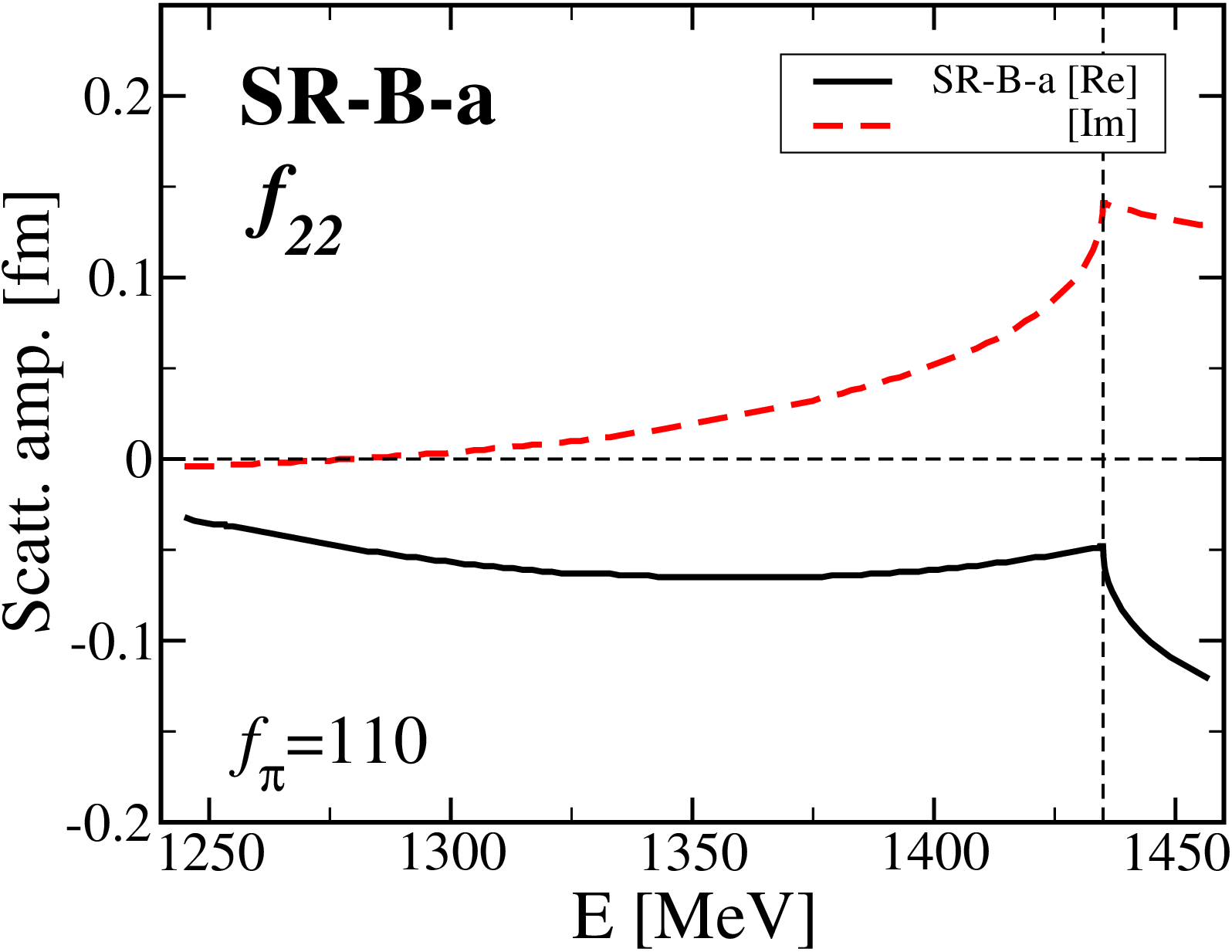}
  \hspace{0.3cm}
  \includegraphics[width=.47\textwidth]{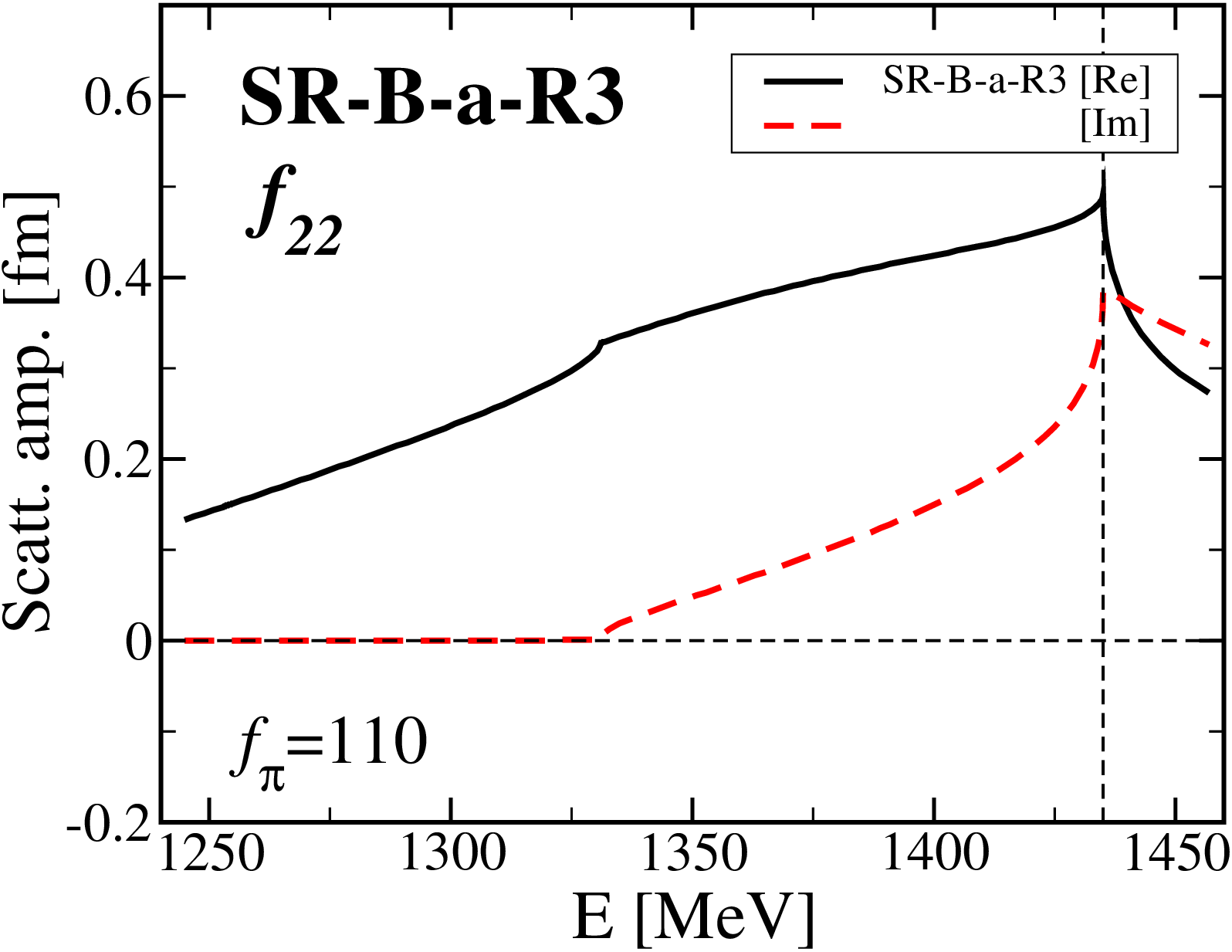}
  \includegraphics[width=.47\textwidth]{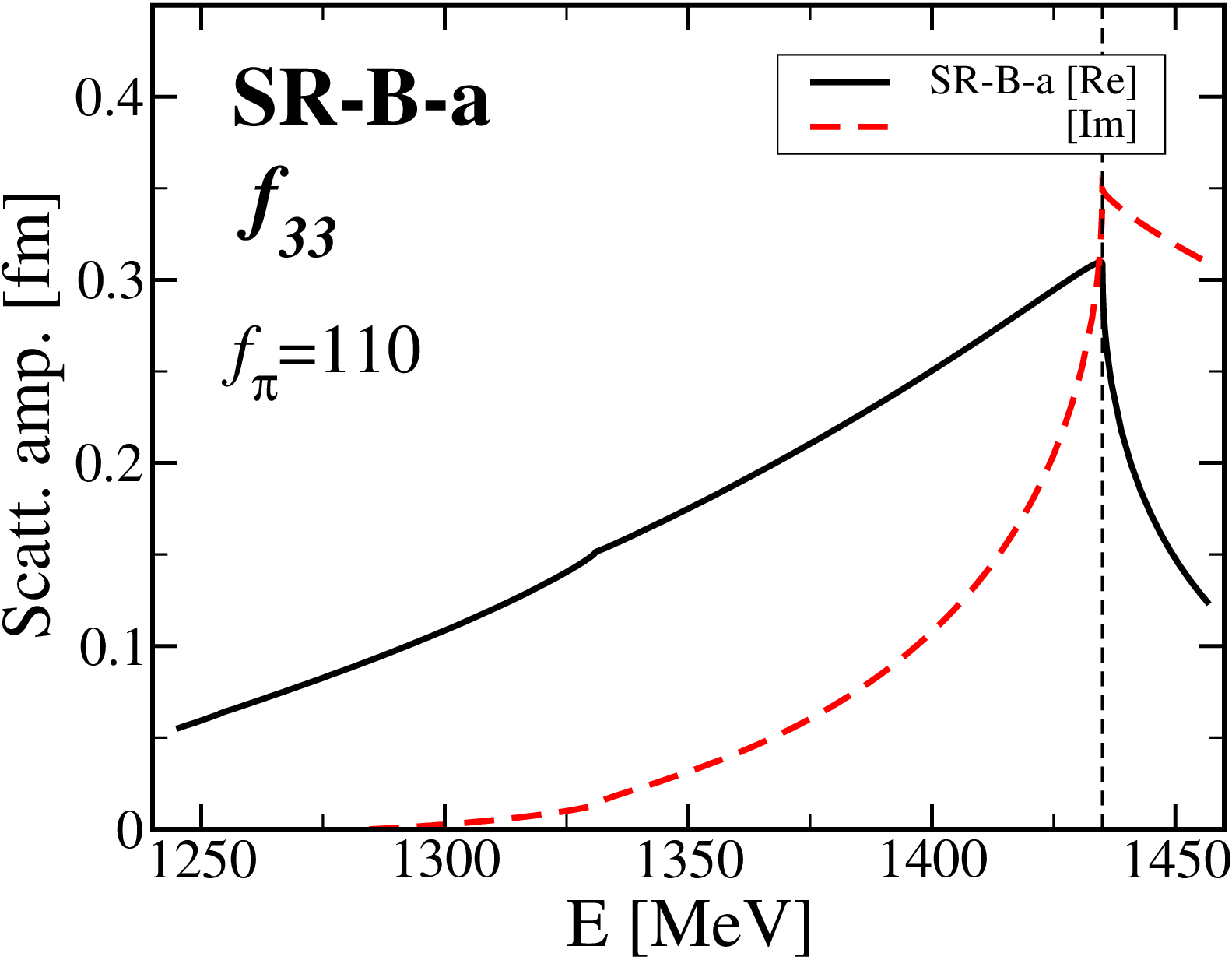}
  \hspace{0.5cm}
  \includegraphics[width=.47\textwidth]{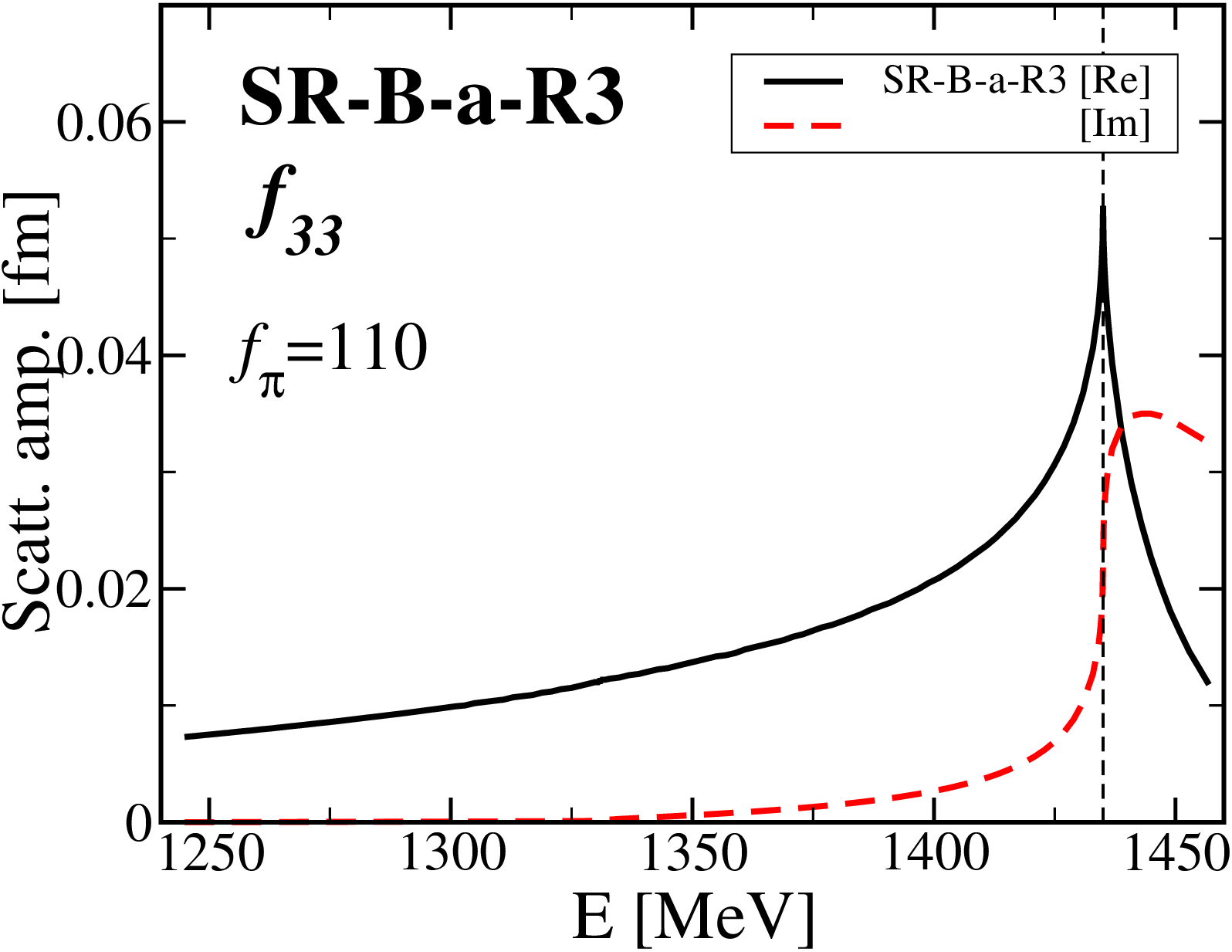}
  \caption{$I=1$ scattering amplitudes for SR-B case with $f_\pi=110$ MeV. 
Similar to Fig.~\ref{KSW_I=1_NR}. 
Left (right) panels are calculated with the condition (a) (condition (a)-R3). 
 \label{KSW_I=1_SR-B}}
\end{figure}

In the other semi-relativistic case, SR-B, a set of range parameters 
to satisfy the conditions 
is found only in the condition (a), but is not found in the conditions 
(b) and (c). 
The best parameter sets of the conditions (b) and (c) are listed 
in Table~\ref{RangePara_I=1_SR-B}. But the scattering $\bar{K}N$ 
length calculated with these parameters is far from the Martin's value. 
The scattering amplitudes calculated with the condition (a) are shown 
in the left column of Fig.~\ref{KSW_I=1_SR-B}. 
Compared with those of SR-A (a) in the Fig.~\ref{KSW_I=1_SR-A},
they are quantitatively the same as each other. 
In particular, also in the SR-B (a) the $\pi\Sigma$ scattering amplitude 
indicates repulsive nature of $\pi\Sigma$-$\pi\Sigma$ channel.

\begin{table}
\caption{Range parameters for the $I=1$ $\bar{K}N$-$\pi\Sigma$-$\pi\Lambda$ 
system with the semi-relativistic kinematics (SR-B). $f_\pi=110$ MeV. 
Same as Table~\ref{RangePara}. 
Columns ``(a)-R1'' to ``(a)-R3'' show the results obtained with the condition (a) relaxed. Details are explained in the text.
 \label{RangePara_I=1_SR-B}}
\begin{tabular*}{\textwidth}{l@{\hspace{0.5cm}}|@{\hspace{0.5cm}} r@{\hspace{0.6cm}}r@{\hspace{0.6cm}}r@{\hspace{0.5cm}}|@{\hspace{0.5cm}}r@{\hspace{0.6cm}}r@{\hspace{0.6cm}}r}
\hline
Case  & \multicolumn{6}{c}{SR-B}   \\
\hline
Kinematics & \multicolumn{6}{c}{Semi-rela.} \\
Potential  & \multicolumn{6}{c}{KSW-type}    \\
\hline
Condition  & (a) & (b) & (c) & (a)-R1 & (a)-R2 & (a)-R3 \\
\hline
&&&&&& 
\\
$d_{\bar{K}N,\, \bar{K}N}$    & 0.369 & 0.369 & 0.369 & 0.369 & 0.369 & 0.369\\
$d_{\pi\Sigma,\, \pi\Sigma}$  & 0.248 & 0.348 & 0.348 & 1.800 & 1.050 & 0.630\\ 
$d_{\bar{K}N,\, \pi\Lambda}$  & 0.276 & 0.157 & 0.104 & 0.480 & 0.530 & 0.960\\ &&&&&& 
\\
Re $a_{\bar{K}N\,(I=1)}$      & 0.369 & 0.075 & $-0.134$ & 0.858 & 0.812 & 0.738\\
Im $a_{\bar{K}N\,(I=1)}$      & 0.600 & 0.154 & 0.337 & 0.600 & 0.600 & 0.600\\
\hline
\end{tabular*}
\end{table}

We investigate other $f_\pi$ values such as 90, 100 and 120 MeV. 
Range parameters and $\bar{K}N$ scattering length for these $f_\pi$'s 
are listed in Table~\ref{RangePara_I=1_NR_SR-A_mfpi}. 
It is confirmed that the scattering amplitudes for these $f_\pi$'s 
are essentially the same as those for $f_\pi=110$ MeV case above mentioned. 

In the remaining part of this section, we consider the cases where the conditions are slightly relaxed. 
We vary the value of the range parameter $d_{\bar{K}N, \bar{K}N}$ 
slightly in the condition (a), since it may be too strict constraint 
that $d_{\bar{K}N, \bar{K}N}$ is fixed to that of the $I=0$ sector. 
With $10\%$ modification of the $d_{\bar{K}N, \bar{K}N}$, 
essential property of the scattering amplitude is found to be unchanged; 
In both cases of SR-A and SR-B, the $\pi\Sigma$ scattering amplitude 
still indicates 
the repulsive nature, and in the NR a resonance structure is kept 
to appear around the $\pi\Sigma$ threshold. 

We make an investigation of the SR-B model with a relaxed condition, 
since this model has a satisfactory range-parameter set only 
for the condition (a). 
We relax the condition (a) to give up the simultaneous reproduction 
of the real and imaginary parts of Martin's value. 
We examine a single constraint with the imaginary part; Im $a_{\bar{K}N(I=1)}$ = 0.600 fm. (Call ``relaxed condition (a)''.)  
In $0.1 < d_{\pi\Sigma,\, \pi\Sigma}, \, d_{\bar{K}N,\, \pi\Lambda} < 2.0$,  
some sets of $\{d_{\pi\Sigma,\, \pi\Sigma}, \, d_{\bar{K}N,\, \pi\Lambda} \}$ 
are obtained under the relaxed condition (a). 
Typical examples of the parameter sets are listed 
on the right three columns in Table~\ref{RangePara_I=1_SR-B}.  
The parameter sets, (a)-R1, (a)-R2 and (a)-R3,   
give upper, middle and lower values of Re $a_{\bar{K}N(I=1)}$ 
ranging from 0.738 to 0.858 fm, respectively.  
The right panels of Fig.~\ref{KSW_I=1_SR-B} show 
the scattering amplitudes of the SR-B with a parameter set (a)-R3. 
We note that the $\pi\Sigma$ scattering amplitude calculated with 
the relaxed condition (a) indicates attractive nature. 
The scattering amplitudes of the SR-B (a)-R3 are found to 
be quite similar to those of NRv2 and SR-A with the condition (c) 
in all channels, 
though the magnitude of $\pi\Lambda$ amplitude is one-order smaller 
than other cases. (See also right panels of 
Figs.~\ref{KSW_I=1_NR} and \ref{KSW_I=1_SR-A}.)
Thus, the scattering amplitudes in $I=1$ sector are 
determined almost independently of the kinematics and potential type, 
when Im $a_{\bar{K}N(I=1)}$ is constrained with the Martin's value. 

Note the case that the real part of $a_{\bar{K}N(I=1)}$ is fixed 
to the Martin's estimation. 
We can obtain range-parameter sets with this constraint for SR-B (a). 
However, it is found that these parameter sets give the imaginary part of $a_{\bar{K}N(I=1)}$ largely deviated from the Martin's value: 
Im $a_{\bar{K}N(I=1)}$ is obtained to be $1.2 \sim 1.9$ fm, when 
Re $a_{\bar{K}N(I=1)}$ is fixed to 0.37 fm. Similar results have been obtained 
also in NRv2 (b) and SR-A (b) as shown in Table \ref{RangePara_I=1_NR_SR-A}. 
We conclude that within our model it is difficult to fix the real part of 
$a_{\bar{K}N(I=1)}$ to be the value estimated by Martin, compared to fixing 
its imaginary part. Thus, we have opted the imaginary part of $a_{\bar{K}N(I=1)}$ to constrain our model of $\bar{K}N$-$\pi Y$ potential in $I=1$ channel.

\section{Summary and future plan}
\label{}

We have studied a $\bar{K}N$-$\pi Y$ system with a coupled-channel 
complex scaling method (ccCSM) \cite{CSM:Myo} using a chiral SU(3) potential. 
In our study, scattering states as well as resonant states 
are investigated within a single framework of ccCSM.   
Resonance poles are obtained by diagonalizing 
a complex-scaled Hamiltonian with Gaussian base, similarly 
to bound states calculation. 
Scattering problem is solved with an advanced use of ccCSM, 
``CS-WF'' method \cite{CSM:Kruppa}. 
In the CS-WF, due to Cauchy's theorem 
scattering amplitudes are calculated using complex-scaled 
wave functions which are also described with Gaussian base. 
Thus, both of resonance and scattering problems can be solved 
with Gaussian base and therefore they can be treated 
with small and straightforward extension of 
the bound-state calculation.
This is the most advantageous point of ccCSM. 

Based on Ref. \cite{ChU:KSW} where a meson-baryon potential 
is derived from a chiral SU(3) theory, we have constructed 
a meson-baryon potential (KSW-type potential) which is 
a local potential with Gaussian form in $r$-space. 
Since in the present study it is necessary to deal with pion whose mass 
is very light, we have examined semi-relativistic kinematics 
(SR) as well as non-relativistic one (NR). 
In the NR case, non-relativistically approximated versions of the KSW-type 
potential are used. 

By using the CS-WF method, 
range parameters of our Gaussian-form potential are determined 
for both the NR and SR kinematics, so as to reproduce 
the value of $\bar{K}N$ scattering length obtained by Martin's analysis 
\cite{Exp:ADMartin}. 
The scattering amplitudes are investigated with the CS-WF method 
using the determined potentials. 
It is interesting that in the SR case 
we find  two sets of the range parameters which give different 
types of scattering amplitudes. 
One, denoted as SR-A, is considered to be the relativistic version 
of the non-relativistic solution, because 
the scattering amplitudes and the resonance pole have 
similar properties as those obtained in the NR kinematics. 
The other one, denoted as SR-B, is unique to the SR kinematics since 
it shows quite different properties from the NR case. 


In the $I=0$ sector, a resonance structure is seen 
below the $\bar{K}N$ threshold in the $I=0$ $\bar{K}N$ and 
$\pi\Sigma$ scattering amplitudes. It appears at $1405\sim1420$ MeV 
in the $\bar{K}N$ amplitude.  
It is found that the $\bar{K}N$ scattering amplitude 
near the $\bar{K}N$ threshold is well constrained by 
the $\bar{K}N$ scattering length, since its $f_\pi$ dependence is 
small and both kinematics give similar results in this region. 
However, far below the $\bar{K}N$ threshold the model dependence 
of scattering amplitudes becomes prominent; 
The NR and SR-A cases give qualitatively similar scattering amplitudes, 
which strongly depends on the $f_\pi$ value. 
The SR-B case gives rather different amplitudes from them, especially 
in the $\pi\Sigma$ amplitude. 
We consider that further data  
far below $\bar{K}N$ threshold, such as the $\pi\Sigma$ scattering length, are 
necessary to reduce such an uncertainty in the deep 
$\bar{K}N$ bound region, as pointed out in Ref. \cite{KN-pS:IHJ}. 

Similarly, the $I=1$ sector has also been investigated. 
When the potential of our model is constrained by 
the complex value of the $\bar{K}N$ scattering length with $I=1$, 
it is found that the scattering amplitude has a resonance structure 
slightly below the $\pi\Sigma$ threshold in the NR case, and that 
$\pi\Sigma$ scattering amplitude shows repulsive nature 
in the semi-relativistic cases. However, if the constraint condition 
for the potential is relaxed so that only the imaginary part of 
the $\bar{K}N$ scattering length is reproduced, such a resonance 
structure disappears in the NR case and the $\pi\Sigma$ scattering 
amplitude becomes attractive in the SR cases. 
We found that it is difficult to constrain our potential model 
with the real part of the $\bar{K}N$ scattering length. Under such a constraint condition, the imaginary part is obtained to be largely deviated from 
the value estimated by Martin.  

Compared with other studies, the $I=0$ scattering amplitudes obtained 
in our calculation are a little different.
In particular, near the $\pi\Sigma$ threshold 
the amplitudes in the present results have 
larger magnitude than those in Refs. \cite{KN-pS:IHJ} 
and \cite{ChU:HW}. 
One of reasons for this discrepancy is confirmed to be 
the small difference in the interaction kernel. 
As another reason, we consider that the Gaussian-form factor in 
our potential may cause such the enhancement of the scattering amplitudes.

Properties of the resonance pole in the $I=0$ sector corresponding to 
the $\Lambda(1405)$ have been studied with the usual ccCSM. 
For $f_\pi$ = 90 $\sim$ 120 MeV, resonance pole is found around 
\begin{eqnarray*}
(E_R, -\Gamma/2)=
\left\{
\begin{tabular}{cl}
$(\, 1418.5\pm1.5, \, -19.5\pm5.5\, )$ & {\rm in NR case}, \\
$(\, 1420.5\pm3, \, -24.5\pm2\, )$ & {\rm in SR-A case}, \\
$(\, 1419\pm1, \, -13\pm2\, )$ & {\rm in SR-B case}, 
\end{tabular}
\right.
\end{eqnarray*}
on the complex-energy plane. 
The real energy of the pole is well determined to be 1420 MeV, 
independently of models. The imaginary energy, namely the decay width, 
depends on the cases. 
The NR and the SR-A give large value of $\Gamma/2$, while 
it is small in the other semi-relativistic case SR-B. 
As for the $f_\pi$ dependence of the pole position,  
it is found that the NR and the SR-A have similar tendency, but 
that the SR-B shows different behavior that 
the pole energy is quite stable for the change of the $f_\pi$ value.
We have estimated the size of the resonance pole, evaluating  
a root-mean-square distance with a bi-orthogonal set of 
complex-scaled wave function.
This quantity is not the mean of the physical meson-baryon distance 
but is expected to give us a guide of it.
The obtained meson-baryon mean ``distance''
is $1.3 - i0.3$ fm for the NR case and $1.2 - i0.5$ fm for the two SR cases. 

As a result of the present study with the NR and SR kinematics,  
it is found that $\bar{K}N$ quantities ($\bar{K}N$ scattering amplitude, 
the real energy of $I=0$ pole state and its size) near the $\bar{K}N$ 
threshold are essentially the same in both the kinematics,  
when we constrain the model by the $\bar{K}N$ scattering length 
that is the quantity at the threshold. However, these two kinematics give 
largely different results on the quantities far 
below the $\bar{K}N$ threshold and those related to $\pi\Sigma$, 
where the relativistic effect becomes important. 

In other studies, two poles are reported in 
$I=0$ sector and are related to $\Lambda(1405)$. 
The pole discussed above is considered to be the higher pole of the two poles. 
As mentioned in the section~\ref{Sec:Resonace_I=0}, 
we found a signature of the lower pole around the complex energies of 
$(\sim 1360, -90 \sim -40)$
MeV in a non-relativistic case and $(1350\sim1390, -100\sim-30)$ MeV 
in a semi-relativistic case. However, we can't conclude 
that this is the lower pole, because 
its $\theta$ trajectory is somehow unstable in CSM. 
We consider that this is due to limitation of numerical accuracy of 
the CSM with finite number of the Gaussian base 
for the resonances involving large decay widths. 
The poles of broad resonances can be investigated 
by applying an analytic continuation in the coupling constant 
to the complex scaling method (ACCC+CSM) \cite{ACCC+CSM:Aoyama}. 
It is one of our future plans to carry out ACCC+CSM and 
clarify whether our potential leads to the double-pole structure or not.


Thus, we have a $\bar{K}N$-$\pi Y$ potential for both isospin channels, 
which is based on a chiral SU(3) theory and is a local Gaussian 
form in $r$-space. In our future plan, 
we will investigate the three-body system of $K^-pp$ 
($\bar{K}NN$-$\pi YN$ system with $J^\pi=0^-, \,T=1/2$) which is 
the most essential kaonic nuclei. 
Since the ccCSM can adequately deal with resonant states 
of a multi-channel system in principle and the CSM is known to 
be effective for the nuclear many-body study \cite{CSM-8He:Myo}, 
we expect that 
the ccCSM will give a pole position of the $K^-pp$ accurately 
and reveal its structure. We are interested in the role 
of $\pi YN$ three-body dynamics, 
because its implicit/explicit treatment may cause a large difference 
in the binding of $K^-pp$ as pointed out in Ref. \cite{piSigN:IS}. 
It is expected that the contribution of $\pi YN$ three-body 
dynamics will be investigated with the ccCSM. 

In our analysis, it is worthwhile to use the updated value of 
the $\bar{K}N$ scattering length based on the latest data, 
instead of the Martin's value obtained from old data. 
The SIDDHARTA group measured quite precisely the shift and width of the 
$1s$ atomic level energy of kaonic hydrogen atom \cite{KpX:SIDDHARTA}. 
With the $K^-p$ scattering length based on this data, 
$K^-n$ scattering length is estimated with the coupled-channel 
chiral dynamics \cite{ChU:IHW}. 
These values of $K^-p$ and $K^-n$ scattering lengths are available 
in our calculation. Furthermore, the SIDDHARTA group is planning 
to perform experiments on kaonic deuterium in the SIDDHARTA-2 experiment. 
These forthcoming experiment will constrain more strictly  
the $\bar{K}N$ scattering length for 
both isospin channels \cite{Exp:SIDDHARTA-2}, though it is pointed out 
that the analysis with the improved Deser-Trueman relation involves 
about 10\% error on the $K^-p$ scattering length \cite{K-p_anal:Cieply}.  


Furthermore, the ccCSM approach can be applied to other hadronic systems. 
For instance, with this method it seems interesting to investigate 
a few-body system involving $D$ meson in the charm sector, 
which is an analogous system with $\bar{K}$ meson 
in the strangeness sector \cite{DNN}.

\section*{Acknowledgment}
\label{}

One of authors (A. D.) is thankful to Prof. Kato for his advice on 
the scattering-amplitude calculation based on the complex scaling method, 
and to Dr. Hyodo for fruitful discussion on the $\bar{K}N$ system.
He is grateful also to Prof. Oset and Prof. Gal for their useful comments 
to improve our paper. 

\appendix

\section{Scattering amplitude for a multi-channel system in non-relativistic kinematics}
\label{app_scatt_m}

We consider a Schr\"odinger equation for a multi-channel system; 
$H | \Psi \rangle = E | \Psi \rangle$. 
Hamiltonian $H$ and total wave function $| \Psi \rangle$ are 
given as 
\begin{eqnarray}
H= \sum_c H^0_c |c\rangle \langle c| 
+ \sum_{c,c'} V_{cc'} |c\rangle \langle c'|, \quad \quad 
| \Psi \rangle = \sum_c | \Psi_c \rangle | c \rangle \label{def_Psi}, 
\end{eqnarray}
where $H^0_c$ is the kinetic-energy operator $\bm{p}_c^2/2\mu_c$ 
for the channel $c$ 
and $V_{cc'}$ is a potential between channels $c$ and $c'$.  
$| \Psi_c \rangle$ is a wave function of a channel $c$. 
Projecting the Schr\"odinger equation onto a channel $c$, we get 
such an equation for the $c$-channel wave function as 
\begin{eqnarray}
H^0_c | \Psi_c \rangle + \sum_{c,c'} V_{cc'} | \Psi_{c'} \rangle = E | \Psi_c 
\rangle. 
\end{eqnarray}
When the incident channel is $c_0$ and the incoming wave function is given 
as $| \phi_{c_0,\bm{k}_{c_0}} \rangle$, the above equation can be modified 
formally as 
\begin{eqnarray}
| \Psi^{(c_0)}_c \rangle = | \phi_{c_0,\bm{k}_{c_0}} \rangle \delta_{cc_0} 
+ \frac{1}{E-H^0_c +i\epsilon} \sum_{c'} V_{cc'} | \Psi^{(c_0)}_{c'} \rangle, 
\end{eqnarray}
taking into account the outgoing boundary condition appropriately. 
Here, we write the incoming channel $c_0$ on the wave function 
of each channel explicitly. The above equation is expressed in $r$-space as 
\begin{eqnarray}
&& \Psi^{(c_0)}_c (\bm{r}_c) = \phi_{c_0,\bm{k}_{c_0}} (\bm{r}_{c_0}) \delta_{cc_0} 
\nonumber\\
&& \quad \quad \quad \quad
+ \sum_{c'} \int d \bm{r}'_{c} \; G_c(\bm{r}_{c}, \bm{r}'_{c}; E)
\, \langle \bm{r}'_{c} | V_{cc'} | \Psi^{(c_0)}_{c'} \rangle,  
\end{eqnarray}
where $G_c(\bm{r}_{c}, \bm{r}'_{c}; E)$ is a Green function in 
the channel $c$. By performing complex integral as shown in many textbooks, 
it becomes 
\begin{eqnarray}
G_c(\bm{r}_{c}, \bm{r}'_{c}; E) &=& \langle \bm{r}_{c} | \frac{1}{E-H^0_c +i\epsilon}|\bm{r}_{c'} \rangle \; 
= \; -\frac{1}{4\pi} \frac{2\mu_c}{\hbar^2} 
\frac{e^{ik_c|\bm{r}_{c}- \bm{r}'_{c}|}}{|\bm{r}_{c}- \bm{r}'_{c}|} \label{Eq:Green_NR}\\
&\simeq& -\frac{1}{4\pi} \frac{2\mu_c}{\hbar^2} 
\frac{e^{ik_c r_c}}{r_c} e^{-i \bm{k}_c \cdot \bm{r}'_c} 
\quad \quad (|\bm{r}_c| >> |\bm{r}'_c|), 
\end{eqnarray}
where $\bm{k}_c=k_c \bm{r}_c/r_c$ and $r_c=|\bm{r}_c|$.
Then, the channel-$c$ wave function becomes 
\begin{eqnarray}
\Psi^{(c_0)}_c (\bm{r}_c) = \phi_{c_0} (\bm{r}_{c_0}) \delta_{cc_0} 
+ \frac{e^{ik_c r_c}}{r_c} \times \left(-\frac{1}{4\pi}\right) \frac{2\mu_c}{\hbar^2} 
\sum_{c'} 
\langle \phi_{c,\bm{k}_c} | V_{cc'} | \Psi^{(c_0)}_{c'} \rangle, 
\end{eqnarray}
using the fact that the function $\phi_{c,\bm{k}_c}(\bm{r}_c)$
 is $e^{i \bm{k}_c \cdot \bm{r}_c}$. 
Substituting the channel-$c$ wave function in Eq. (\ref{def_Psi}) 
with the above expression, the total wave function is given as 
\begin{eqnarray}
|\Psi \rangle = \phi_{c_0} (\bm{r}_{c_0}) |c_0\rangle
+ \sum_c \frac{e^{ik_c r_c}}{r_c} |c \rangle 
\cdot \left(-\frac{1}{4\pi}\right) \frac{2\mu_c}{\hbar^2} 
\sum_{c'} \langle \phi_{c,\bm{k}_c} | V_{cc'} | \Psi^{(c_0)}_{c'} \rangle. 
\end{eqnarray}
Therefore, the scattering amplitude between the initial channel $c_0$ 
and the final channel $c$ is 
\begin{eqnarray}
f_{cc_0} (\bm{k}_c, \bm{k}_{c_0}) =  -\frac{1}{4\pi} \frac{2\mu_c}{\hbar^2} 
\sum_{c'} \langle \phi_{c,\bm{k}_c} | V_{cc'} | \Psi^{(c_0)}_{c'} \rangle. \label{scatt_amp}
\end{eqnarray}

The wave functions 
$\phi_{c,\bm{k}_c}(\bm{r})=e^{i \bm{k}_c \cdot \bm{r}}$ and 
$\Psi^{(c_0)}_{c'}(\bm{r})$ are expanded on partial waves $l$ as 
\begin{eqnarray}
\phi_{c,\bm{k}_{c}} (\bm{r})  
&=& 4\pi \sum_{l,m} i^l \, \frac{\hat{j}_l (k_{c} r)}{k_{c} r} \;
Y^*_{lm} (\Omega_{\bm{k}_{c}}) Y_{lm} (\Omega_{\bm{r}}), \label{P-expand1} \\
\Psi^{(c_0)}_{c} (\bm{r}) 
&=& 4\pi \sum_{l,m} \, i^l \frac{\psi^{(c_0)}_{l,c} (r)}{k_{c_0} r} \; 
Y^*_{lm} (\Omega_{\bm{k}_{c_0}}) Y_{lm} (\Omega_{\bm{r}}) \label{P-expand2}
\end{eqnarray}
and  the scattering amplitude is expanded on the orbital angular 
momentum as 
\begin{eqnarray}
f_{cc_0} (\bm{k}_c, \bm{k}_{c_0})
&=& \sum_l (2l+1) P_l (\cos \theta_{(\bm{k}_c, \bm{k}_{c_0})}) 
f_{l,cc_0} (k_c, k_{c_0}). \label{P-expand3} 
\end{eqnarray}
Thus, expanding Eq. (\ref{scatt_amp}) for the partial waves using Eqs. 
(\ref{P-expand1})-(\ref{P-expand3}), 
the scattering amplitude for the partial wave $l$ is given as
\begin{eqnarray}
f_{l,cc_0} (k_c, k_{c_0}) &=& -\frac{2\mu_c}{\hbar^2 k_c k_{c_0}} 
\sum_{c'} \; \langle \hat{j}_l(k_c r) | \, V_{cc'} \, | \psi^{(c_0)}_{l,c'} (r) \rangle,
\end{eqnarray}
if the potential $V_{cc'}$ is a central potential.

\section{Scattering amplitude in semi-relativistic kinematics}
\label{app_scatt_SR}

We give the formula of the scattering amplitude 
for a semi-relativistic Hamiltonian. We derive it 
by using Green function in the same way as the non-relativistic case 
explained in~\ref{app_scatt_m}. In the semi-relativistic case, 
the kinetic term of Hamiltonian is 
\begin{eqnarray}
H^0_c = \sqrt{m_c^2+\hat{\bm p}_c^2}+\sqrt{M_c^2+\hat{\bm p}_c^2}. 
\end{eqnarray}
We consider the Green function for this $H^0_c$: 
\begin{eqnarray}
\left\{E- \left(\sqrt{m^2+\hat{\bm p}^2}+\sqrt{M^2+\hat{\bm p}^2}\right)\right\} G^{(+)}({\bm r}, {\bm r}'; E) = \delta^3({\bm r}- {\bm r}') . \label{eq:Green1}
\end{eqnarray}
Hereafter, we drop the channel suffix ``$c$'' because the following calculation is considered just in the channel $c$. 

By the Fourier transformation, $G^{(+)}({\bm r}, {\bm r}')$ and 
$\delta^3({\bm r}- {\bm r}')$ are expressed as  
\begin{eqnarray}
G^{(+)}({\bm r}, {\bm r}') &=& (2\pi)^{-3} \int d{\bm k} \, \tilde{G}({\bm k}) \, e^{i{\bm k}\cdot({\bm r}- {\bm r}')}, \label{eq:Green2}\\
\delta^3({\bm r}- {\bm r}') &=& (2\pi)^{-3} \int d{\bm k} \,  \, e^{i{\bm k}\cdot({\bm r}- {\bm r}')}.
\end{eqnarray}
Using these equations, Eq. (\ref{eq:Green1}) becomes 
\begin{eqnarray}
\left\{E- \left(\sqrt{m^2+\hbar^2{\bm k}^2}+\sqrt{M^2+\hbar^2{\bm k}^2}\right)\right\} \tilde{G}({\bm k}) = 1 . 
\end{eqnarray}
Then, the Green function in the momentum space, $\tilde{G}({\bm k})$, is  
\begin{eqnarray}
&& \tilde{G}({\bm k}) = 
\left\{E- \left(\omega({\bm k})+\Omega({\bm k})\right)\right\}^{-1}, \\
&& \quad \quad \quad \omega({\bm k}) = \sqrt{m^2+\hbar^2{\bm k}^2}, \quad
\Omega({\bm k}) = \sqrt{M^2+\hbar^2{\bm k}^2}. 
\end{eqnarray}
Inserting this equation into Eq. (\ref{eq:Green2}) and 
integrating on angular directions, we obtain 
\begin{eqnarray}
G^{(+)}({\bm r}, {\bm r}') &=& (2\pi)^{-2} (i|{\bm r}- {\bm r}'|)^{-1} \nn \\
&& \times
\int_{-\infty}^{\infty} kdk \, 
\left\{E- \left(\omega(k)+\Omega(k)\right)\right\}^{-1} \, e^{ik|{\bm r}- {\bm r}'|}. \label{eq:Green3}
\end{eqnarray}

We rationalize $\left\{E- \left(\omega(k)+\Omega(k)\right)\right\}^{-1}$ 
in terms of $\hbar^2 k^2$. After tiresome algebraic calculation, we obtain 
\begin{eqnarray}
&&\frac{1}{E- \left(\omega(k)+\Omega(k)\right)} = \frac{(Num.)}{(Den.)}, \\
&&(Num.) = \{E+(\omega+\Omega)\} \{E-(\omega-\Omega)\} \{E+(\omega-\Omega)\}, \\
&&(Den.) = -4 E^2 \hbar^2 \left[ k^2 - \frac{E^2 - (m-M)^2}{2E\hbar} 
\frac{E^2 - (m+M)^2}{2E\hbar}\right].
\end{eqnarray}
When we define a variable $k_0$ as  
\begin{eqnarray}
k_0 \equiv \sqrt{ \frac{E^2 - (m-M)^2}{2E\hbar} 
\frac{E^2 - (m+M)^2}{2E\hbar}}, \label{eq:def_K0}
\end{eqnarray}
then 
\begin{eqnarray}
&&\frac{1}{E- \left(\omega(k)+\Omega(k)\right)} = 
\frac{(Num.)}{-4 E^2 \hbar^2} \, \left( \frac{1}{k-k_0} + \frac{1}{k+k_0}\right) 
\frac{1}{2k}.
\end{eqnarray}
Inserting this into Eq. (\ref{eq:Green3}), 
\begin{eqnarray}
G^{(+)}({\bm r}, {\bm r}') &=& (2\pi)^{-2} (i|{\bm r}- {\bm r}'|)^{-1} 
(-8 E^2 \hbar^2)^{-1} \nn \\
&& \times
\int_{-\infty}^{\infty} dk \, 
(Num.) \, \left( \frac{1}{k-(k_0+i\epsilon)} + \frac{1}{k+(k_0+i\epsilon)}\right) \, \nn \\ &&\quad \quad \quad \quad\times e^{ik|{\bm r}- {\bm r}'|}. 
\end{eqnarray}
Here, we add $i\epsilon$ to $k_0$ in order to satisfy the outgoing 
boundary condition in the later calculation. 
By the principal integration, we pick up the pole $k=k_0+i\epsilon$. 
At the limit of $\epsilon \rightarrow 0$, then the Green function is  
\begin{eqnarray}
G^{(+)}({\bm r}, {\bm r}') 
&=& \frac{e^{ik_0|{\bm r}- {\bm r}'|}}{|{\bm r}- {\bm r}'|} \times 
(-16 \pi E^2 \hbar^2)^{-1} (Num.)_{k=k_0} \label{eq:Green4}
\end{eqnarray}

Finally, we consider the last term $(Num.)_{k=k_0}$. 
After bothersome calculation using the definition of 
$k_0$ (Eq. (\ref{eq:def_K0})), we obtain  
\begin{eqnarray}
\omega(k_0) = \frac{E^2 + m^2 - M^2}{2E}, \quad  
\Omega(k_0) = \frac{E^2 - m^2 + M^2}{2E}.   
\end{eqnarray}
Here we notice that 
$\omega(k_0) + \Omega(k_0) = E$.   
This is a trivial equation because it indicates energy conservation. 
Using this fact, we can simplify the term $(Num.)_{k=k_0}$ to be 
\begin{eqnarray}
(Num.)_{k=k_0} &=& 
\{E+(\omega+\Omega)\}_{k=k_0} \{E-(\omega-\Omega)\}_{k=k_0} \{E+(\omega-\Omega)\}_{k=k_0} \nn \\
&=& 8E \,\omega(k_0)\Omega(k_0) 
\end{eqnarray}
Inserting this result into Eq. (\ref{eq:Green4}), then
\begin{eqnarray}
G^{(+)}({\bm r}, {\bm r}') 
&=& \frac{e^{ik_0|{\bm r}- {\bm r}'|}}{|{\bm r}- {\bm r}'|} \times 
\left(-\frac{1}{2 \pi}\right) \frac{1}{\hbar^2} \,\frac{\omega(k_0)\Omega(k_0)}{E}. 
\end{eqnarray}
By remembering $\omega(k=k_0) + \Omega(k=k_0) = E$ and the definition of 
the reduced energy as 
$\tilde{\omega} \equiv \omega \Omega /(\omega + \Omega)$, 
we obtain the Green function for the channel $c$ in the semi-relativistic kinematics: 
\begin{eqnarray}
G_c^{(+)}({\bm r}_c, {\bm r}_c') = 
-\frac{1}{4 \pi} \frac{2\tilde{\omega}_c(k_c)}{\hbar^2} \, 
\frac{e^{ik_c|{\bm r}_c- {\bm r}_c'|}}{|{\bm r}_c- {\bm r}_c'|},  
\end{eqnarray}
where the suffix of channel $c$ is recovered and $k_c$ is $k_0$ 
given as Eq. (\ref{eq:def_K0}) in the channel $c$ . 
Compared with the non-relativistic case (Eq. (\ref{Eq:Green_NR})), 
it is found that the reduced mass $\mu_c$ is simply replaced with 
the reduced energy $\tilde{\omega}_c$. 
Since the remaining part of calculation is completely the same as 
that for the non-relativistic case, 
the scattering amplitude for the semi-relativistic 
kinematics is obtained to be 
\begin{eqnarray}
f_{l,cc_0} (k_c, k_{c_0}) &=& -\frac{2\tilde{\omega}_c}{\hbar^2 k_c k_{c_0}} 
\sum_{c'} \; \langle \hat{j}_l(k_c r) | \, V_{cc'} \, | \psi^{(c_0)}_{l,c'} (r) \rangle,  
\end{eqnarray}
by replacing $\mu_c$ in Eq. (\ref{scatt_amp}) with $\tilde{\omega}_c$.

\section{Matrix element of the kinetic term in the semi-relativistic case}
\label{Mat_Kin_SR}

In the semi-relativistic kinematics, the kinetic energy and mass terms in 
the Hamiltonian are of the form of $\sqrt{m^2+\hat{{\bm p}}^2}+
\sqrt{M^2+\hat{{\bm p}}^2}$ as shown in Eq. (\ref{HamSR}). 
In this article, a wave function is expanded 
in terms of partial waves; $\Psi({\bm r}) = \sum_{lm} r^{-1}\psi_l (r) Y_{lm} (\Omega)$. Furthermore its radial part is 
expanded with a Gaussian base, as Eq.(\ref{Eq:GaussE_CSM}) and Eq.(\ref{Eq:GaussE_CSWF}) explained in non-relativistic case; 
\begin{eqnarray}
\psi_l (r) / r = \sum_j C^l_j \,G^l_j (r) /r, \quad \quad
G^l_j (r) = N_l (b_j) r^{l+1} \exp[-r^2/2b_j^2],  
\end{eqnarray}
where $N_l (b_j)$ means a normalization factor.
We need to calculate the matrix element $\langle r^{-1}G^l_i Y_{lm} | \sqrt{m^2+\hat{{\bm p}}^2} | r^{-1}G^{l'}_j Y_{l'm'} \rangle$. 
This matrix element is calculated as follows: 
\begin{eqnarray}
&&\langle \frac{1}{r}G^l_i Y_{lm} | \sqrt{m^2+\hat{{\bm p}}^2} \;| \frac{1}{r}G^{l'}_j Y_{l'm'} \rangle \nonumber \\
 &=& \int d{\bm r} d{\bm r'} d{\bm q} d{\bm q'} 
\; \langle \frac{1}{r}G^l_i Y_{lm} | {\bm r} \rangle \langle {\bm r} | {\bm q} \rangle 
\langle {\bm q} |
\sqrt{m^2+\hat{{\bm p}}^2} 
| {\bm q'} \rangle \langle {\bm q'} | {\bm r'} \rangle \langle {\bm r'}
| \frac{1}{r}G^{l'}_j Y_{l'm'} \rangle \nonumber\\
 &=& \delta_{ll'} \delta_{mm'}\; \frac{2}{\pi} \int^\infty_0 dq q^2 \sqrt{m^2+q^2} \nonumber\\
&& \quad \quad \quad \quad \quad
\times \int^\infty_0  dr G^l_i(r) j_l(qr) 
\times \int^\infty_0  dr' G^l_j(r') j_l(qr'), 
\end{eqnarray}
where $\langle {\bm r} | {\bm q} \rangle = e^{i {\bm q}\cdot{\bm r}}$ and 
its expansion (Eq. (\ref{P-expand1})) is used. 
In case of $s$-wave ($l=0$) which we are considering in this article, 
the last integration for $r$ and $r'$ variables can be performed analytically. 
Finally, the above matrix element for a complex-scaled $\hat{{\bm p}}$ is expressed as 
\begin{eqnarray}
&&\langle \frac{1}{r}G^0_i Y_{00} | \sqrt{m^2+{(\hat{{\bm p}} e^{-i\theta}})^2} \;| \frac{1}{r}G^{0}_j Y_{00} \rangle \nonumber\\
&=&\frac{4}{\sqrt{\pi}} (b_i b_j)^{3/2} \, \int^\infty_0 dq q^2 
\sqrt{m^2+q^2e^{-2i\theta}} \,
\exp\left[-\frac{1}{2}(b^2_i+b^2_j)q^2\right] . \label{SR_Kin_mat}
\end{eqnarray}
The integration for the variable $q$ is carried out numerically. 
The matrix element of Eq. (\ref{SR_Kin_mat}) is used when we diagonalize 
the complex-scaled Hamiltonian to find resonance poles as explained in 
section~\ref{CSM:usual} and we 
calculate a complex-scaled wave function as shown in Eq. (\ref{Eq:CSwfnc}) 
to obtain scattering amplitudes. 

Note that in a semi-relativistic case any wave number $k$ 
is calculated from $\sqrt{m^2+{\hbar k}^2}+\sqrt{M^2+{\hbar k}^2}=E$ 
and that any reduced mass $\mu$ in a non-relativistic case is 
replaced with corresponding reduced energy 
$\omega=E_M E_B /(E_M + E_B)$ where 
$E_{M}$ and $E_{B}$ mean meson and baryon energies, respectively; 
$E_{M}=\sqrt{m^2+{\hbar k}^2}$ and $E_{B}=\sqrt{M^2+{\hbar k}^2}$.

\section{Detailed results of $f_\pi=90, 100, 120$ MeV cases} 
\label{app:other_fpi}

We show the results for the cases of $f_\pi=90, 100$ and 120 MeV. 
Tables~\ref{RangePara_100-120},~\ref{ResPole_100-120},  
\ref{RangePara_I=1_NR_SR-A_mfpi} and~\ref{RangePara_I=1_SR-B_mfpi} 
correspond 
to Tables~\ref{RangePara},~\ref{Poles_I=0},~\ref{RangePara_I=1_NR_SR-A} 
and~\ref{RangePara_I=1_SR-B} 
which are for the case of $f_\pi=110$ MeV, respectively.  

\begin{table}[h]
\caption{Range parameters for $I=0$ $\bar{K}N$-$\pi\Sigma$ system. 
All quantities are in unit of fm. 
$f_\pi=90, 100$ and $120$ MeV cases. 
Corresponding to Table~\ref{RangePara}. 
\label{RangePara_100-120}}
\begin{tabular*}{\textwidth}{ll@{\hspace{0.5cm}}|@{\hspace{0.5cm}} r@{\hspace{1cm}}r@{\hspace{1cm}}r@{\hspace{1cm}}r}
\hline
&Case  & NRv1 & NRv2 & SR-A & SR-B  \\
\hline
&Kinematics & \multicolumn{2}{c}{Non-rela.} & \multicolumn{2}{c}{Semi-rela.} \\
&Potential  & NRv1 & NRv2 & \multicolumn{2}{c}{KSW-type}   \\
\hline
$f_\pi=90$ &&&&\\
&$d_{\bar{K}N,\, \bar{K}N}$  & 0.576 & 0.574 & 0.635 & 0.487 \\
&$d_{\pi\Sigma,\, \pi\Sigma}$  & 0.725 & 0.751 & 0.830 & 0.457 \\ 
&Re $a_{\bar{K}N\,(I=0)}$ & $-1.700$ & $-1.700$ & $-1.702$& $-1.703$ \\
&Im $a_{\bar{K}N\,(I=0)}$ & 0.677 & 0.687 & 0.685 & 0.677 \\
\hline
$f_\pi=100$ &&&&\\
&$d_{\bar{K}N,\, \bar{K}N}$  &  0.503& 0.501 & 0.561 &  0.421\\
&$d_{\pi\Sigma,\, \pi\Sigma}$ &0.665  & 0.695 & 0.770 & 0.395 \\ 
&Re $a_{\bar{K}N\,(I=0)}$ & $-1.702$ & $-1.702$ & $-1.698$ & $-1.703$ \\
&Im $a_{\bar{K}N\,(I=0)}$ & 0.680 & 0.683 & 0.679 & 0.673 \\
\hline
$f_\pi=120$ &&&&\\
&$d_{\bar{K}N,\, \bar{K}N}$  & 0.386 & 0.384 & 0.446 & 0.327 \\
&$d_{\pi\Sigma,\, \pi\Sigma}$ & 0.549 & 0.581 & 0.659 & 0.310 \\ 
&Re $a_{\bar{K}N\,(I=0)}$ & $-1.699$ & $-1.700$ & $-1.697$ & $-1.690$ \\
&Im $a_{\bar{K}N\,(I=0)}$ & 0.674 & 0.669 & 0.672 & 0.669 \\
\hline
\end{tabular*}
\end{table}


\begin{table}[h]
\caption{Pole position of the $I=0$ $\bar{K}N$-$\pi\Sigma$ system 
and the meson-baryon distance in the pole state.
$f_\pi=90, 100$ and $120$ MeV cases. 
Corresponding to Table~\ref{Poles_I=0}. 
\label{ResPole_100-120}}
\begin{tabular*}{\textwidth}{l@{\hspace{0.3cm}}|@{\hspace{0.3cm}} r@{\hspace{0.5cm}}r@{\hspace{0.8cm}}r@{\hspace{0.5cm}}r}
\hline
Case & NRv1 & NRv2 & SR-A & SR-B  \\
\hline
Kinematics& \multicolumn{2}{c}{Non-rela.} & \multicolumn{2}{c}{Semi-rela.} \\
Potential & NRv1 & NRv2 & \multicolumn{2}{c}{KSW-type}   \\
\hline
$f_\pi=90$ &&&&\\
$E_R$ & 1419.8 & 1419.9 & 1423.6 &1419.0 \\
$\Gamma/2$ & 26.0 & 23.1 & 26.4 &14.4 \\
($B_{\bar{K}N}$) & (15.2) & (15.1) & (11.4) &(16.0) \\
$\sqrt{\langle r^2 \rangle}_{\bar{K}N+\pi\Sigma}$ &$1.20-0.28i$& $1.25-0.29i$& 
$1.20-0.42i$ &$1.21-0.49i$\\
\hline
$f_\pi=100$ &&&&\\
$E_R$ & 1417.3 & 1418.0 & 1421.5 &1419.6 \\
$\Gamma/2$ & 23.1 & 19.8 & 26.3 &13.2 \\
($B_{\bar{K}N}$) & (17.7) & (17.0) & (13.5) &(15.4) \\
$\sqrt{\langle r^2 \rangle}_{\bar{K}N+\pi\Sigma}$ &$1.31-0.29i$&$1.36-0.30i$& 
$1.22-0.43i$ &$1.21-0.49i$\\
\hline
$f_\pi=120$ &&&&\\
$E_R$ & 1416.9 & 1418.3 & 1417.5 &1418.9 \\
$\Gamma/2$ & 16.7 & 14.0 & 22.9 &11.7 \\
($B_{\bar{K}N}$) & (18.1) & (16.7) & (17.5) &(16.1) \\
$\sqrt{\langle r^2 \rangle}_{\bar{K}N+\pi\Sigma}$ &$1.38-0.38i$&$1.44-0.38i$& 
$1.20-0.49i$ &$1.16-0.42i$\\
\hline
\end{tabular*}
\end{table}

\begin{table}[h]
\caption{Range parameters for $I=1$ $\bar{K}N$-$\pi\Sigma$-$\pi\Lambda$ 
system with non-relativistic kinematics (NRv2, left three columns) and semi-relativistic kinematics (SR-A, right three columns). $f_\pi=90, 100, 120$ MeV cases. Corresponding to Tables~\ref{RangePara_I=1_NR_SR-A}. 
 \label{RangePara_I=1_NR_SR-A_mfpi}}
\begin{tabular*}{\textwidth}{l@{\hspace{0.5cm}}|@{\hspace{0.5cm}} r@{\hspace{0.7cm}}r@{\hspace{0.7cm}}r@{\hspace{1.2cm}}r@{\hspace{0.7cm}}r@{\hspace{0.7cm}}r}
\hline
Case  & \multicolumn{3}{c}{NRv2} & \multicolumn{3}{c}{SR-A}  \\
\hline
Kinematics & \multicolumn{3}{c}{Non-rela.} & \multicolumn{3}{c}{Semi-rela.} \\
Potential  & \multicolumn{3}{c}{NRv2} & \multicolumn{3}{c}{KSW-type}  \\
\hline
Condition  & (a) & (b) & (c) & (a) & (b) & (c) \\
\hline
$f_\pi=90$&&&&&&\\
$d_{\bar{K}N,\, \bar{K}N}$  & 0.574 & 0.574 & 0.574 & 0.635 & 0.635& 0.635\\
$d_{\pi\Sigma,\, \pi\Sigma}$  & 0.221 & 0.751 & 0.751 & 0.334 & 0.830& 0.830\\ 
$d_{\bar{K}N,\, \pi\Lambda}$  & 0.391 & 0.558 & 1.138 & 0.413 & 0.556& 0.942\\ 
Re $a_{\bar{K}N\,(I=1)}$ & 0.371 & 0.370 & 0.621 & 0.365 & 0.368& 0.624\\
Im $a_{\bar{K}N\,(I=1)}$ & 0.601 & 1.358 & 0.600 & 0.600 & 1.353& 0.600\\
\hline
$f_\pi=100$&&&&&&\\
$d_{\bar{K}N,\, \bar{K}N}$  & 0.501 & 0.501 & 0.501 & 0.561 & 0.561& 0.561\\
$d_{\pi\Sigma,\, \pi\Sigma}$  & 0.186 & 0.695 & 0.695 & 0.293 & 0.770& 0.770\\ 
$d_{\bar{K}N,\, \pi\Lambda}$  & 0.289 & 0.394 & 0.663 & 0.336 & 0.431& 0.631\\ 
Re $a_{\bar{K}N\,(I=1)}$ & 0.367 & 0.368 & 0.626 & 0.372 & 0.366& 0.629\\
Im $a_{\bar{K}N\,(I=1)}$ & 0.601 & 1.452 & 0.600 & 0.598 & 1.438& 0.600\\
\hline
$f_\pi=120$&&&&&&\\
$d_{\bar{K}N,\, \bar{K}N}$  & 0.384 & 0.384 & 0.384 & 0.446 & 0.446& 0.446\\
$d_{\pi\Sigma,\, \pi\Sigma}$  & 0.139 & 0.581 & 0.581 & 0.236 & 0.659& 0.659\\ 
$d_{\bar{K}N,\, \pi\Lambda}$  & 0.173 & 0.240 & 0.324 & 0.242 & 0.301& 0.370\\ 
Re $a_{\bar{K}N\,(I=1)}$ & 0.382 & 0.369 & 0.689 & 0.359 & 0.365& 0.691\\
Im $a_{\bar{K}N\,(I=1)}$ & 0.608 & 1.539 & 0.601 & 0.603 & 1.542& 0.600\\
\hline
\end{tabular*}
\end{table}

\begin{table}
\caption{Range parameters for the $I=1$ $\bar{K}N$-$\pi\Sigma$-$\pi\Lambda$ 
system with the semi-relativistic kinematics (SR-B). 
$f_\pi=90, 100, 120$ MeV cases. Corresponding to Table~\ref{RangePara_I=1_SR-B}. \label{RangePara_I=1_SR-B_mfpi}}
\begin{tabular*}{\textwidth}{l@{\hspace{0.5cm}}|@{\hspace{0.5cm}} r@{\hspace{0.6cm}}r@{\hspace{0.6cm}}r@{\hspace{0.5cm}}|@{\hspace{0.5cm}}r@{\hspace{0.6cm}}r@{\hspace{0.6cm}}r}
\hline
Case  & \multicolumn{6}{c}{SR-B}   \\
\hline
Kinematics & \multicolumn{6}{c}{Semi-rela.} \\
Potential  & \multicolumn{6}{c}{KSW-type}    \\
\hline
Condition  & (a) & (b) & (c) & (a)-R1 & (a)-R2 & (a)-R3 \\
\hline
$f_\pi=90$&&&&&& \\
$d_{\bar{K}N,\, \bar{K}N}$    & 0.487 & 0.487 & 0.487 & 0.487 & 0.487 & 0.487\\
$d_{\pi\Sigma,\, \pi\Sigma}$  & 0.313 & 0.457 & 0.457 & 0.930 & 0.960 & 1.340\\
$d_{\bar{K}N,\, \pi\Lambda}$  & 0.422 & 0.187 & 0.100 & 1.880 & 1.510 & 0.930\\
Re $a_{\bar{K}N\,(I=1)}$      & 0.374 & 0.072 & $-0.145$ & 0.966 & 0.928 & 0.869\\
Im $a_{\bar{K}N\,(I=1)}$      & 0.602 & 0.268 & 0.404 & 0.600 & 0.600 & 0.600\\
\hline
$f_\pi=100$&&&&&& \\
$d_{\bar{K}N,\, \bar{K}N}$    & 0.421 & 0.421 & 0.421 & 0.421 & 0.421 & 0.421\\
$d_{\pi\Sigma,\, \pi\Sigma}$  & 0.277 & 0.395 & 0.395 & 1.870 & 0.740 & 0.850\\
$d_{\bar{K}N,\, \pi\Lambda}$  & 0.335 & 0.174 & 0.100 & 0.610 & 1.440 & 0.910\\
Re $a_{\bar{K}N\,(I=1)}$      & 0.366 & 0.057 & $-0.153$ & 0.855 & 0.825 & 0.793\\
Im $a_{\bar{K}N\,(I=1)}$      & 0.601 & 0.189 & 0.351 & 0.599 & 0.600 & 0.599\\
\hline
$f_\pi=120$&&&&&& \\
$d_{\bar{K}N,\, \bar{K}N}$    & 0.327 & 0.327 & 0.327 & 0.327 & 0.327 & 0.327\\
$d_{\pi\Sigma,\, \pi\Sigma}$  & 0.223 & 0.310 & 0.310 & 1.400 & 0.810 & 0.520\\ 
$d_{\bar{K}N,\, \pi\Lambda}$  & 0.236 & 0.140 & 0.105 & 0.400 & 0.450 & 0.910\\
Re $a_{\bar{K}N\,(I=1)}$      & 0.379 & 0.123 & $-0.097$ & 0.859 & 0.799 & 0.698\\
Im $a_{\bar{K}N\,(I=1)}$      & 0.600 & 0.141 & 0.360 & 0.601 & 0.600 & 0.599\\
\hline
\end{tabular*}
\end{table}




\begin{thebibliography}{00}
\bibitem{GaussDiv} This is based on the original discussion between 
Dr. Hyodo and Prof. Yazaki.
\end{thebibliography}



\end{document}